\documentclass{aa}  

\usepackage{array}
\usepackage{graphicx}
\usepackage{multirow}
\usepackage{color}
\usepackage{hyperref}

\usepackage[capitalize]{cleveref}
\usepackage{txfonts}

\usepackage{hyperref}
\begin{document}

   \title{Effects of ultra-fast outflows on X-ray time lags in AGN}

   
   \titlerunning{Effects of ultra-fast outflows on X-ray time lags in AGN}
   \authorrunning{Yerong Xu et al.}
   \author{Yerong Xu
          \inst{1,2,3}\fnmsep\thanks{yerong.xu@inaf.it},
          Ciro Pinto
          \inst{1},
          Erin Kara\inst{4},
          Stefano Bianchi\inst{5},
          William Alston\inst{6},
          \and
          Francesco Tombesi\inst{7,8,9}
          }

   \institute{INAF - IASF Palermo, Via U. La Malfa 153, I-90146 Palermo, Italy
         \and
            Universit\`a degli Studi di Palermo, Dipartimento di Fisica e Chimica, via Archirafi 36, I-90123 Palermo, Italy
         \and
            Department of Astronomy \& Physics, Saint Mary's University, 923 Robie Street, Halifax, NS B3H 3C3, Canada
        \and
        MIT Kavli Institute for Astrophysics and Space Research, Massachusetts Institute of Technology, Cambridge, MA 02139, USA 
        \and
        Dipartimento di Matematica e Fisica, Università degli Studi Roma Tre, via della Vasca Navale 84, I-00146 Roma, Italy 
        \and
        Centre for Astrophysics Research, University of Hertfordshire, College Lane, Hatfield AL10 9AB, UK  
        \and
        Department of Physics, University of Rome Tor Vergata, Via della Ricerca Scientifica 1, I-00133 Rome, Italy
        \and
        INAF – Astronomical Observatory of Rome, Via Frascati 33, 00040 Monte Porzio Catone, Italy
        \and
        INFN - Rome Tor Vergata, Via della Ricerca Scientifica 1, 00133 Rome, Italy
             }

   \date{Received XXX; accepted XXX}

 
  \abstract
   {The time lag between soft (e.g. $0.3\mbox{--}1$\,keV) and hard (e.g. $1\mbox{--}4$\,keV) X-ray photons has been observed in many active galactic nuclei (AGN) and can reveal the accretion process and geometry around supermassive black holes (SMBHs). High-frequency Fe K and soft lags are considered to originate from the light-travel distances between the corona and the accretion disk, while the propagation of the inward mass accretion fluctuation usually explains the low-frequency hard lags. Ultra-fast outflows (UFOs), with a velocity range of $\sim0.03\mbox{--}0.3$c, have also been discovered in numerous AGN and are believed to be launched from the inner accretion disk. However, it remains unclear whether UFOs can affect the X-ray time lags.
   }
   {As a pilot work, we aim to investigate the potential influence of UFOs on X-ray time lags of AGN in a small sample.}
   {By performing the UFO-resolved Fourier spectral timing analysis of archival \textit{XMM-Newton} observations of three AGN with transient UFOs: PG 1448+273, IRAS 13224-3809, and PG 1211+143, we compare their X-ray timing products, such as lag-frequency and lag-energy spectra, of observations with and without UFO obscuration.}
   {Our results find that in each AGN, low-frequency hard lags become weak or even disappear when they are accompanied by UFOs. This change is confirmed by the Monte Carlo simulations at a confidence level of at least $2.7\sigma$. In the high-frequency domain, soft lags remain unchanged while the Fe K reverberation lags tentatively disappear. The comparison between timing products of low- and high-flux observations on another three AGN without UFOs (Ark 564, NGC 7469, and Mrk 335) suggests that the disappearance of low-frequency hard lags is likely related to the emergence of UFOs, not necessarily related to the source flux. }
   {The presence of UFOs can affect X-ray time lags of AGN by suppressing the low-frequency hard lags, which can be explained by an additional time delay introduced by UFOs or disk accretion energy, which should transferred to heat the corona, carried away by UFOs.}
   \keywords{Black hole physics -- X-rays: galaxies -- Galaxies: Seyfert}
   \maketitle
%

\section{Introduction}\label{sec:intro}

Many supermassive black holes (SMBHs) in the center of active galactic nuclei (AGN) accrete materials through a geometrically thin and optically thick disk \citep{1973Shakura}. The X-ray spectra of unobscured AGN have three primary broadband components: a power-law-like continuum originating from the inverse Compton scattering of the thermal (UV) disk photons in the central hot plasma \citep[`hot corona',][]{1980Sunyaev,1993Haardt}, a reflection component from the X-ray reprocessing of coronal emission in the accretion disk \citep{1989Fabian,1991George,2010Garc} including a wealth of fluorescent lines \citep[in particular a Fe K line at $\sim6.4$\,keV,][]{1991George,2005Ross} and a Compton hump ($>$10\,keV), and a blackbody-shape-like soft excess whose nature is still under debate, probably originating from a warm Comptonization \citep[e.g.,][]{2012Done,2018Petrucci,2020Petrucci,2024Ballantyne} or relativistically blurred reflection \citep[e.g.,][]{2019Garc,2021XuSE}, or the combination of two scenarios \citep{2018Porquet,2021Porquet,2022Xiang}. 

X-ray variability studies of unobscured AGN have shown that the $0.3\mbox{--}1$\,keV band (referred to as the soft band), dominated by soft excess, and the Fe K band ($5\mbox{--}7$\,keV), dominated by reflection, lag behind the continuum-dominated $1\mbox{--}4$\,keV band (referred to as the hard band) in the high-frequency domain \citep[e.g.,][]{2013DeMarco,2014Uttley,2016Kara}. Despite the complicated soft excess origin (perhaps due to the reflection), the Fe K lag explicitly suggests light-travel distances between the corona and the accretion disk at a few gravitational radii, called the X-ray reverberation lags \citep[e.g.,][]{2009Fabian,2010Zoghbi,2016Kara}. On the other hand, the low-frequency variability of unobscured AGN commonly shows that the hard X-rays lag behind the soft emission with a time delay increasing steadily with energy. This low-frequency hard lag has been observed for several decades \citep[e.g.,][]{2001Papadakis,2013Kara564335,2013Kara1h0707,2016Kara}. The timescales of the low-frequency X-ray lags are orders of magnitude longer than the viscous timescale of the inner accretion flow. The origin of this lag is still not well understood, though in the prevailing model is the multiplicative mass accretion fluctuation generated at large radii propagating inward to the inner accretion disk \citep{1997Lyubarskii,2001Kotov,2006Arevalo,2023Uttley}. These time lags provide critical insights into the geometry and dynamics of the regions very close to the black hole, thus helping to map out the configuration of the inner accretion environment.

Besides accreting a large amount of gas, AGN can release part of the accumulated energy via the ejection of outflows \citep[for review see][]{2012Fabian,2021Laha}, some of which may have sufficient kinetic power ($L_\mathrm{kin}>0.5\mbox{--}5\%L_\mathrm{Edd}$, $L_\mathrm{Edd}$ is the Eddington luminosity) to impact the evolution of the host galaxy and growth of SMBHs \citep[e.g.,][]{2005DiMatteo,2010Hopkins}. Ultra-fast outflows (UFOs) are an extreme subclass of AGN winds with mildly relativistic speeds ($>$10000\,km/s), which are commonly discovered by strongly blueshifted absorption lines in X-ray spectra \citep{2002Chartas,2003Chartas,2003Pounds,2010Tombesi,2012Patrick,2013Gofford,2021Chartas,2023Matzeu,2024Yamada,2024Xu}. The ultra-fast velocity suggests that UFOs are launched from the accretion disk within a few hundred gravitational radii close to the central SMBH \citep[e.g.,][]{2013Tombesi}, the same region probed by the X-ray spectral-timing analysis. Therefore, UFOs are expected to leave imprints on the X-ray variability of AGN. 

The variability study of UFOs has shown that UFOs produce characteristic positive `spikes' of enhanced variability in variance spectra, which exhibit the energy dependence of variability \citep{2017ParkerPCA,2018ParkerPCA,2020Igo,2021Harer,2021Parker}. These features result from the UFO response to the continuum, where absorption lines from the UFO disappear as the X-ray flux rises \citep{2017Parker,2018Pinto}, leading to a higher variability amplitude in energy bands affected by absorption than others dominated by the continuum. 

The contribution of UFOs to the X-ray reverberation lags in the high-frequency domain was studied by \citet{2018Mizumoto,2019Mizumoto}, showing that the high-frequency Fe K lag may arise from the scattering in a highly ionized disk wind instead of the reflection onto the accretion disk. This alternative explanation could also reproduce the observed Fe K lags in 1H 0707-495 and Ark 564. However, the impact of UFOs on low-frequency hard lags in AGN has not yet been investigated well. A transient UFO event was once observed by a large \textit{XMM-Newton} campaign in PG 1211+143 \citep{2016Pounds,2016Pounds2UFO}, wherein \citet{2018Lobban} noticed the low-frequency lag weakened in the presence of the UFO, which was explained by a change or disruption in the inner accretion flow before the re-emergence or an additional soft lag introduced by other physical processes. 

We would like to explore: Is the co-existence of UFOs and weak low-frequency hard lags common? If yes, are they coincidental or causally related? To answer these questions, we revisit the archival \textit{XMM-Newton} dataset and select two variable Type 1 AGN with transient UFOs, i.e. PG 1448+273 and IRAS 13224-3809 (hereafter PG 1448 and IRAS 13224) to perform the X-ray lag analysis. The data of PG 1211+143 are also re-analyzed for reproducing results in \citet{2018Lobban}. The choice of \textit{XMM-Newton} is because it has a higher orbit and thus longer orbital period ($\sim48$h) than other X-ray satellites, enabling the probe for the long-timescale variability of AGN. The targets are required to be Type 1 AGN to avoid influence from the large-scale obscuration \citep[i.e., torus,][]{1988Krolik,2013Netzer} on X-ray variability and have at least two \textit{XMM-Newton} observations with net exposures of $>30$\,ks with and without UFOs separately for comparison.

PG 1448 is a luminous \citep[$L_\mathrm{bol}\sim10^{45.24}\,\mathrm{erg/s}$,][]{2020Rakshit}, nearby ($z=0.0645$), narrow line Seyfert 1 (NLS1) galaxy, hosting a supermassive BH of $M_\mathrm{BH}\sim10^{7}M_\odot$ \citep{2006Vestergaard}, indicating its accretion rate is close to the Eddington limit. PG 1448 was observed by \textit{XMM-Newton} in 2017 ($\sim75$\,ks net EPIC-pn exposure) at a low-flux state ($F_\mathrm{2\mbox{--}10\,keV}=1.3\times10^{-12}\,\mathrm{erg/s/cm^2}$) with the presence of a UFO ($\sim0.09c$), measured by the simultaneous detection of blueshifted soft X-ray and Fe K absorption lines \citep{2020Kosec,2021Laurenti}. Recently, PG 1448 was observed by \textit{NuSTAR} for 130\,ks net exposure (250\,ks duration) joint with \textit{XMM-Newton} for $\sim60$\,ks net exposure. The observation was performed at a bright state ($F_\mathrm{2\mbox{--}10\,keV}=4.8\times10^{-12}\,\mathrm{erg/s/cm^2}$), revealing strong X-ray variability with a generally featureless X-ray spectrum. Although the last 60\,ks of the \textit{NuSTAR} observation (without a simultaneous \textit{XMM-Newton} exposure) exhibited a flux decline and the emergence of a UFO, no UFO was detected during the \textit{XMM-Newton} observation \citep{2023Reeves}. 

IRAS 13224 is a nearby \citep[$z=0.066$,][]{1991Allen}, variable \citep[e.g.,][]{2012Ponti} NLS1 galaxy, hosting a supermassive black hole of $M_\mathrm{BH}\sim2\times10^{6}M_\odot$ \citep{2005Zhou,2020Alston}. This target was extensively observed by \textit{XMM-Newton} for total $\sim2$Ms, including a $\sim1.5$Ms campaign in 2016. UFOs ($\sim0.236c$) found in both soft and hard X-ray bands of IRAS 13224 show a strong dependence on the source flux, where the strength of the absorption decreases as the flux increases, explained by the over-ionization of plasma \citep{2017Parker,2018Pinto}. High-frequency soft lag and low-frequency hard lag were significantly detected \citep[e.g.,][]{2010Ponti,2013Karairas13324,2022Hancock}, where the reverberation lag is positively correlated with luminosity, indicating a variable corona height \citep{2020Alston}.

PG 1211 is a nearby \citep[$z=0.0809$,][]{1996Marziani} and variable \citep[e.g.,][]{2012Ponti} NLS1 galaxy/quasar, hosting a black hole of $M_\mathrm{BH}\sim1.5\times10^{8}M_\odot$ \citep{2004Peterson}. It is bright in both the optical and X-ray bands with an X-ray luminosity of the order of $\sim10^{44}\,$erg/s \citep[$L_\mathrm{bol}\sim5\times10^{45}$\,erg/s,][]{2003Pounds}. The source is the archetypal example of an AGN exhibiting a UFO in both UV \citep{2018Kriss} and X-ray \citep[e.g.,][]{2003Pounds,2009Pounds,2016Pounds,2018Danehkar,2018Reeves,2023Pounds} bands at velocities between $\sim0.066\mbox{--}0.2c$. The large \textit{XMM-Newton} campaign in 2014 ($\sim630$, ks) observed a complete UFO obscuration event from its emergence to its disappearance. The high-frequency reverberation lag and variable low-frequency hard lag were discovered from the same observations \citep{2016Lobban,2018Lobban}.

The paper is organized as follows: Observations and data reduction used in this analysis are described in Section \ref{sec:reduction}. Details of our Fourier analysis and results are shown in Section \ref{sec:results}. Finally, we discuss the results and provide our conclusions in Section \ref{sec:discussion} and Section \ref{sec:conclusion}, respectively.

\section{Data Reduction}\label{sec:reduction}

The basic information about analyzed \textit{XMM-Newton} observations of three AGN is shown in Table \ref{tab:obs}. The observations are classified as those with or without UFO according to the literature \citep{2017Parker,2018Pinto,2018Lobban,2023Reeves}. To increase the signal-to-noise ratio (S/N), we stack observations within the same class for our analysis.

\begin{table*}
\centering
\caption{\textit{XMM-Newton} observations of three AGN analyzed in this work. }
\begin{tabular}{lcccc}
\hline
\hline
Source & Label & Obs. ID & Exposure & Count Rates\\
  &   &  & (ks) & (cts/s) \\
\hline
\multirow{2}{*}{PG 1448+273} & low-flux w/ UFO & 0781430101 & 76 & 2.3 \\
            & high-flux w/o UFO & 0890600101   & 58 & 11.3 \\
\hline
\multirow{2}{*}{IRAS 13224-3809} & low-flux w/ UFO & 0673580301, 0780561501, 0780561701 & 82, 127, 122 & 1.2, 1.7, 2.0\\
            & high-flux w/o UFO & 0780561601, 0792180401, 0792180601  & 124, 115, 118& 4.0, 5.3, 5.2\\ 
\hline
\multirow{2}{*}{PG 1211+143} & low-flux w/ UFO & 0745110201, 0745110301$^\star$, 0745110401, 0745110501 & 89, 75, 87, 51  & 3.1, 3.9, 4.6, 5.8\\
            & high-flux w/o UFO & 0745110101, 0745110601, 0745110701  & 78, 86, 89 & 4.7, 5.7, 4.4\\ 
\hline
\end{tabular}
\tablefoottext{$\star$}{The source spectrum is extracted from an annulus region with an inner radius of 5 arcsec and an outer radius of 30 arcsec. }
\vspace{-2.5mm}
\tablefoot{
The second column shows the label for stacked observations at two states. The observation IDs are shown in the third column. The corresponding net exposure time and count rate of EPIC-pn are listed in the fourth and fifth columns, respectively.
}
\label{tab:obs}
\vspace{-3mm}
\end{table*}
For this analysis, we use the data from the \textit{XMM-Newton} EPIC-pn camera because of its high effective area and fast readout \citep{2001Jansen,2001Struder}. Data are reduced following standard threads with the \textit{XMM-Newton} Science Analysis System (SAS v20.0.0) and calibration files available by February 2023. We reduce the EPIC-pn data using \textsc{epproc} and filter the time intervals affected by the background solar flares, which show the count rates larger than 1 counts/sec in the $10\mbox{--}12$\,keV. This filter is higher than the regular threshold (0.5 count/sec),  reducing interruptions in the extracted light curves caused by background flares, so that we can obtain a long continuous light curve to probe the low-frequency domain. Our results estimated the effect of the introduced background flares to be negligible by comparing results from this filter and those from the stricter filter with interpolated and Poisson-noise-added gaps. The net EPIC-pn exposure time and count rate are listed in Table \ref{tab:obs}. Meanwhile, for flares of duration $<500$\,s, the source light curve is cut out and interpolated between the flare gap by adding Poisson noise using the mean of neighboring points. The interpolation fraction is typically $<1\%$. For gaps longer than $500$\,s the data are treated as separate segments. 
The source and background regions are selected by a circular region with a radius of 30 arcsec centered on and offset but near the source, respectively. The observation affected by the pile-up effect is marked by an asterisk in Table \ref{tab:obs}, and the corresponding source region is replaced by an annulus with an inner radius of 5 arcsec and an outer radius of 30 arcsec. The background-subtracted lightcurves of AGN are extracted by the \textsc{epiclccorr} package and shown in the top panels of Figure \ref{fig:lc_pds}.

\section{Results}\label{sec:results}

\begin{figure*}[htbp]
    \centering
	\includegraphics[width=0.66\columnwidth]{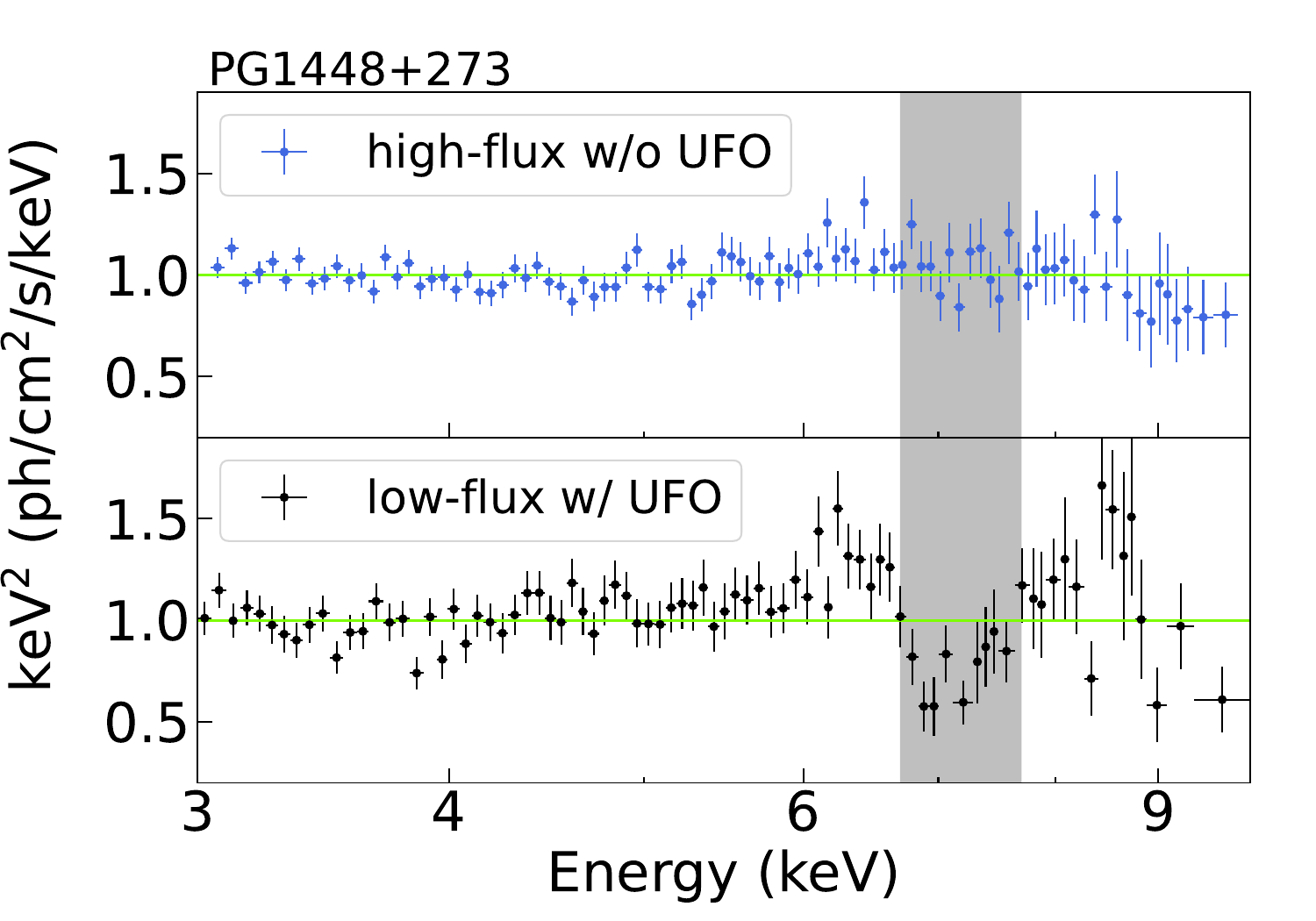}
	\includegraphics[width=0.66\columnwidth]{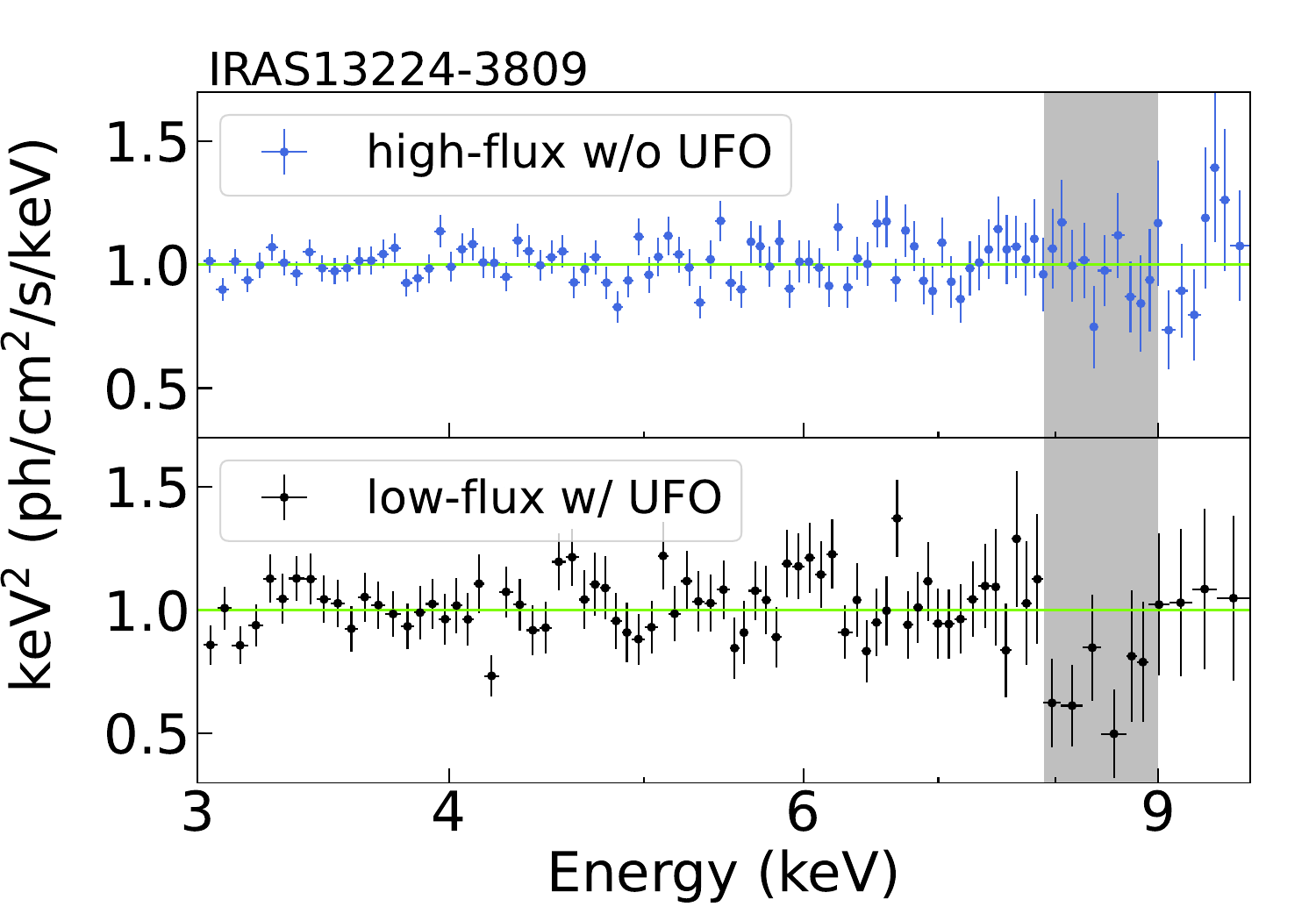}
 	\includegraphics[width=0.66\columnwidth]{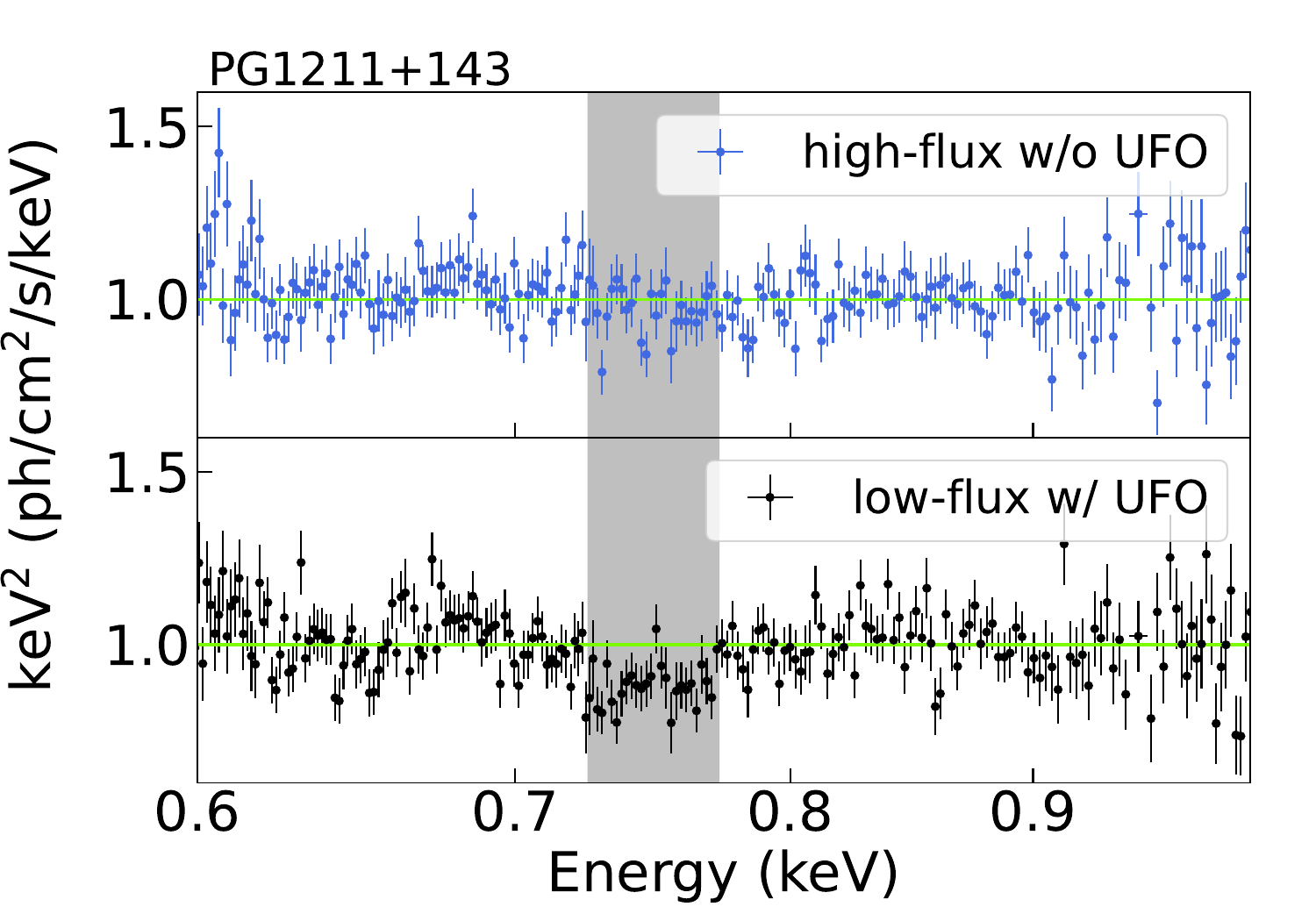}
    \caption{The data-to-model ratios of stacked EPIC (PG 1448 and IRAS 13224) and RGS (PG 1211) spectra to the best-fit broadband continuum model (see details in Section \ref{sec:results}). The `high-flux w/o UFO' and `low-flux w/ UFO' results are marked by \textit{royal blue} and \textit{black} points, respectively. The energy band distinctly affected by the UFO is marked by the \textit{grey} region.}
    \label{fig:ratios}
\end{figure*}

Observations within the same class (i.e. with or without UFOs) are stacked, resulting in two stacked observations for each target, although the classification of IRAS 13224 observations is not explicit in the literature. However, given the flux-dependent UFO detection in IRAS 13224, UFOs are expected to be significantly detected in faint states and become negligible in bright states. Therefore, the three brightest observations are classified as those without UFOs and the three faintest exposures are regarded as observations with UFOs. The presence/absence of UFOs in stacked spectra labeled by `low/high-flux w/wo UFO' are illustrated by the data-to-model ratios to the best-fit broadband continuum model (shown in Figure \ref{fig:ratios}), also justifying our expectation for UFOs in IRAS 13224. The continuum model consists of an absorbed power law, a soft excess, and a reflection component. The unfolded stacked EPIC-pn spectra with respect to a power law with an index of $\Gamma=2$ of each AGN are shown in the first column of Figure \ref{fig:lags}. In PG 1448 and IRAS 13224, the difference between stacked spectra w/ and w/o UFOs comes from both the continuum shape and the UFO absorption features ($>7$\,keV), while that difference in PG 1211 only occurs in the soft X-ray band, mainly dominated by the transient UFO.

\subsection{Lag-frequency spectra}\label{subsec:lag-frequency}

We use the package \texttt{pylag}\footnote{https://github.com/wilkinsdr/pyLag} \citep{2019Wilkins} to compute the X-ray time-lag spectrum \citep[e.g.,][]{2014Uttley}. Briefly, we perform the Fourier transform of the light curves in two different energy bands and multiply the transform from the soft band and the complex conjugate of that from the hard band, called the cross-spectrum. In this paper, we take the conventional $0.3\mbox{--}1.0$\,keV as the soft band and $1\mbox{--}4$\,keV as the hard band, except for the $0.7\mbox{--}1.5$\,keV and $2\mbox{--}10$\,keV bands for PG 1211 for the consistency with the literature \citep{2018Lobban}. The segment lengths are typically 50\,ks for PG 1448, and 70\,ks for IRAS 13224 and PG 1211, respectively, so that we can probe the lowest frequencies possible. The minimum frequency is defined by the inverse of the segment length, and the maximum frequency is set by the frequency at which the power spectral density (PSD, see details in Appendix \ref{app:sec:transient}) becomes dominated by Poisson noise. The argument of the
cross-spectrum is the phase lag between the two light curves. The time lag is $\tau(f)=\phi(f)/2\pi f$, where $\phi(f)$ is the frequency-dependent phase lag. The uncertainty of the time lag depends on the coherence of two light curves (see details in \citet{2014Uttley}). We bin the lag-frequency spectrum into equal logarithmic frequency bins and the number of frequency bins depends on the quality of the data (i.e. more bins for higher-quality data). 

\begin{figure*}[htbp]
    \centering
	\includegraphics[width=0.5\columnwidth]{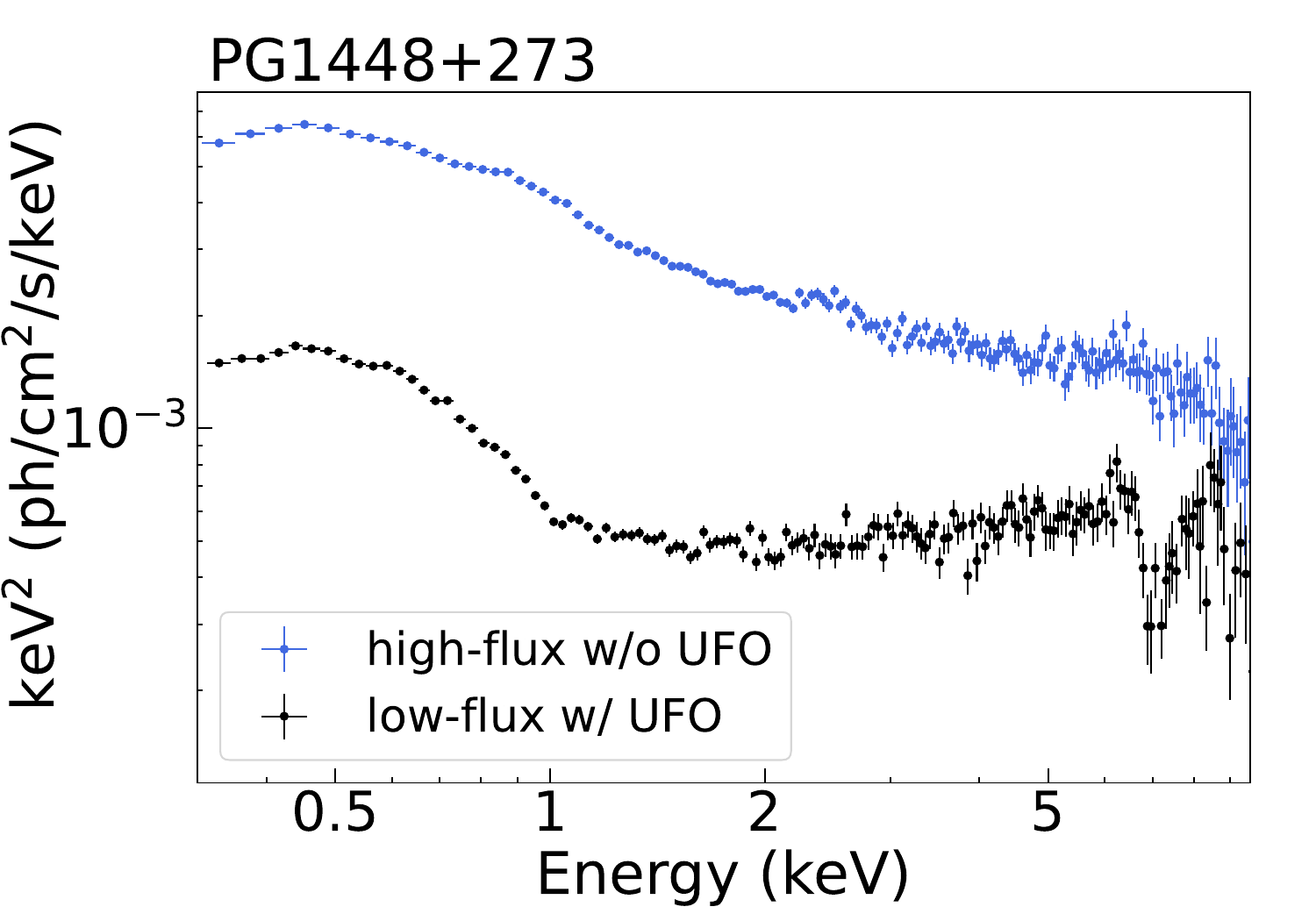}
	\includegraphics[width=0.5\columnwidth]{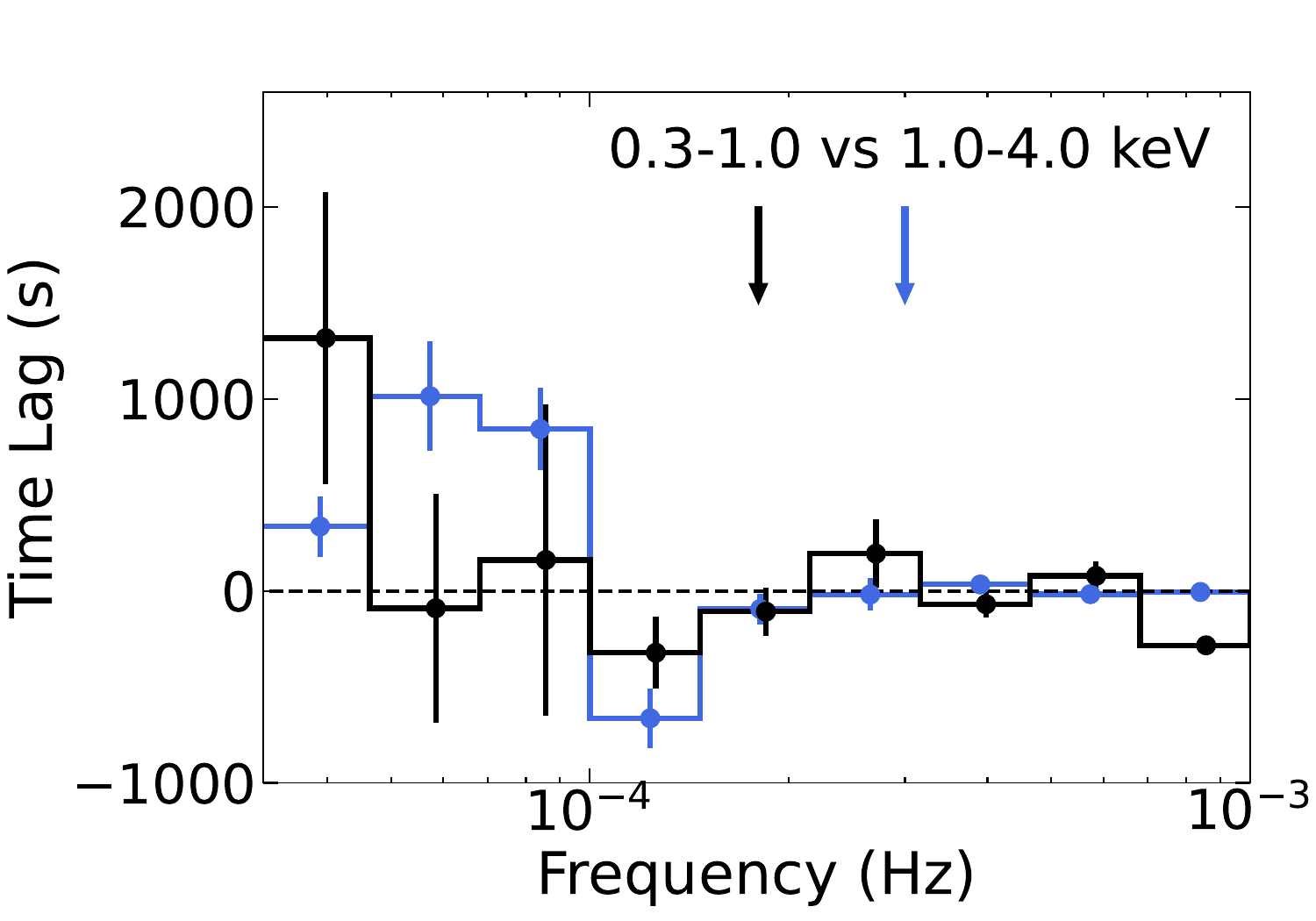}
	\includegraphics[width=0.5\columnwidth]{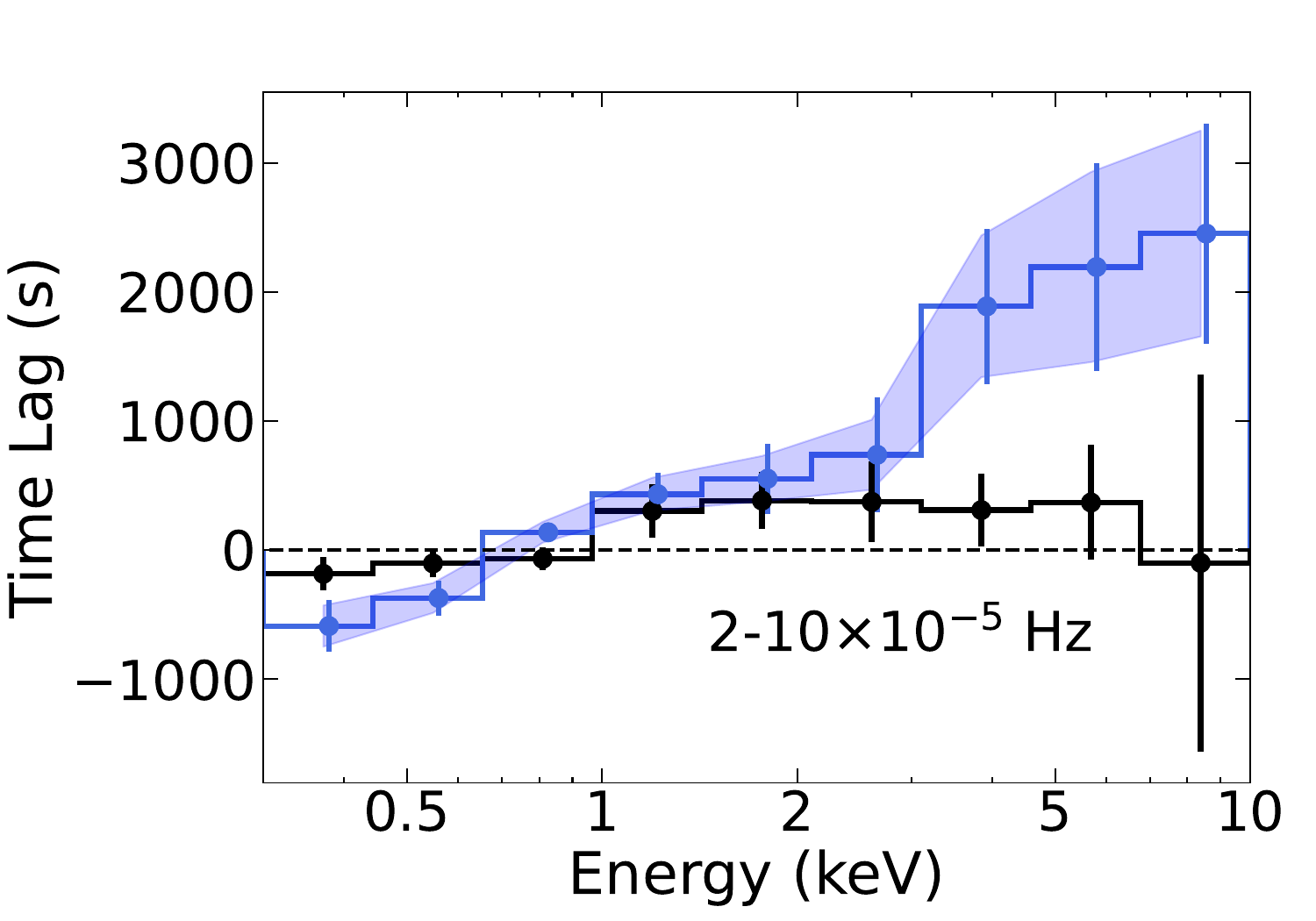}
	\includegraphics[width=0.5\columnwidth]{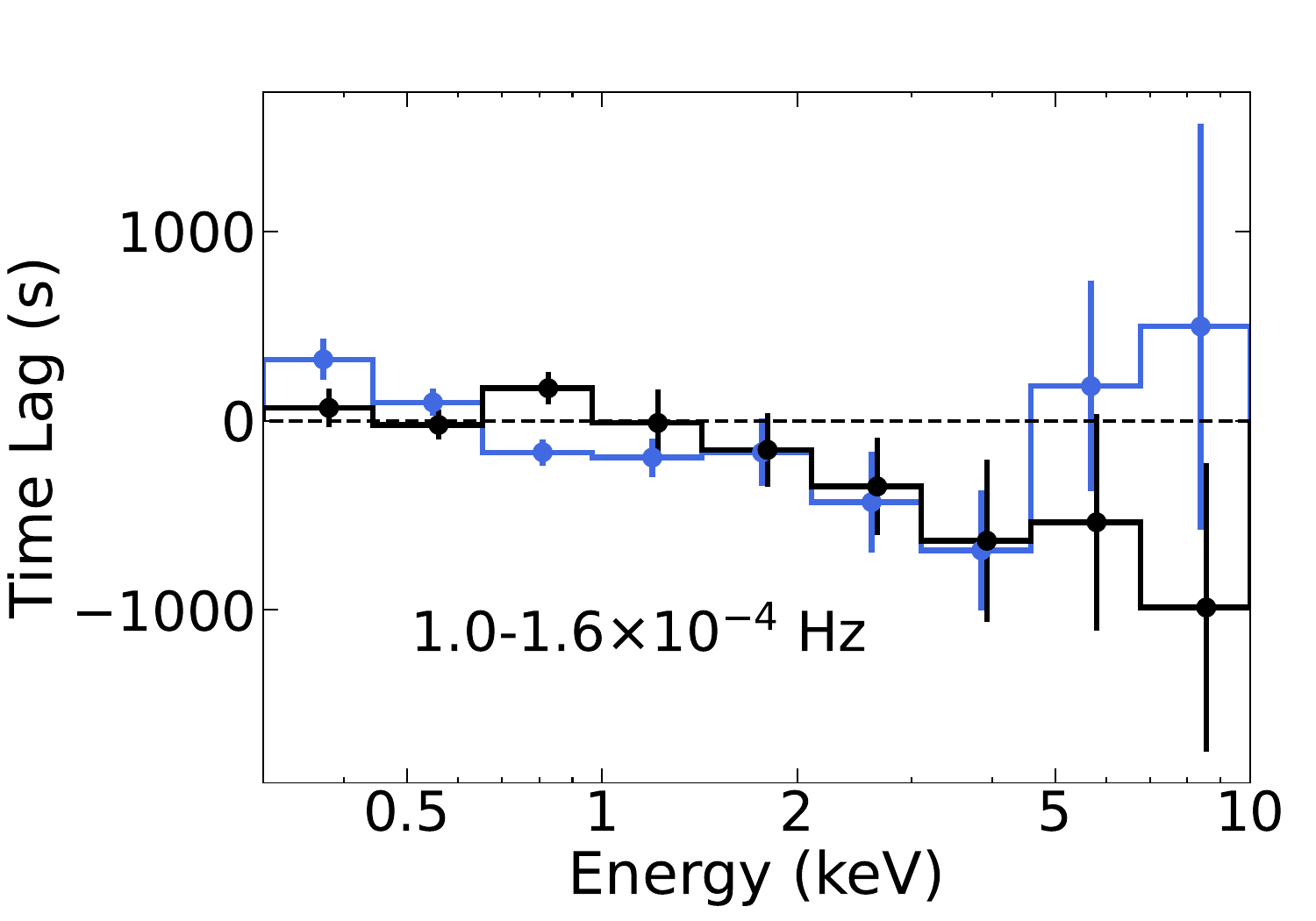}
	\includegraphics[width=0.5\columnwidth]{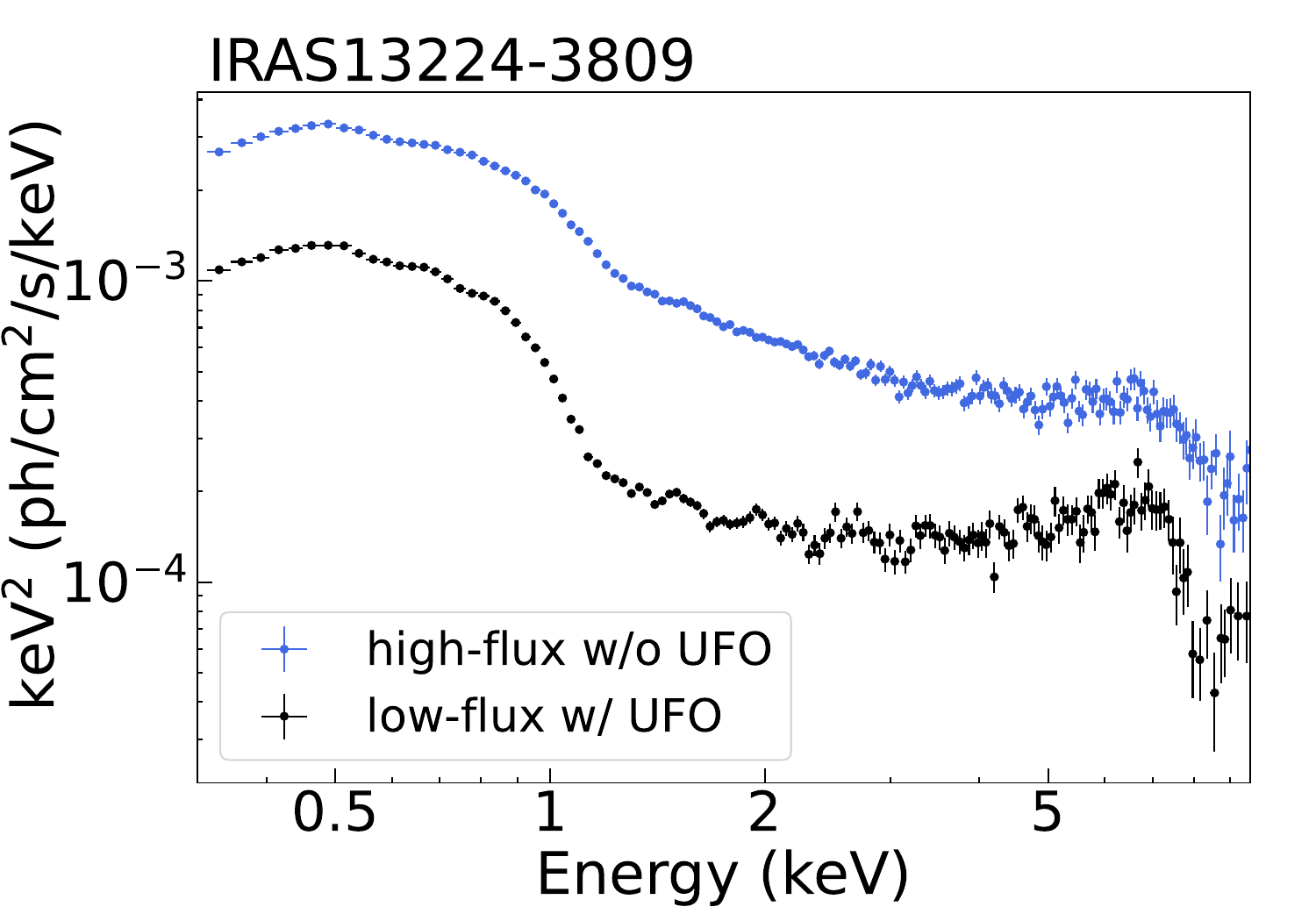}
 	\includegraphics[width=0.5\columnwidth]{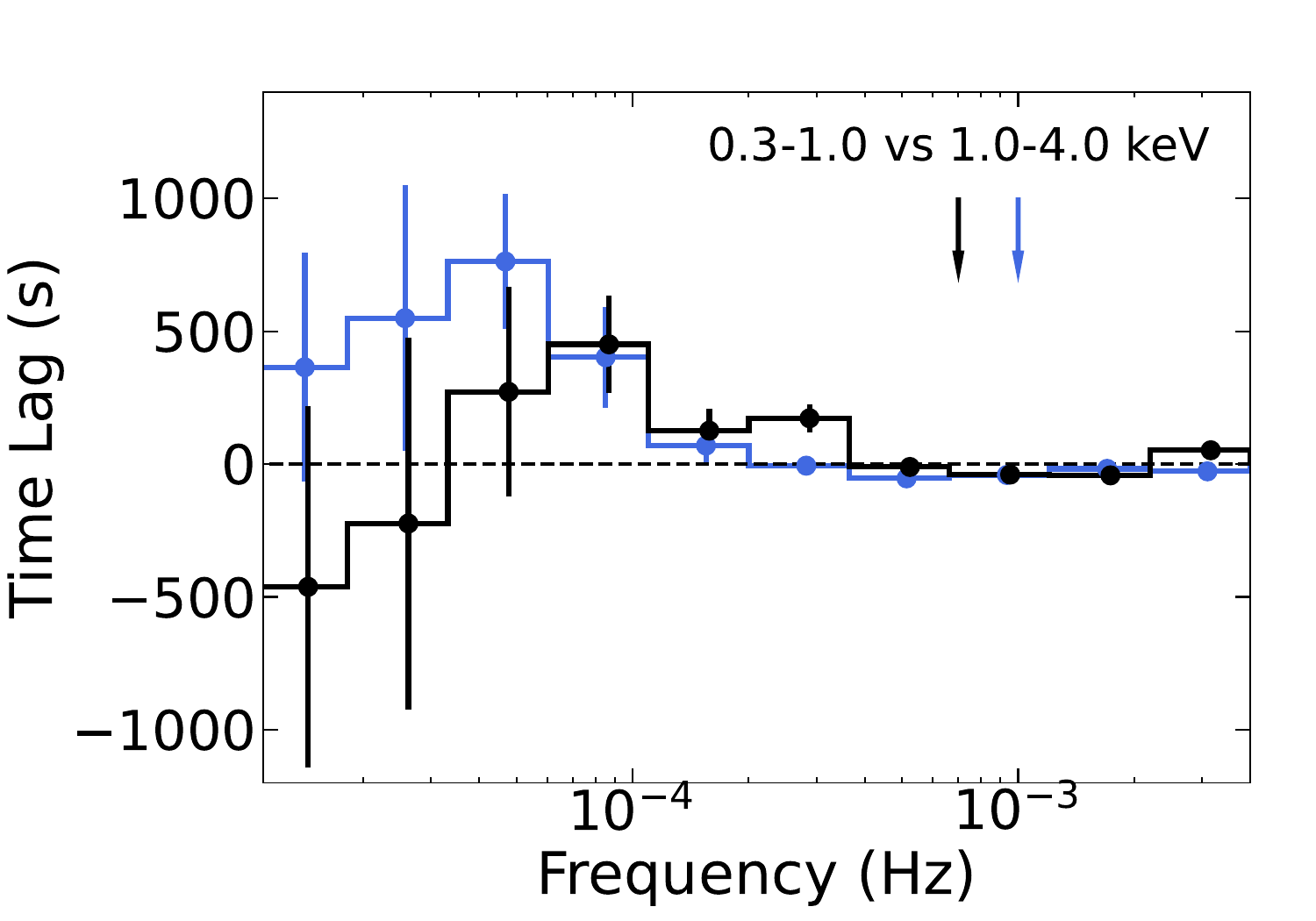}
  	\includegraphics[width=0.5\columnwidth]{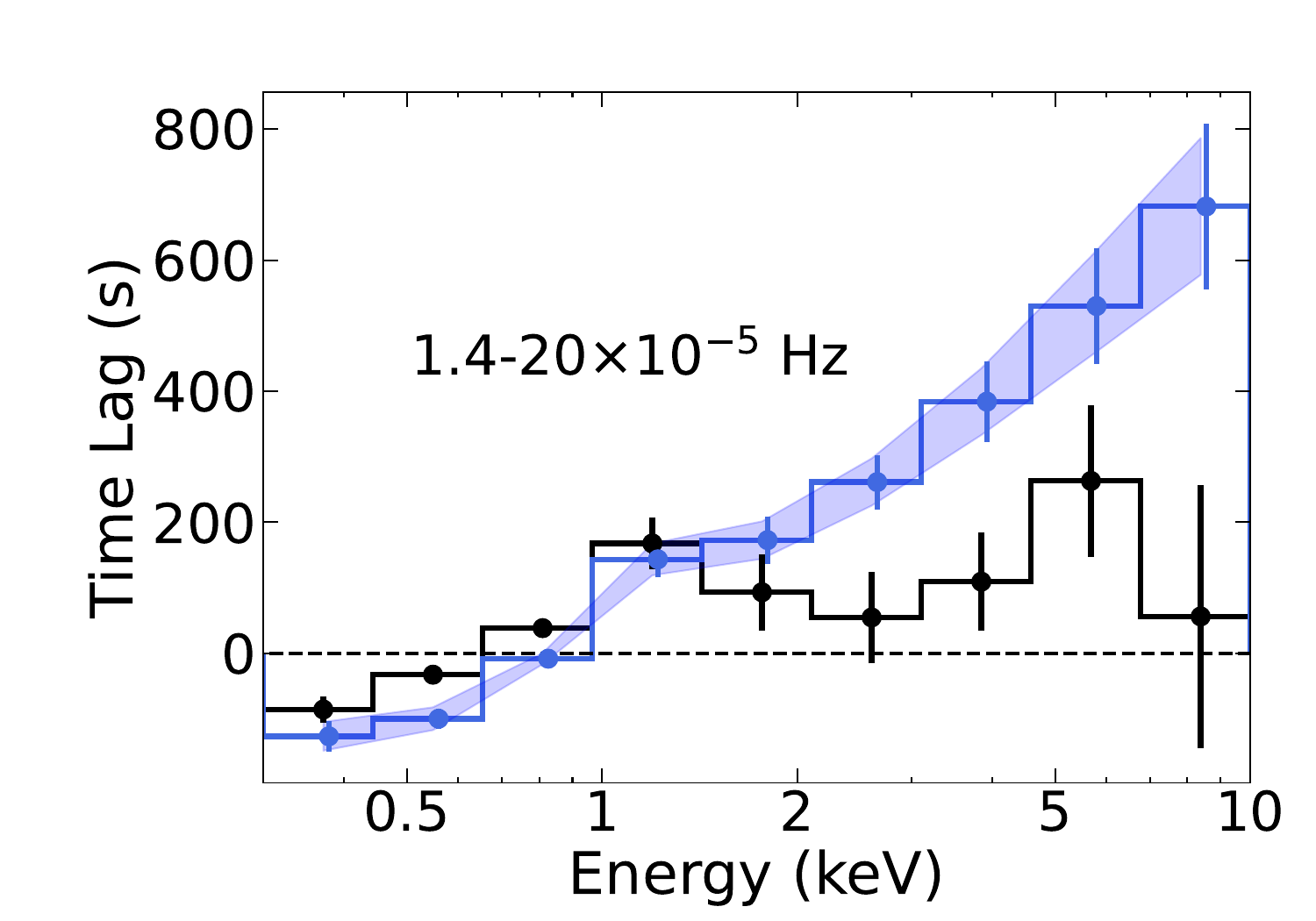}
   	\includegraphics[width=0.5\columnwidth]{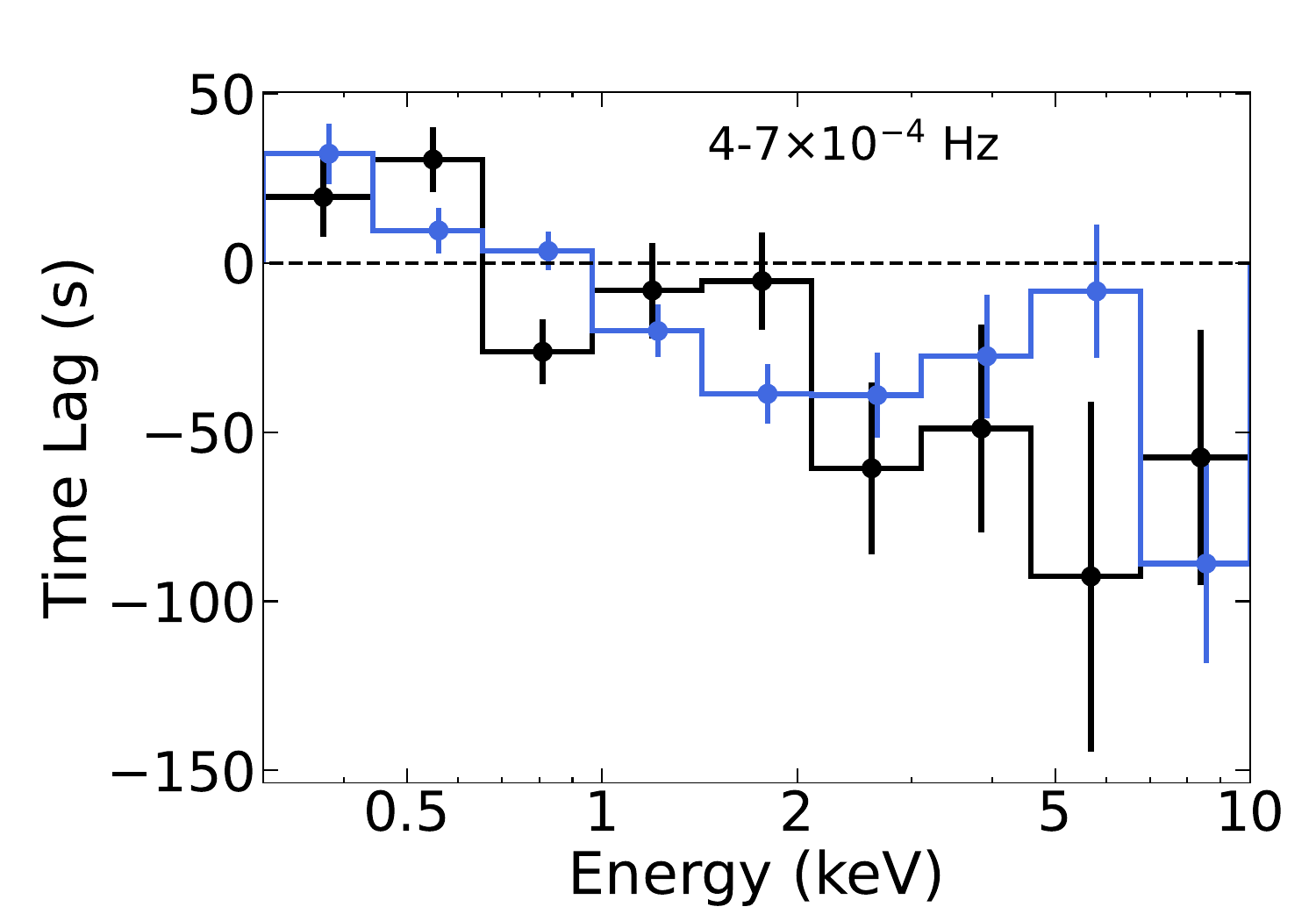}
   	\includegraphics[width=0.5\columnwidth]{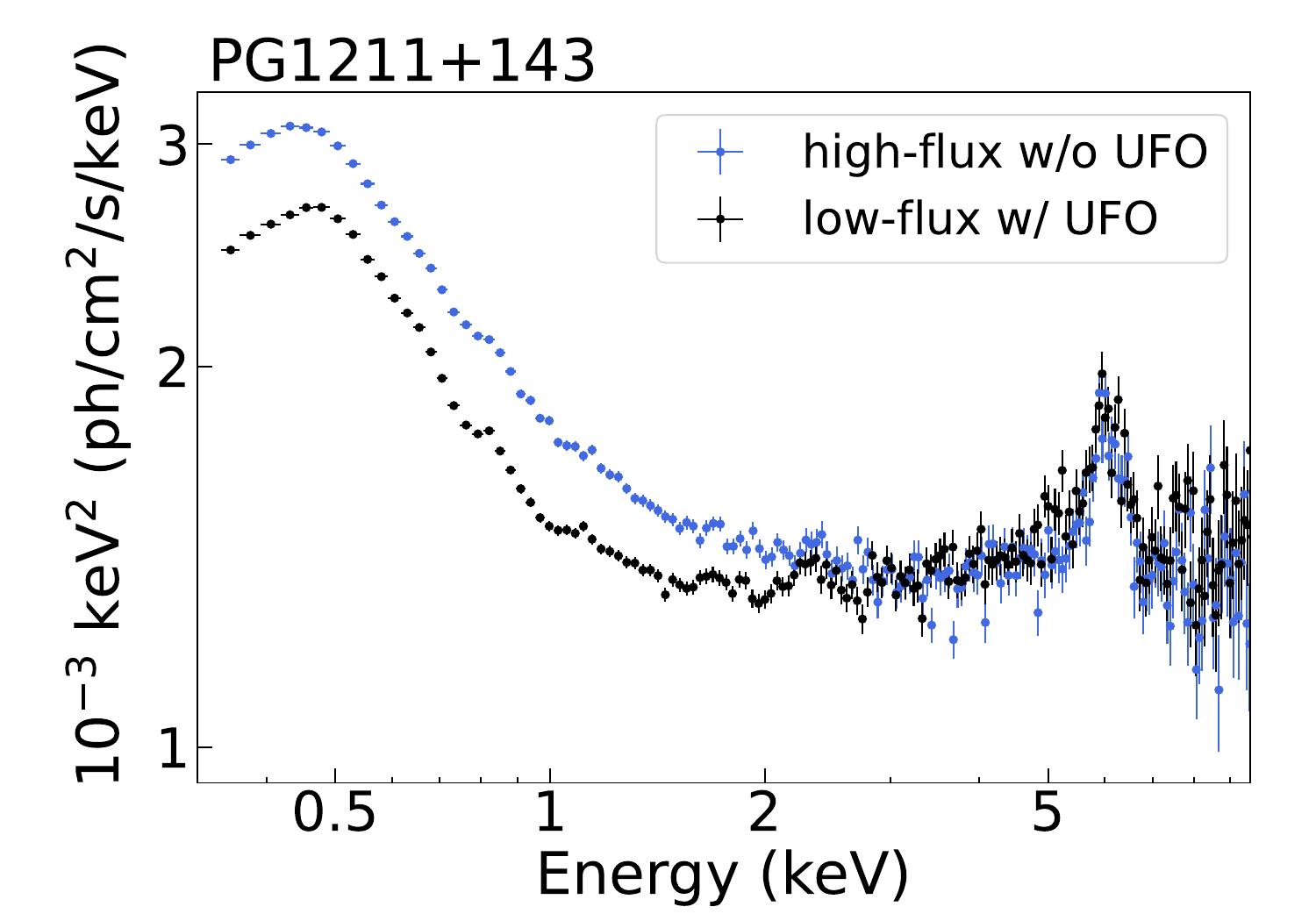}
   	\includegraphics[width=0.5\columnwidth]{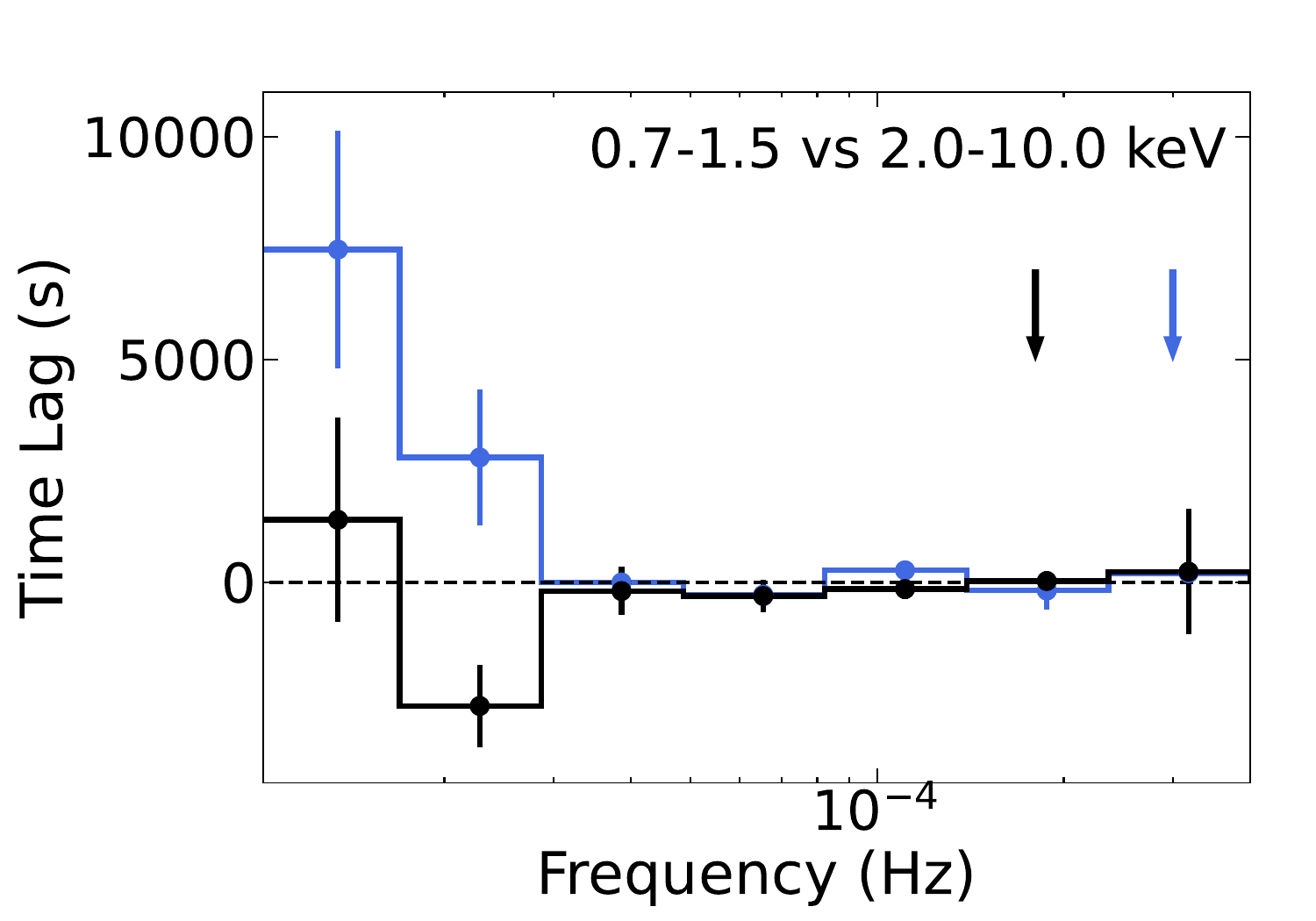}
   	\includegraphics[width=0.5\columnwidth]{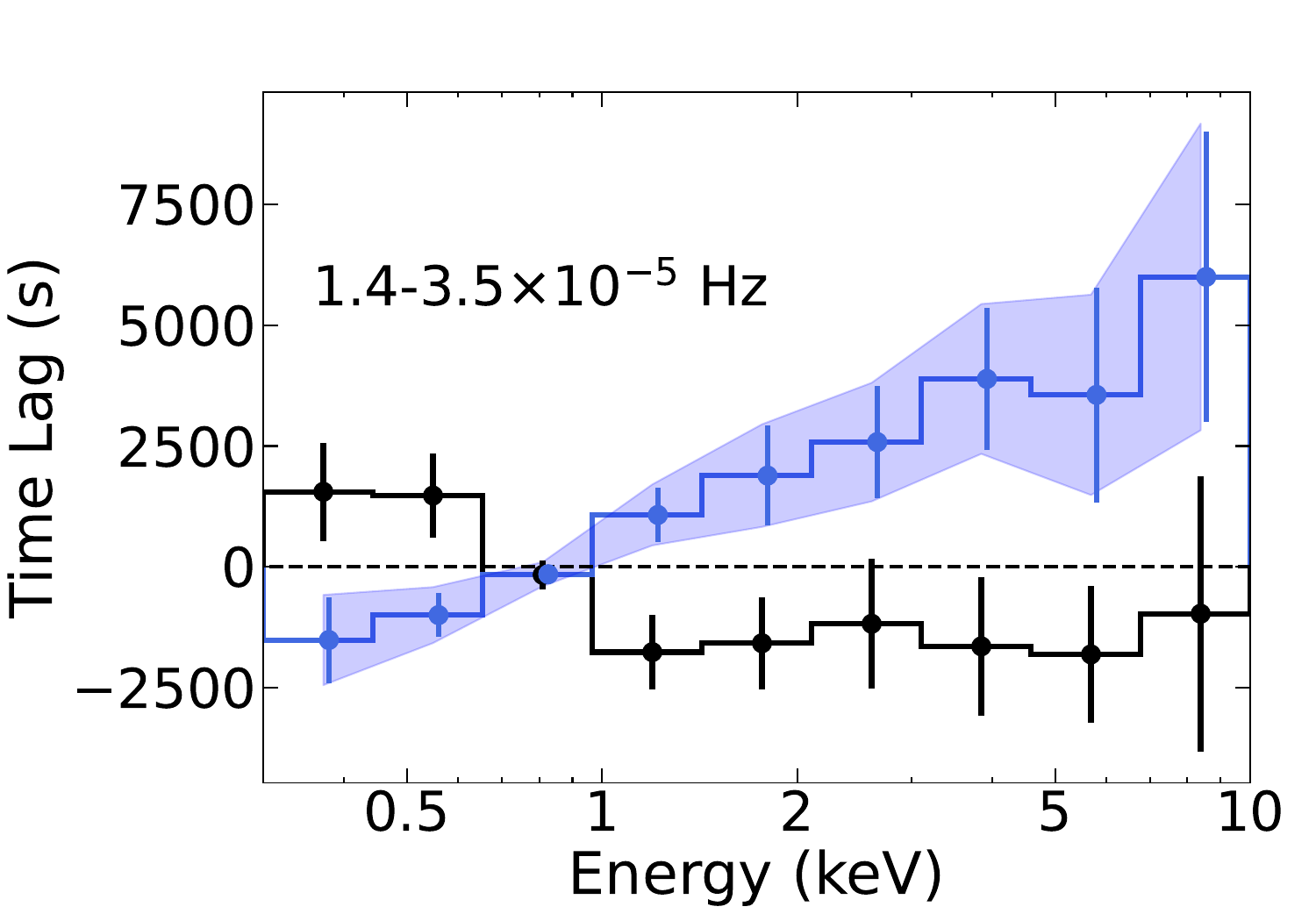}
   	\includegraphics[width=0.5\columnwidth]{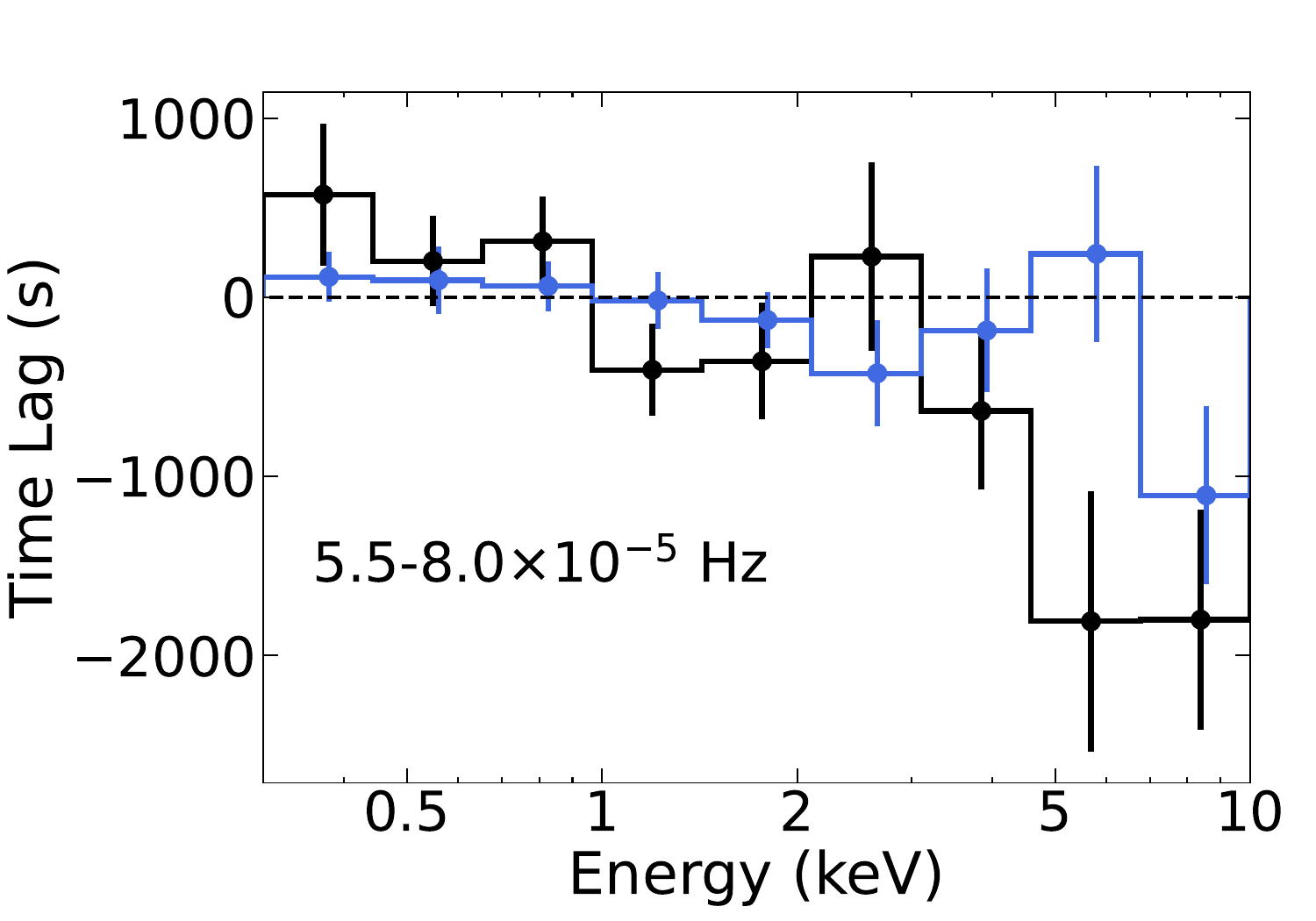}
    \caption{The X-ray time lags for PG 1448 (\textit{top}), IRAS 13224 (\textit{middle}), and PG 1211 (\textit{bottom}). The `low-/high-flux w/wo UFO' observations are marked by the \textit{black} and \textit{royalblue}, respectively. The \textit{first} column shows the unfolded stacked flux energy spectrum with respect to a power law of $\Gamma=2$ of each AGN. The \textit{second} column shows the lag-frequency spectrum between soft and hard energy bands. The arrows separately refer to the frequencies, at which the PSD of the highest energy bin (in the lag-energy spectra) becomes dominated by Poisson noise, for the corresponding stacked observations. The \textit{third} and \textit{fourth} columns present the low- (\textit{third}) and high-frequency (\textit{fourth}) lag-energy spectra separately at particular frequencies, marked in the plots. The \textit{blue} shaded regions in the low-frequency lag-energy spectra show the $1\sigma$ spread in lags from Monte Carlo simulations (see details in Section \ref{subsec:lag-energy}).
    }
    \label{fig:lags}
\end{figure*}

The lag-frequency spectra of the stacked observations in each AGN are shown in the second column of Figure \ref{fig:lags}. The \textit{black} and \textit{royalblue} arrows separately refer to frequencies, where the PSD of the highest energy bin (in lag-energy spectra) becomes dominated by Poisson noise, for the `low/high-flux w/wo UFO' observations. All lag-frequency spectra in Figure \ref{fig:lags} successfully reveal the high-frequency soft reverberation lag \citep[e.g.,][]{2009Fabian,2013Karairas13324,2013DeMarco,2016Kara} and low-frequency hard continuum lag. The soft reverberation lag in IRAS 13224 is not obvious due to its tiny lag amplitude ($\sim100$\,s), aligned with the literature \citep[e.g.,][]{2013Karairas13324,2020Alston}. The lag-frequency spectra of `low-/high-flux w/wo UFO' observations are more or less consistent within uncertainties. However, it generally indicates that the `high-flux w/o UFO' observations have a larger low-frequency hard lag than `low-flux w/ UFO' observations. 

\subsection{Lag-energy spectra}\label{subsec:lag-energy}
The lag-energy spectra are computed in a similar way to the lag-frequency spectrum. The only difference is that, at this moment, the lag is calculated in a particular frequency range between a narrow band of interest and a large reference band. To obtain the best S/N, we choose the reference band to be the entire $0.3\mbox{--}10$\,keV, and remove the band of interest so that the noise is not correlated. The frequency ranges for the lag-energy spectra are chosen based on the lag-frequency spectra and marked in the plots. As a conservative approach, the upper bound of the high-frequency lag-energy spectrum is set by the frequency at which the PSD of the hardest energy band becomes dominated by Poisson noise (indicated by arrows in Figure \ref{fig:lags}).

For observations unaffected by obscuration, the lag-energy spectra present an increasing lag with energy in the low frequencies and a soft excess lag plus an iron K reverberation lag in the high frequencies, which is consistent with previous results of IRAS 13224 and PG 1211 \citep{2013Karairas13324,2018Lobban,2022Hancock}. On the other hand, this is also the first time these typical AGN X-ray time lags have been detected in PG 1448. Regarding observations obscured by UFOs, we reproduce the same results of PG 1211 discovered in \citet{2018Lobban} that the low-frequency ($1.4\mbox{--}3.5\times10^{-5}$\,Hz) hard lag changes to flatter. Furthermore, we find this phenomenon also appears in the other two AGN, which is compatible with the results of lag-frequency spectra. In the high-frequency domain, the soft lags still exist while the iron K reverberation lag seems to disappear or become weak. However, due to the fewer counts in the low-flux state, especially in the Fe K band, we cannot conclude that the origin of the disappearance or weakness of the Fe K reverberation lag is physical or just limited by data quality.

\begin{table}[htbp]
\setlength{\tabcolsep}{3pt}
\centering
\caption{Table of the significance of the deviation between low-frequency lag-energy spectra of two stacked observations in AGN with a transient UFO. Two kinds of null hypothesis probability $P_\mathrm{null}$ and the significance $\sigma$ of deviation are calculated by the $\chi^2$ statistics of lags with uncertainties using the method in \citet{1996Nowak} and the MC method, respectively. Both approaches confirm that the low-frequency lag-energy spectra in the `low-flux w/ UFO' state are different from those of the `high-flux w/o UFO' state.
}
\begin{tabular}{lcccc}
\hline
\hline
Names & $P_\mathrm{null}^\mathrm{Cal}$ & $\sigma_\mathrm{Cal}$ & $P_\mathrm{null}^\mathrm{MC}$& $\sigma_\mathrm{MC}$ \\
\hline
PG 1448+273 &  $0.006$  & $2.7$ & $0.003$  & $2.8$  \\
IRAS 13224-3809 &  $6.6\times10^{-8}$  & $5.4$  & $1.9\times10^{-8}$   & $5.6$  \\
PG 1211+143 &  $3.4\times10^{-7}$   & $5.1$  & $1.7\times10^{-6}$  & $4.7$  \\
\hline
\end{tabular}
\label{tab:UFO_lag_CL}
\vspace{-3mm}
\end{table}

To quantify the significance of the deviation between the low-frequency lag-energy spectra of two states, we adopt the $\chi^2$ statistics by assuming one spectrum as expected values and the other as observed values. The calculated null hypothesis probability $P_\mathrm{null}^\mathrm{Cal}$ and corresponding significance $\sigma_\mathrm{Cal}$ are shown in Table \ref{tab:UFO_lag_CL}, where all cases exhibit the deviation significance larger than $2.7\sigma$. The deviation is also confirmed by Monte Carlo (MC) simulations. Following the method described in \citet{1995Timmer}, we use the \texttt{GaussianResponse} task in \texttt{pylag} package to simulate light curve pairs based on the observed PSD in each energy bin with a time lag described by a Gaussian response function. The Gaussian response function centers on the observed lag with the width of the lag uncertainty. The \textit{blue} shaded regions in Figure \ref{fig:lags} represent the $1\sigma$ uncertainties in lag from 1000 MC simulated light curve pairs in each energy band, which are comparable with our calculated uncertainties \citep{2014Uttley}. The null hypothesis probability $P_\mathrm{null}^\mathrm{MC}$ and significance $\sigma_\mathrm{MC}$ of deviation are then calculated based on uncertainties obtained from MC simulations (see Table \ref{tab:UFO_lag_CL}), where the significance is at least larger than $2.8\sigma$. Our results show that the lag-energy spectra of the `low-flux w/ UFO' observations remarkably differ from those of `high-flux w/o UFO' observations.

\section{Discussion}\label{sec:discussion}

Through the spectral-timing analysis of archival \textit{XMM-Newton} observations on three AGN with transient UFOs, we find suppression of low-frequency hard lags during low-flux UFO-affected states is not unique in PG 1211, since the other two AGN manifest it with a transient UFO obscuration event. However, whether the suppressed continuum lags are related to UFOs is still unknown because UFOs and low-flux states commonly appear simultaneously. Perhaps the lower lag amplitude results purely from the low flux, related to the continuum variation, as suggested by \citet{2013Alston} that X-ray time lags are flux-dependent.

\subsection{Are UFOs causally related to suppressed low-frequency hard lags?}\label{subsec:causal-relation}

We therefore search for another three bright targets with large X-ray variability but without UFOs in archival \textit{XMM-Newton} observations: Ark 564, NGC 7469, and Mrk 335. Their X-ray time properties have been well explored, and typical low-frequency hard lags and high-frequency iron K reverberation lags have been found in all of them \citep[e.g.,][]{2013Kara564335,2016Kara,2020Mastroserio,2022Lewin}. It should also be noted that despite the absence of UFOs in these targets, slower and less ionized absorbers were discovered in all of them \citep{2015Giustini,2013Longinotti,2016Khanna,2018Mehdipour,2020Grafton-Waters,2021Liu}.

Following the same data reduction and analysis routines, we select archival \textit{XMM-Newton} observations of these three targets with distinct fluxes and stack them into `low-flux' and `high-flux' subclasses. The basic information on selected observations is listed in Table \ref{app:tab:obs}. The light curves in the EPIC-pn band and the corresponding PSD are shown in Figure \ref{app:fig:lc_pds}, exhibiting frequencies above which the Poisson noise becomes dominant. The segment lengths are 50\,ks for Ark 564, and 80\,ks for NGC 7469 and Mrk 335, respectively. According to the literature, the origins of the flux variation among these targets are different: the variation in Ark 564 originates mainly from the variable normalization of the broadband continuum \citep{2015Giustini}; the variation in NGC 7469 results from the variable normalization of the soft excess component \citep{2018Peretz,2018Middei}; the variation in Mrk 335 is contributed by both the variable intrinsic continuum and ionized absorbers \citep{2015Gallo,2020Komossa}.

The stacked flux-energy, lag-frequency, and lag-energy spectra are shown in Figure \ref{app:fig:lags}. The details of the lag-frequency and high-frequency lag-energy spectra are described in Appendix \ref{app:sec:wotransient} and, briefly speaking, the flux variability has negligible influence on them. For the low-frequency lag-energy spectra, the typical hard lags that increase with energy are observed and are consistent with each other within spectra of two flux states for every AGN. The null hypothesis probability and the significance of deviation are listed in Table \ref{tab:woUFO_lag_CL}, where no case exhibits a deviation significance greater than $1.2\sigma$. This behavior is quite different from previous results of AGN with UFOs, suggesting that flux changes, originating from the continuum variations, do not affect the low-frequency hard lags, and the suppression of low-frequency hard lags might be linked with UFOs. We note a tentative deviation in the low-frequency lag-energy spectrum of Mrk 335, where the deviation occurs in the highest energy bin, leading to a $\sim1\sigma$ significance of the deviation. However, it should be noted that the absorbers in Mrk 335 have large velocities of $\sim5000\mbox{--}7000$\,km/s \citep{2019Longinotti}, making them distinct from typical warm absorbers \citep[WAs, $v<1000$\,km/s,][]{2021Laha} but closer to UFOs ($v>10000$\,km/s). Therefore, they are called obscuring outflows. If the hypothesis that the presence of UFOs can suppress the low-frequency hard lags is true, this weak indication may be related to the similarity between obscuring outflows in Mrk 335 and typical UFOs. 

Similar fast obscuring outflows were also observed in NGC 3783 and Mrk 817 and expected from the outer disk or broad line region (BLR). In \citet{2020DeMarco}, NGC 3783 shows negative low-frequency lags in obscured observations and positive low-frequency lags in unobscured exposures, which is consistent with what we observed in targets with variable UFOs. In \citet{2024Lewin}, Mrk 817 exhibits low-frequency hard lags up to 800s when the source is partially obscured but relatively bright compared with other heavily obscured long \textit{XMM-Newton} observations. As an SMBH with a mass of $3.85\times10^{7}M_\odot$, its amplitude of the low-frequency lag is smaller than those at similar masses (e.g. PG 1448) and comparable with results of IRAS 13224 ($M_\mathrm{BH}\sim2\times10^{6}M_\odot$), if we assume low-frequency lags scale with SMBH masses. This indicates that obscuring outflows might have a moderate influence (between UFOs and WAs) on suppressing low-frequency hard lags, pointing the suppression strength is related to outflow properties (awaiting for further works).

\begin{table}[htbp]
\setlength{\tabcolsep}{3pt}
\centering
\caption{Table of the significance of the deviation between low-frequency lag-energy spectra of two stacked observations in AGN without UFOs (Similar to Figure \ref{tab:woUFO_lag_CL}). Both two approaches confirm that, in AGN without UFOs, the low-frequency lag-energy spectra of the `low-flux' observations are consistent with those of the `high-flux' observations.
}
\begin{tabular}{lcccc}
\hline
\hline
Names & $P_\mathrm{null}^\mathrm{Cal}$ & $\sigma_\mathrm{Cal}$ & $P_\mathrm{null}^\mathrm{MC}$& $\sigma_\mathrm{MC}$ \\
\hline
Ark 564 &  $0.90$ & $0.12$  &  $0.96$  & $0.05$ \\
NGC 7469 &  $0.96$ & $0.05$  & $0.97$  & $0.03$   \\
Mrk 335 &  $0.26$ & $1.12$  & $0.27$ & $1.09$\\
\hline
\end{tabular}
\label{tab:woUFO_lag_CL}
\vspace{-3mm}
\end{table}
As a result, by comparing X-ray time lags at different flux states of AGN without UFOs, we find no evidence that the suppressed low-frequency hard lags originate from the faint continuum flux. The results of Mrk 335, combined with archival discoveries of Mrk 817 and NGC 3783, tentatively reveal that obscuring outflows with velocities close to those of UFOs might have a moderate influence on low-frequency hard lags. In fact, our discoveries are compatible with the results of NGC 4051 discovered by \citet{2013Alston}, because some ionized winds in NGC 4051 are also close to UFOs with high velocities \citep[$v>3000$\,km/s,][]{2013Pounds} and the major differences between the high and low flux X-ray spectra comes from the varying ionized winds \citep{2004Pounds}.

\subsection{Explanations for varying low-frequency hard lags}\label{subsec:explanation}

Based on our results and discussions, the suppressed hard continuum lags are likely related to the emergence of fast-moving outflows, not necessarily related to the source flux, while the underlying mechanism remains unclear.

\subsubsection{Scenario I: additional lags associated with outflows}\label{subsubsec:recombination}
One possible interpretation is that different physical processes dominate in different timescales. \citet{2016Silva} once investigated the NGC 4051 data using a time-dependent photoionization code to estimate the influence of outflowing ionized absorbers on X-ray time lags. They found that WAs in NGC 4051 can produce low-frequency soft X-ray lags, where the delay arises from the long radiative recombination time of WAs when the plasma varies in response to the ionizing continuum. The phenomenon of low-frequency soft X-ray lags was also observed in NGC 1365 and was explained by different light-crossing timescales of hard and soft X-ray photons in WAs \citep{2015Kara}. As a result, in this scenario, low-frequency X-ray lags comprise both signals of hard X-ray continuum lags related to the accretion flow and soft X-ray lags associated with the ionized absorbers. Similar effects might also arise in our results if soft X-ray UFOs (detected in all three targets) can enhance the time lags in the soft X-ray bands, leading to a seemingly flat lag-energy spectrum.

In general, UFOs are closer to SMBHs than WAs \citep[e.g.,][]{2021Laha}, and are expected to have a higher number density. Therefore, X-ray time lags introduced by the recombination of UFOs should have shorter timescales than those of WAs since the recombination timescale is inversely proportional to the plasma number density \citep{1995Krolik}. The recombination timescales of UFOs are calculated using the \texttt{rec\_time} tool in the SPEX package \citep{1996Kaastra} with assumed number densities of $n_\mathrm{e}=10^{10}\,\mathrm{cm}^{-3}$ for PG 1448 \citep{2023Reeves}, $n_\mathrm{e}=10^{11}\,\mathrm{cm}^{-3}$ for IRAS 13224, and $n_\mathrm{e}>10^{8}\,\mathrm{cm}^{-3}$ for PG 1211 \citep{2018Reeves}. All densities are estimated by the definition of the ionization parameter $\xi=L_\mathrm{ion}/n_\mathrm{e}R^2$, where the location of outflow $R$ is constrained by the transient time scale of the UFO $\Delta t$ and the variation of the column density $\Delta N_\mathrm{H}$ reported in the literature,
    $R^{5/2}=(GM)^{1/2}(L_\mathrm{ion}\Delta t/\Delta N_\mathrm{H}\xi)$.
Recombination timescales are listed in Table \ref{tab:timescale}. Our estimations show that delays from recombination are negligible and cannot contribute to the flat lag-energy spectra in the low-frequency domain.



\begin{table}[htbp]
\setlength{\tabcolsep}{3pt}
\centering
\caption{Estimations on the recombination time, absorber size, and light-crossing time of UFOs in AGN. The absorber size is calculated by assuming the transverse velocity is equal to the Keplerian velocity. The light-crossing time of absorbers is conservatively estimated by assuming a transparent absorber for hard X-ray photons. The recombination timescale is estimated by assuming the number density of $10^{10}\,\mathrm{cm}^{-3}$ for PG 1448 \citep{2023Reeves}, $10^{11}\,\mathrm{cm}^{-3}$ for IRAS 13224 \citep{2018Pinto} and $10^{8}\,\mathrm{cm}^{-3}$ for PG 1211 \citep{2018Reeves}.
}
\begin{tabular}{lccc}
\hline
\hline
Names & Recombination & Absorber size & Light-crossing   \\
 & time (s) & (cm) & time (s) \\
\hline
PG 1448 & $<10$ &  $<2\times10^{13}$   &  $<700$   \\
IRAS 13224 & $<1$ &  $<1.4\times10^{12}$   & $<50$  \\
PG 1211 & $<100$  &  $<5\times10^{13}$   & $<1700$ \\
\hline
\end{tabular}
\label{tab:timescale}
\vspace{-3mm}
\end{table}
The other possible contributor is the delay resulting from different light-crossing timescales for soft and hard X-rays in UFOs. We estimate the light-crossing time by assuming an extreme situation that UFOs are transparent to hard X-rays but optically thick to soft X-rays. The light-crossing time is then related to the absorber size $\Delta R$, which could be derived through $\Delta R=(GM/R)^{1/2}\Delta t$ by assuming the transverse velocity is equal to the Keplerian velocity. The resulting absorber size and light-crossing time are listed in Table \ref{tab:timescale}. The light-crossing times are on an order of magnitude comparable with observed low-frequency lags, indicating that the additional lags associated with different light-crossing timescales of photons are promising for explaining the observed phenomenon. However, we caution that those values are generally several times lower than observed lags under a conservative condition, which weakens the likelihood of this mechanism.

\subsubsection{Scenario II: inhibited disk-corona energy transfer}\label{subsubsec:communication}
\begin{figure*}[htbp]
    \centering
	\includegraphics[width=0.66\columnwidth]{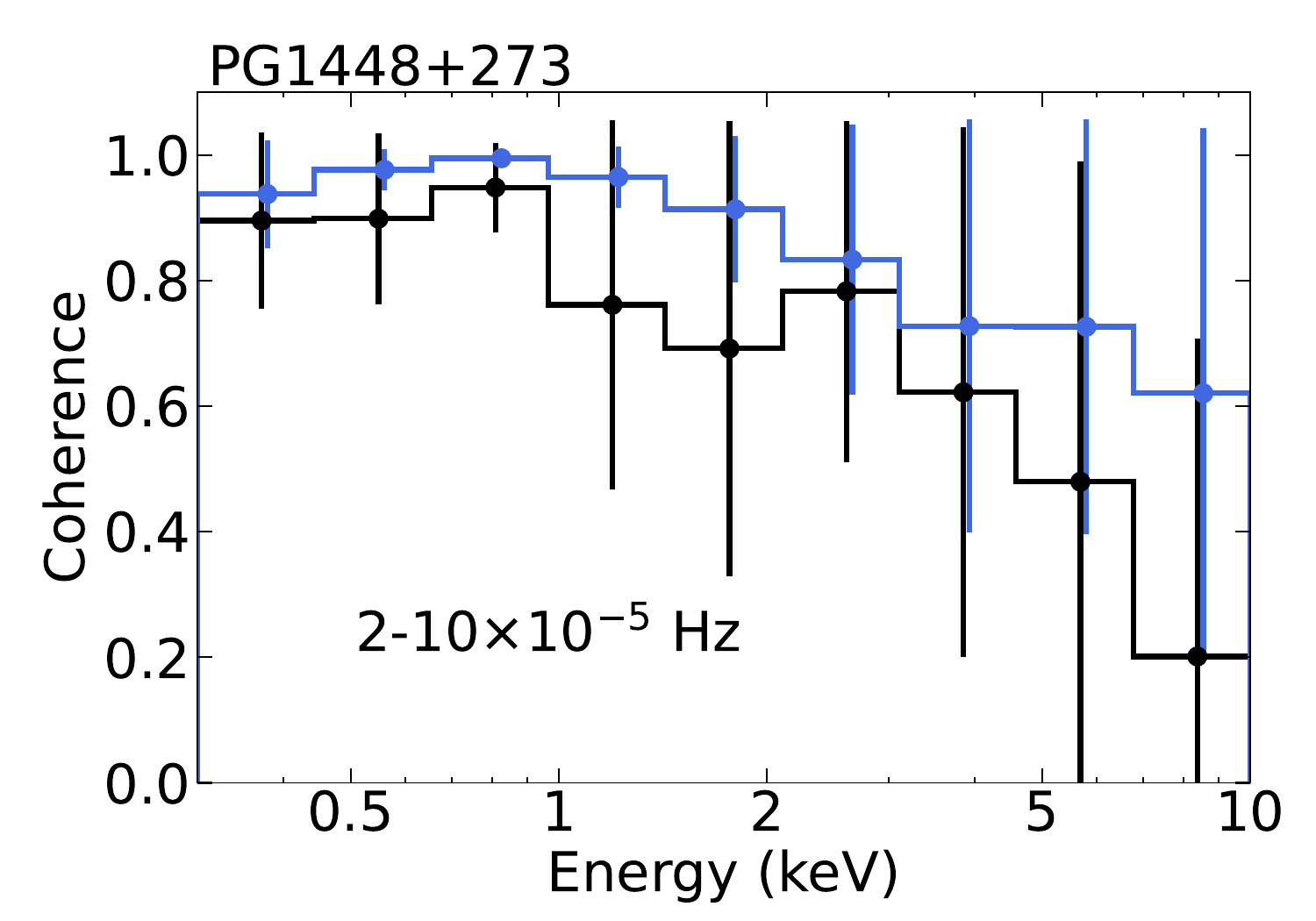}
	\includegraphics[width=0.66\columnwidth]{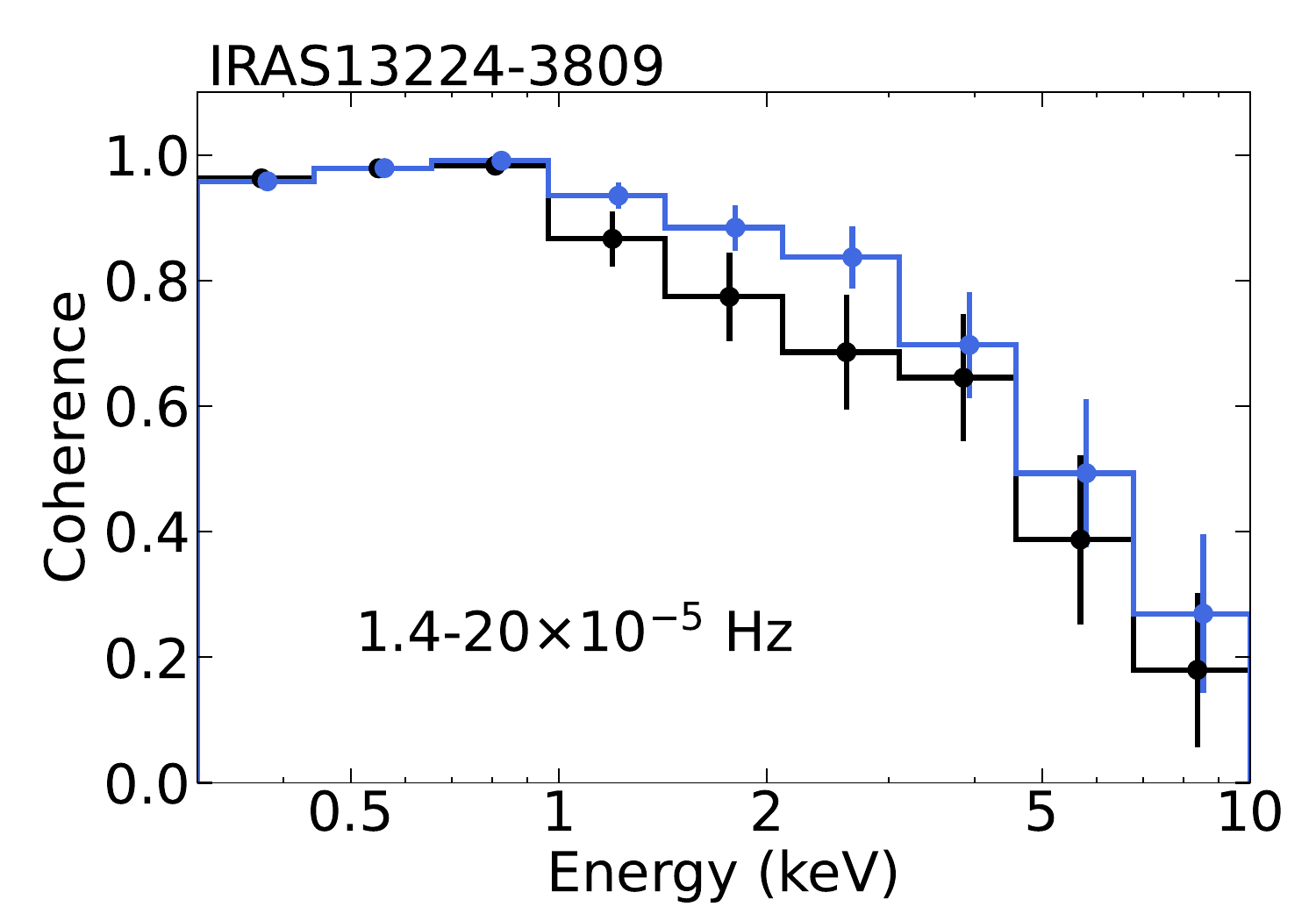}
	\includegraphics[width=0.66\columnwidth]{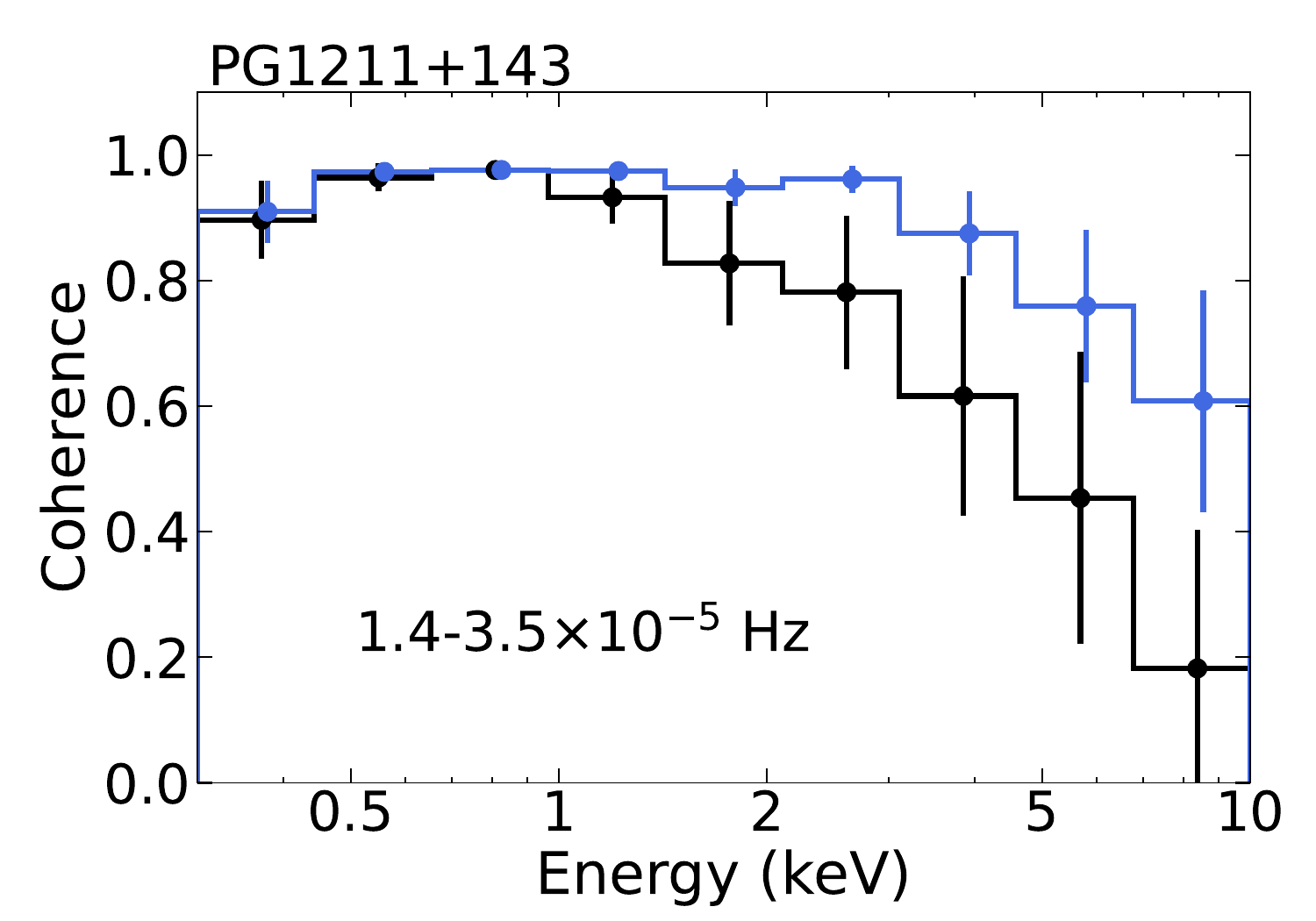}
	\includegraphics[width=0.66\columnwidth]{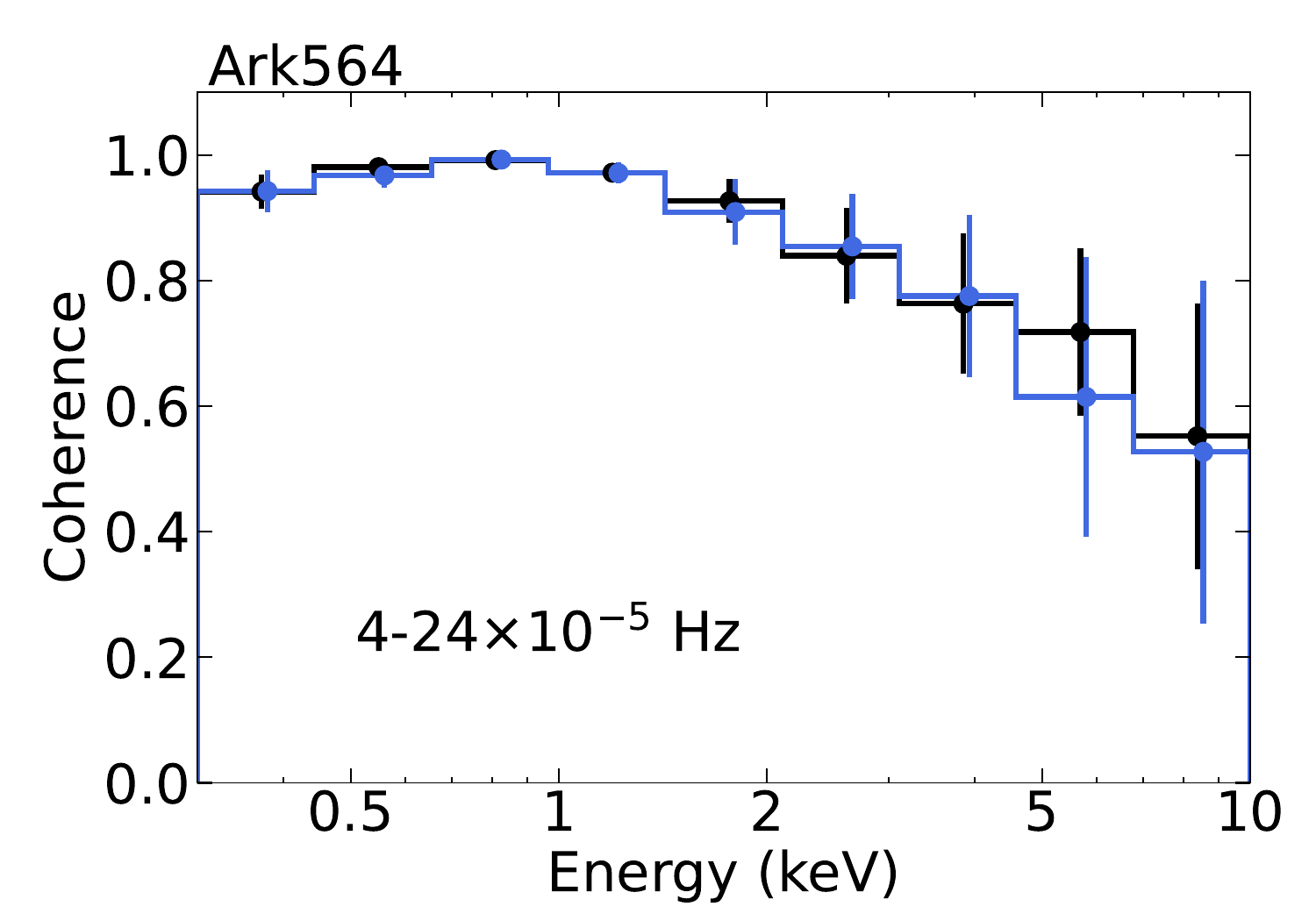}
	\includegraphics[width=0.66\columnwidth]{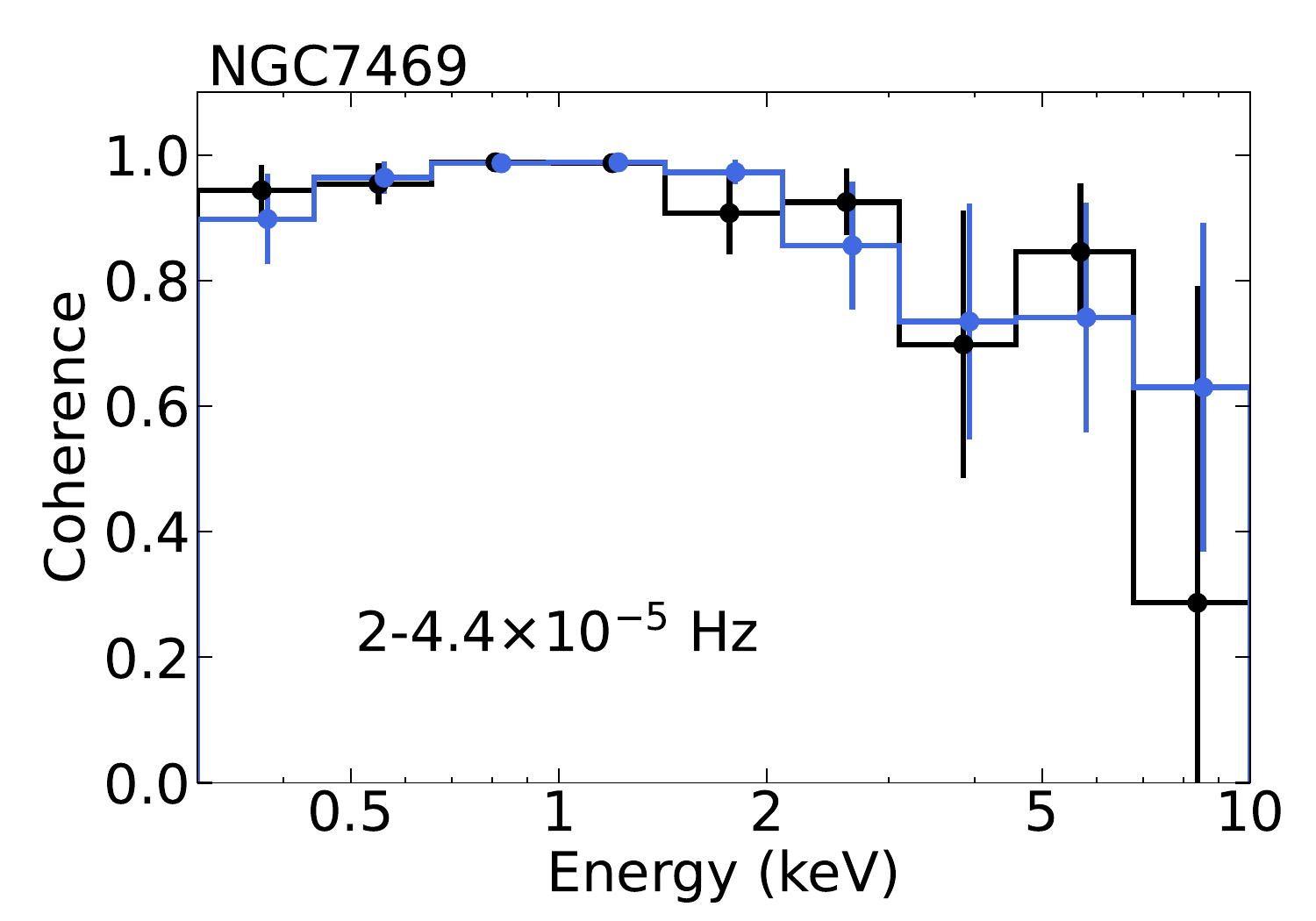}
	\includegraphics[width=0.66\columnwidth]{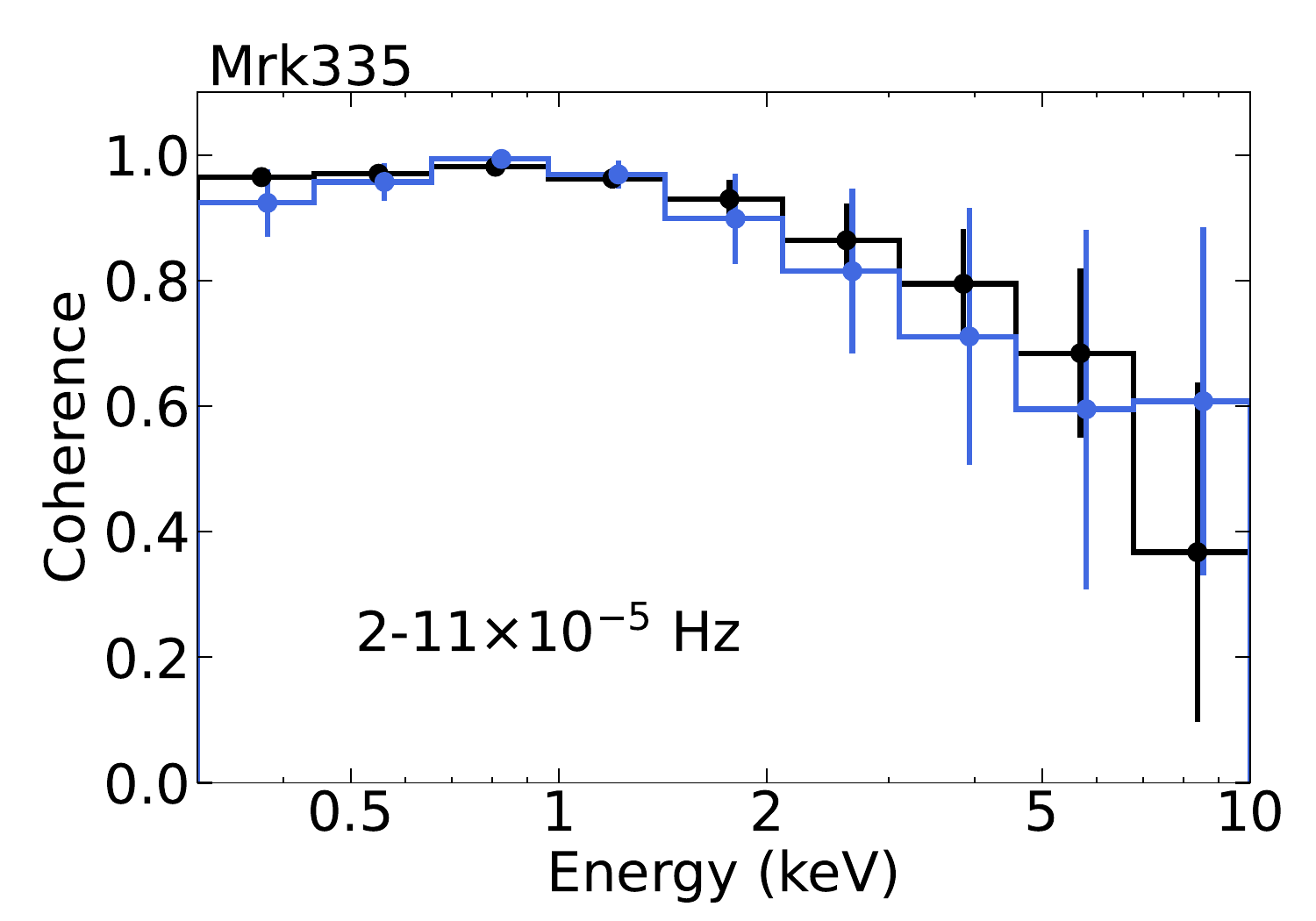}
    \caption{The Poisson noise corrected coherence-energy spectra of UFO-(\textit{upper}) or flux-resolved (\textit{lower}) observations for targets analyzed in this work at low frequencies. In AGN with transient UFOs ((\textit{upper})), the `low-flux w/ UFO' (\textit{black}) coherence spectra are in general lower than `high-flux w/o UFO' coherence (\textit{royalblue}), while the coherence spectra remain unchanged at different flux levels for AGN without UFOs (\textit{lower}).
    }
    \label{fig:coherence}
\end{figure*}

Another intriguing possible scenario is that the energy transfer between the disk and corona is inhibited by outflows, preventing the propagation of accretion fluctuations if low-frequency hard lags originate from the inward propagating mass accretion fluctuation generated in large radii \citep[e.g.,][]{2023Uttley}. Since UFOs, with their extreme velocities, are expected to originate from the inner accretion disk, their launching can carry away a portion of the inflowing materials from the inner disk. In this scenario, the reduction of lags results from the removal of the UV seed photons carrying fluctuations and/or the obscuration of seed photons by UFOs, leading to fewer fluctuation signals propagating into the hot corona. This explanation is supported by the differences between UFO-resolved coherence spectra at low frequencies, where the band choices are the same as those in Figure \ref{fig:lags} and \ref{app:fig:lags}, shown in Figure \ref{fig:coherence}. In AGN with transient UFOs (i.e. upper panels in Figure \ref{fig:coherence}), the entire coherence spectra of UFO-obscured observations are tentatively weaker than those without UFOs, although the results of PG 1448 are compatible within uncertainties due to limited exposures, while in AGN without UFOs (lower panels), the coherence spectra at different fluxes remain consistent. These results suggest that UFOs can prevent the energy transportation between the disk and the hot corona, thus reducing time lags.

In this scenario, X-ray lags in the high-frequency domain are also expected to be affected by UFOs because coronal X-ray emission is correlated with mass accretion \citep[e.g.,][]{2010lusso,2017Svoboda}. The reduction of accretion mass by outflows would decrease the coronal X-ray emission, leading to weaker signals of X-ray reprocessing in the accretion disk. Our results seem to agree with this expectation that in AGN with UFOs, the Fe K reverberation lags of absorbed observations are less significant than those in unobscured states (see Figure \ref{fig:lags}), while they remain almost unchanged in AGN without UFOs (see Figure \ref{app:fig:lags}). However, it is still unclear whether their disappearance is physically related to UFOs or depends on chosen frequency bands or specific observations, and the soft lags remain constant. In addition, it is still possible that the Fe K reverberation lags occur in absorbed states if the coronal variability, in some cases, has not started to be affected by the removal of propagating fluctuation generated in larger radii. 

However, supporting this hypothesis requires a dedicated theoretical simulation, incorporating accretion flows with mass fluctuation propagation and outflows launched from the inner accretion disk. This work has not been accomplished yet and should be done in the future for comparison with observational results. 

\section{Conclusions}\label{sec:conclusion}
Working on archival \textit{XMM-Newton} data, we perform the UFO-resolved spectral-timing analysis of three AGN with transient UFOs to investigate the potential influence of UFOs on X-ray time lags. As a result, we have shown that low-frequency hard lags become weak accompanied by UFOs in all three targets. Our investigations of another three variable AGN without UFOs indicate that suppressed low-frequency hard lags might be unrelated to flux variability. Results of Mrk 335 with fast-moving obscuring outflows, combined with archival results in Mrk 817 and NGC 3783, even indicate an intermediate state of effects between UFOs and slow WAs. We propose an additional time delay associated with UFOs or the inhibited energy and variability transfer between the disk and corona for explanations. Our work indicates that any analysis of X-ray time lags in AGN must account for the potential effects of ionized absorbers if they are present. The same investigation like this work should be extended to more sources, particularly with various black hole masses and accretion rates, to justify proposed explanations.

\section{Acknowledgements}
The authors thank Phil Uttley, Matteo Lucchini, Peter Kosec, and the anonymous referee for constructive suggestions. This research has made use of data obtained with the \textit{XMM-Newton}, an ESA science mission funded by ESA Member States and the USA (NASA). YX, CP, and SB acknowledge support for PRIN MUR 2022 SEAWIND 2022Y2T94C (funded by NextGenerationEU) and INAF LG 2023 BLOSSOM. EK acknowledges the XRISM Participating Scientist Program for support under NASA grant 80NSSC20K0733. FT acknowledges funding from the European Union - Next Generation EU, PRIN/MUR 2022 (2022K9N5B4).

\bibliographystyle{aa}
\bibliography{ref.bib}

\begin{appendix}
\onecolumn

\section{Supplementary results of AGN with a transient UFO}\label{app:sec:transient}

\begin{figure*}[htbp]
    \centering
	\includegraphics[width=0.33\textwidth]{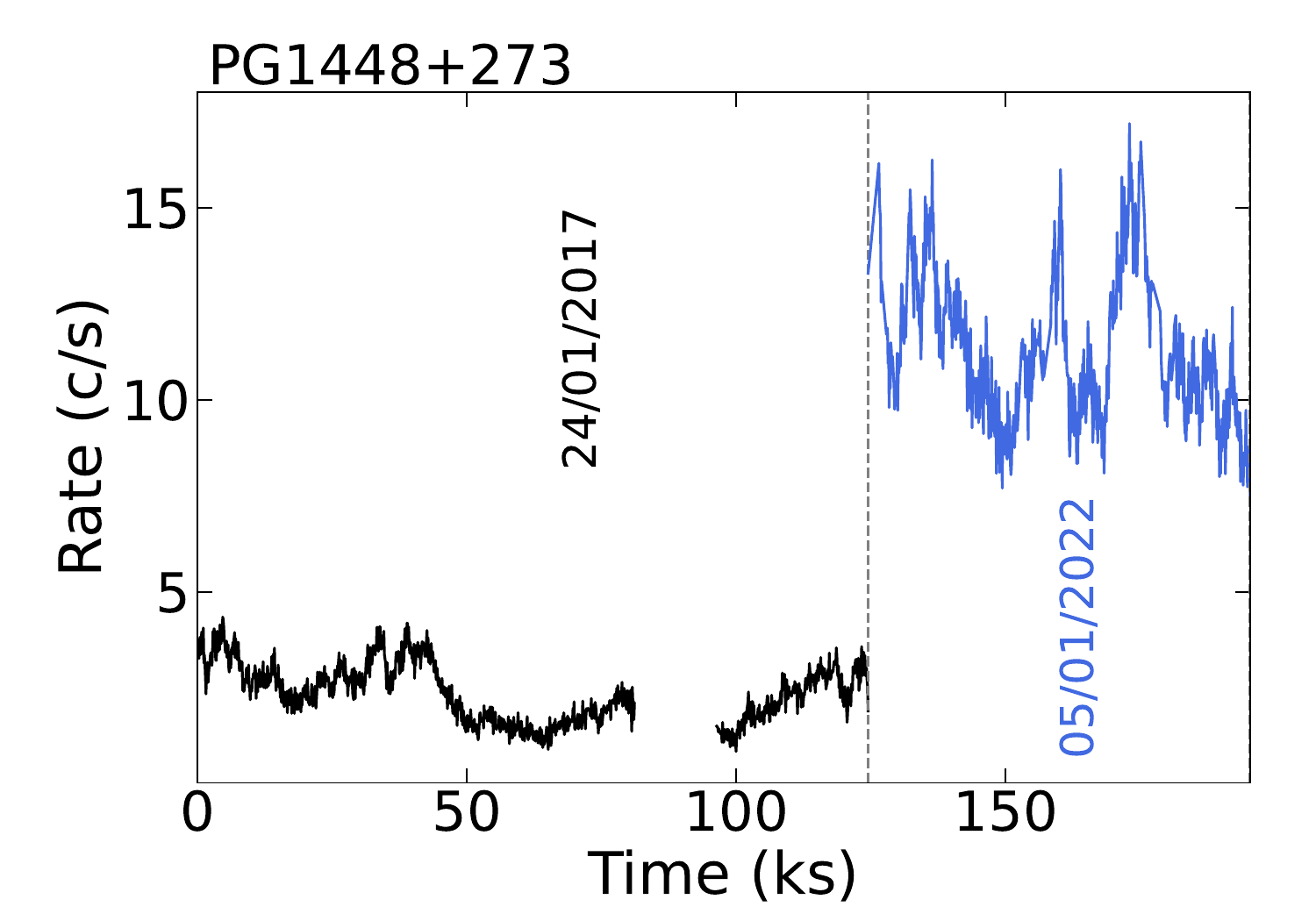}
 	\includegraphics[width=0.33\textwidth]{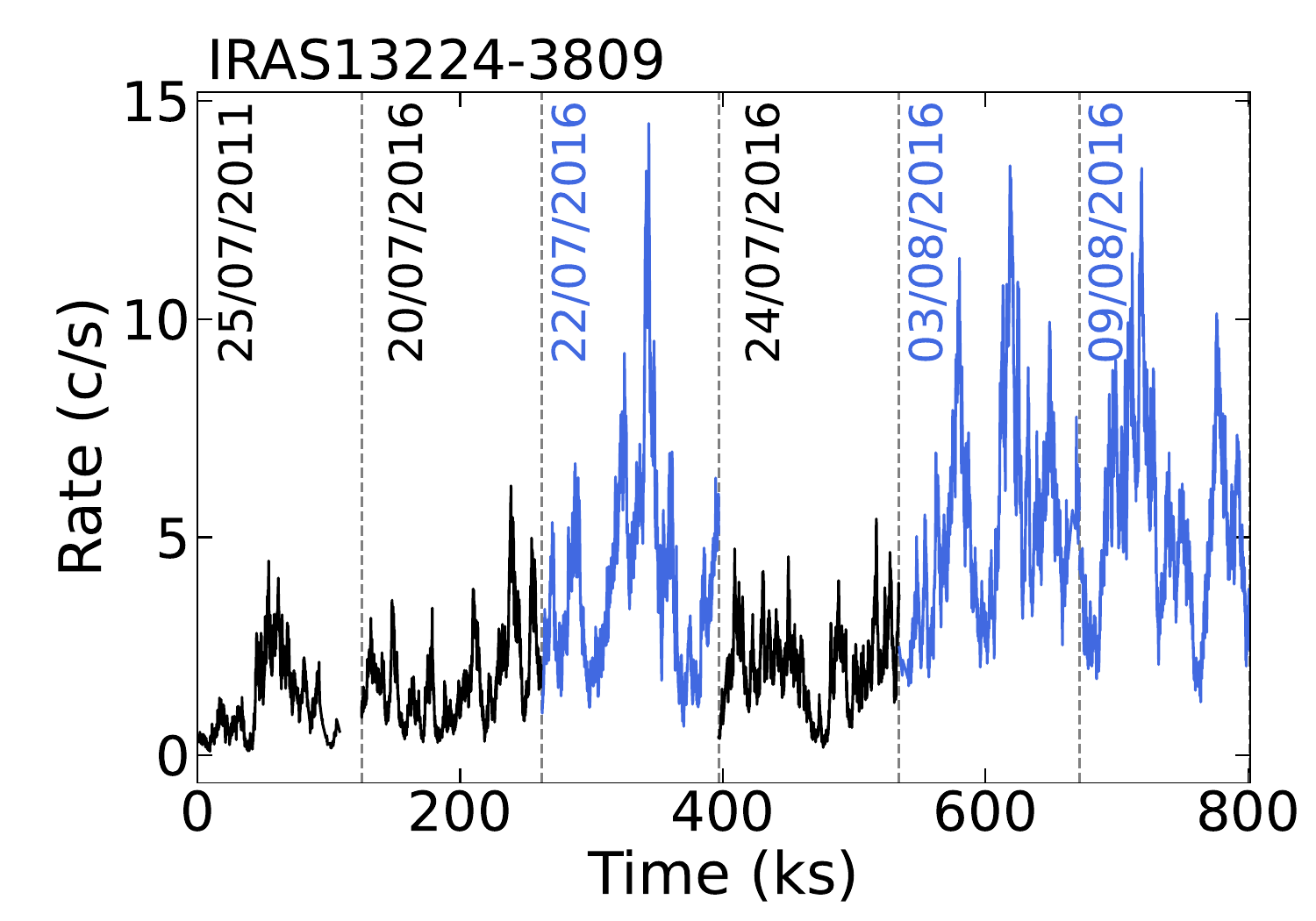}
 	\includegraphics[width=0.33\textwidth]{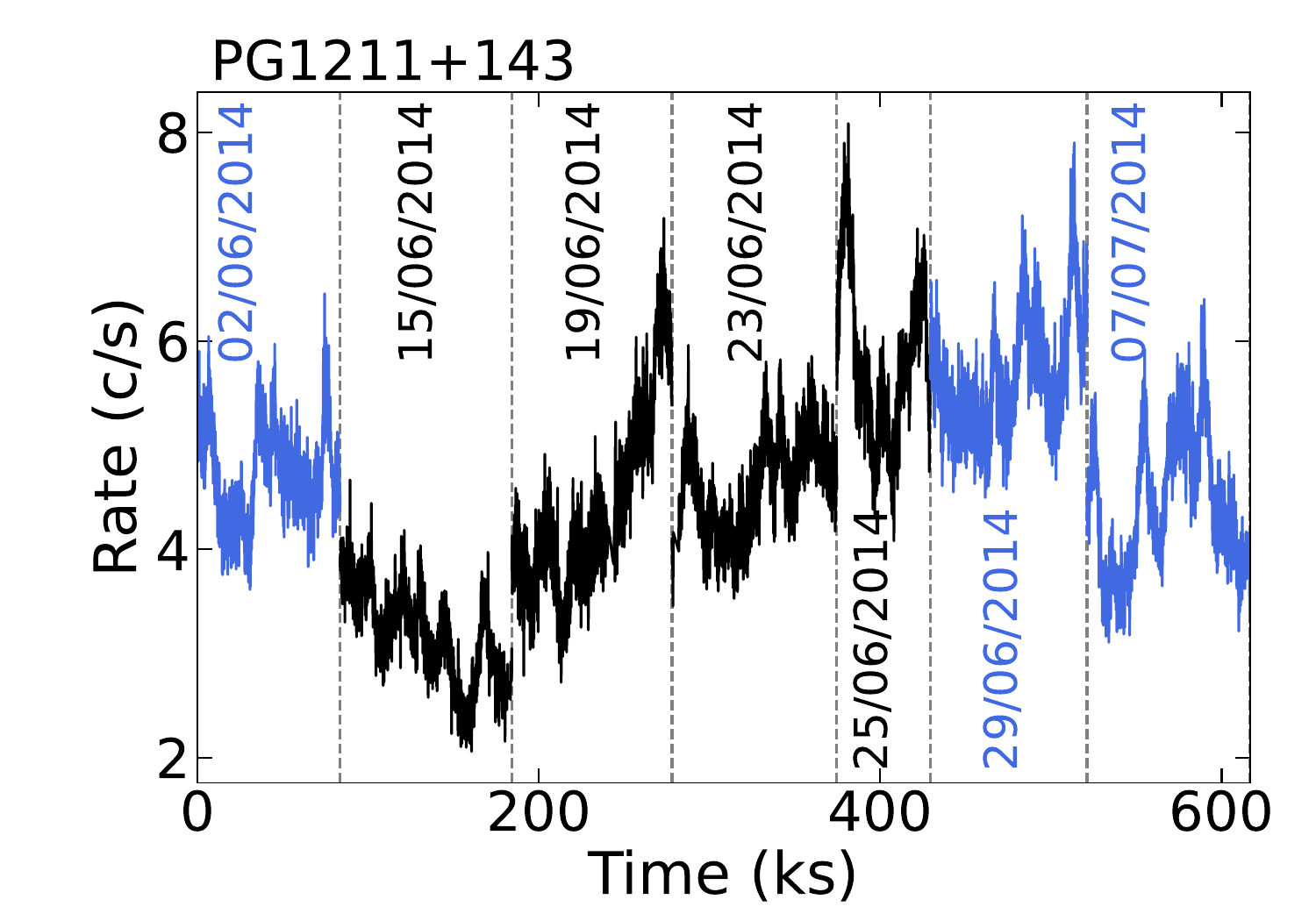}
	\includegraphics[width=0.33\textwidth]{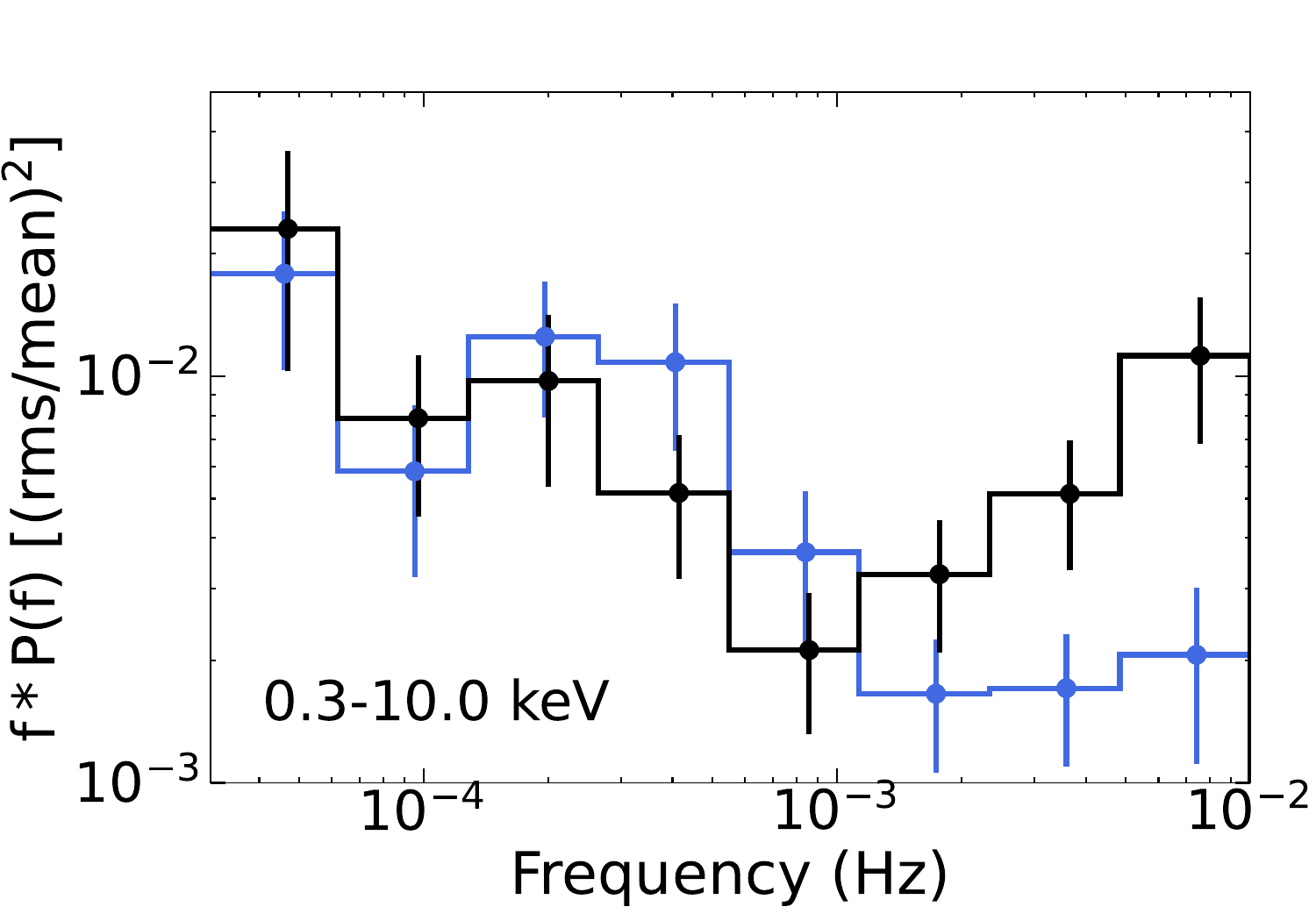}
 	\includegraphics[width=0.33\textwidth]{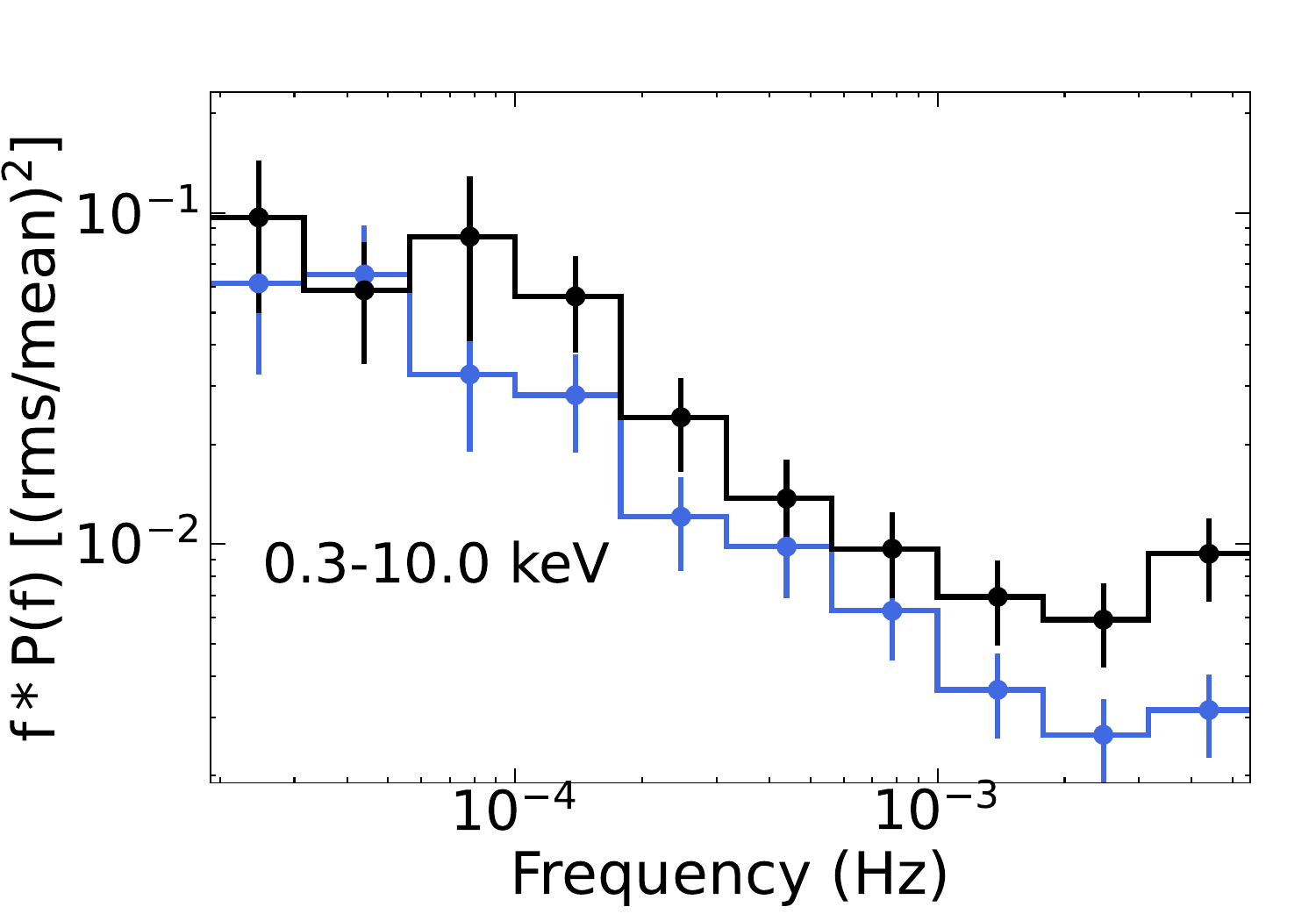}
   	\includegraphics[width=0.33\textwidth]{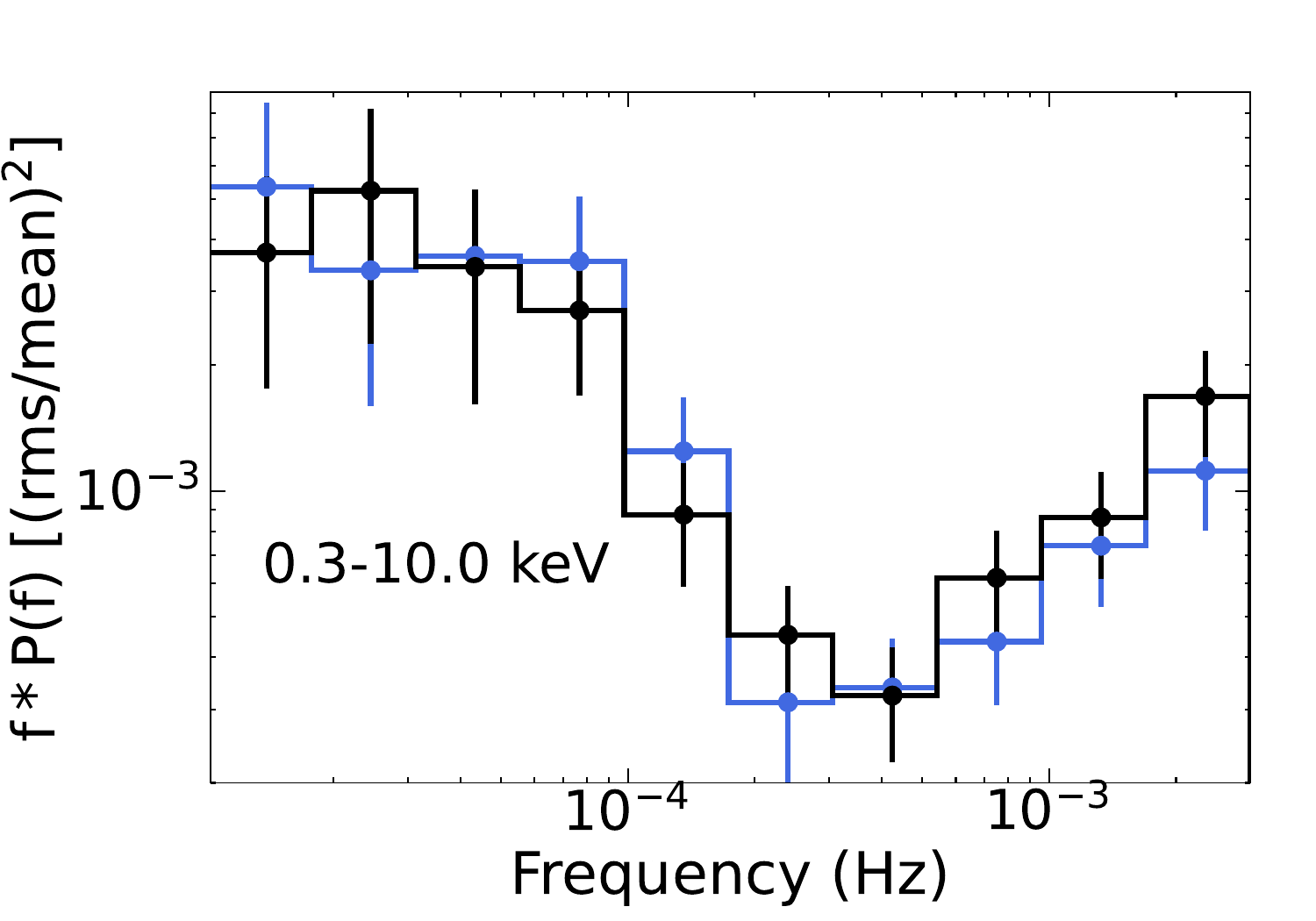}
    \caption{The EPIC-pn ($0.3\mbox{--}10\,keV$) light curves (\textit{top}) and power spectral density (PSD, \textit{bottom}) of PG 1448 (\textit{left}), IRAS 13224 (\textit{middle}), and PG 1211 (\textit{right}). The `low-/high-flux w/wo UFO' observations are marked by the \textit{black} and \textit{royalblue}, respectively.}
    \label{fig:lc_pds}
\end{figure*}

The background-subtracted lightcurves are extracted by the package \textsc{epiclccorr} and are shown in the top panels of Figure \ref{fig:lc_pds}. Observations are classified as the `low-flux w/ UFO' and `high-flux w/o UFO' states, marked in \textit{black} and \textit{royalblue} respectively.

For each AGN with UFOs, the stacked PSD spectra of each classification (in $0.3\mbox{--}10$\,keV energy band) are computed by the \texttt{pylag} package \citep[e.g.,][]{1948Bartlett,1989van.der.Klis,2014Uttley}, and are normalized to units of fractional variance per Hz \citep{1990Belloni}. The minimum frequency is defined by the inverse of the length of the observation (though we do not include the lowest-frequency bin as this is often biased by red noise leakage), and the maximum frequency is set by the frequency at which the PSD becomes dominated by Poisson noise.  We bin the PSD in equal logarithmic frequency bins, and the number of frequency bins depends on the quality of the data (i.e. more bins for higher-quality data). The computed PSD functions of each AGN are shown in the bottom panels of Figure \ref{fig:lc_pds}. Here, the Y-axis is in units of frequency*periodogram to illustrate the region dominated by Poisson noise, where the quantity of the Y-axis increases with frequency.

\section{Supplementary results of targets without a transient UFO}\label{app:sec:wotransient}

The basic information on the selected AGN observations without transient UFOs is listed in Table \ref{tab:woUFO_lag_CL}, including the observation classification, observation IDs, net EPIC-pn exposure time, and the corresponding count rates. Observations in each AGN are classified as `low-flux' and `high-flux' states for the investigation discussed in Section \ref{subsec:causal-relation} that whether the varying low-frequency hard lags are related to the source luminosity.

\begin{table*}[!htbp]
\centering
\caption{\textit{XMM-Newton} observations of three AGN without a transient UFO mentioned in Section \ref{sec:discussion}. }
\begin{tabular}{lcccc}
\hline
\hline
Source & Label & Obs. ID & Exposure & Count Rates\\
  &   &  & (ks) & (cts/s) \\
\hline
\multirow{2}{*}{Ark 564} & Low-flux & 0206400101, 0670130701, 0830540101 & 69, 36, 79 & 37.9, 29.5, 31.4 \\
            & High-flux & 0670130201, 0670130501, 0670130901   & 41, 47, 39 & 61.8, 49.7, 57.1 \\
\hline
\multirow{2}{*}{NGC 7469} & Low-flux  & 0760350401, 0760350501 & 52, 60.0 & 18.7, 19.7\\
            & High-flux  & 0207090101, 0760350201  & 59, 61 & 25.2, 26.1\\ 
\hline
\multirow{2}{*}{Mrk 335} & Low-flux  & 0600540601, 0600540501$^\star$, 0842761301$^\star$, 0842760201, 0842761101 & 97, 69, 76, 74, 74  & 3.1, 4.3, 5.3, 3.2, 3.6 \\
            & High-flux & 0306870101  & 92 & 26.3 \\ 
\hline
\end{tabular}
\tablefoottext{$\star$}{The source spectrum is extracted from an annulus region with an inner radius of 5 arcsec and an outer radius of 30 arcsec. }
\vspace{-2.5mm}
\tablefoot{Similar to Table \ref{tab:obs}. The second column shows the label for stacked observations at two states. The observation IDs are shown in the third column. The corresponding net exposure time and count rate of EPIC-pn are listed in the fourth and fifth columns, respectively.
}
\label{app:tab:obs}
\vspace{-3mm}
\end{table*}

\begin{figure*}[htbp]
    \centering
	\includegraphics[width=0.33\textwidth]{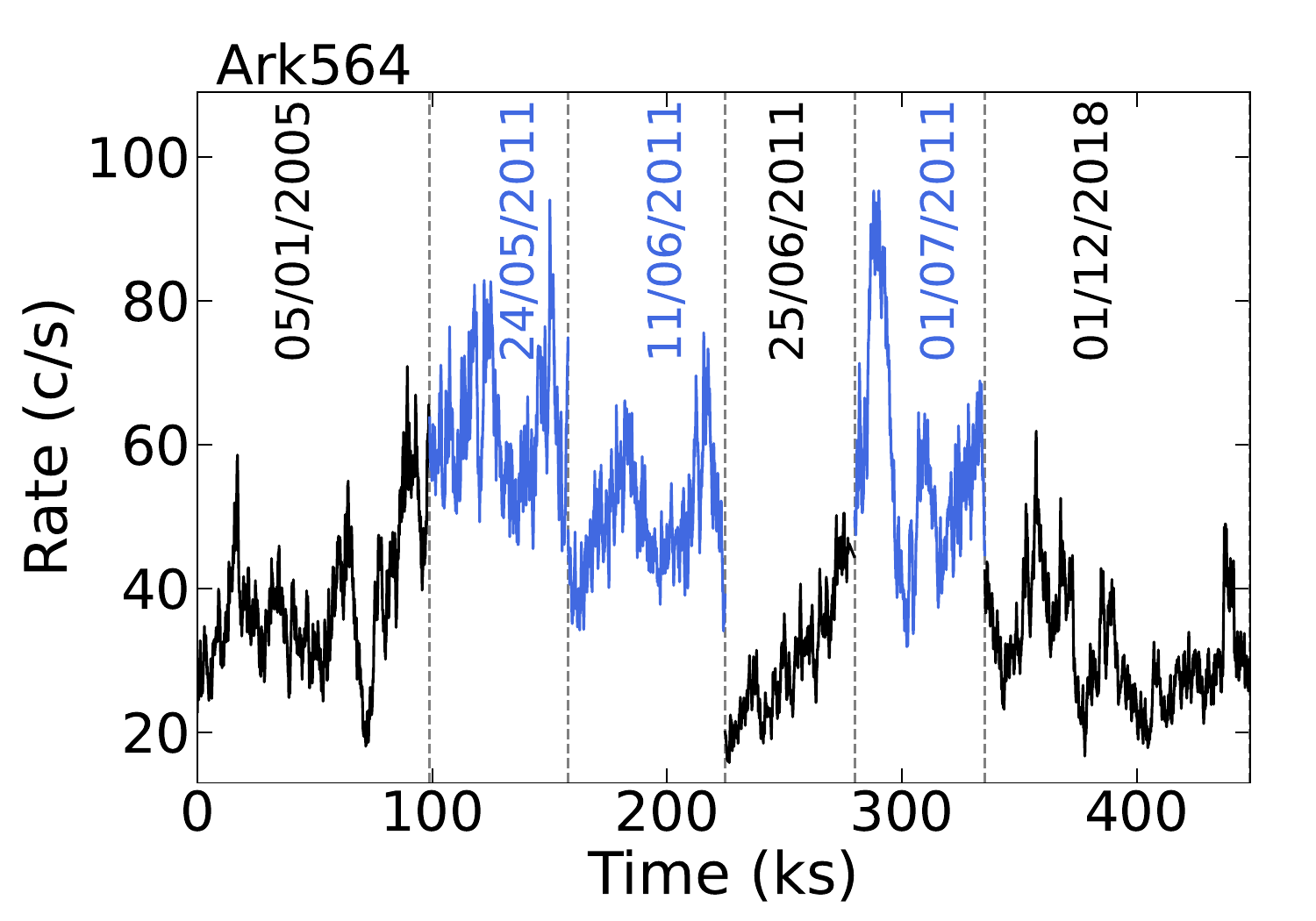}
 	\includegraphics[width=0.33\textwidth]{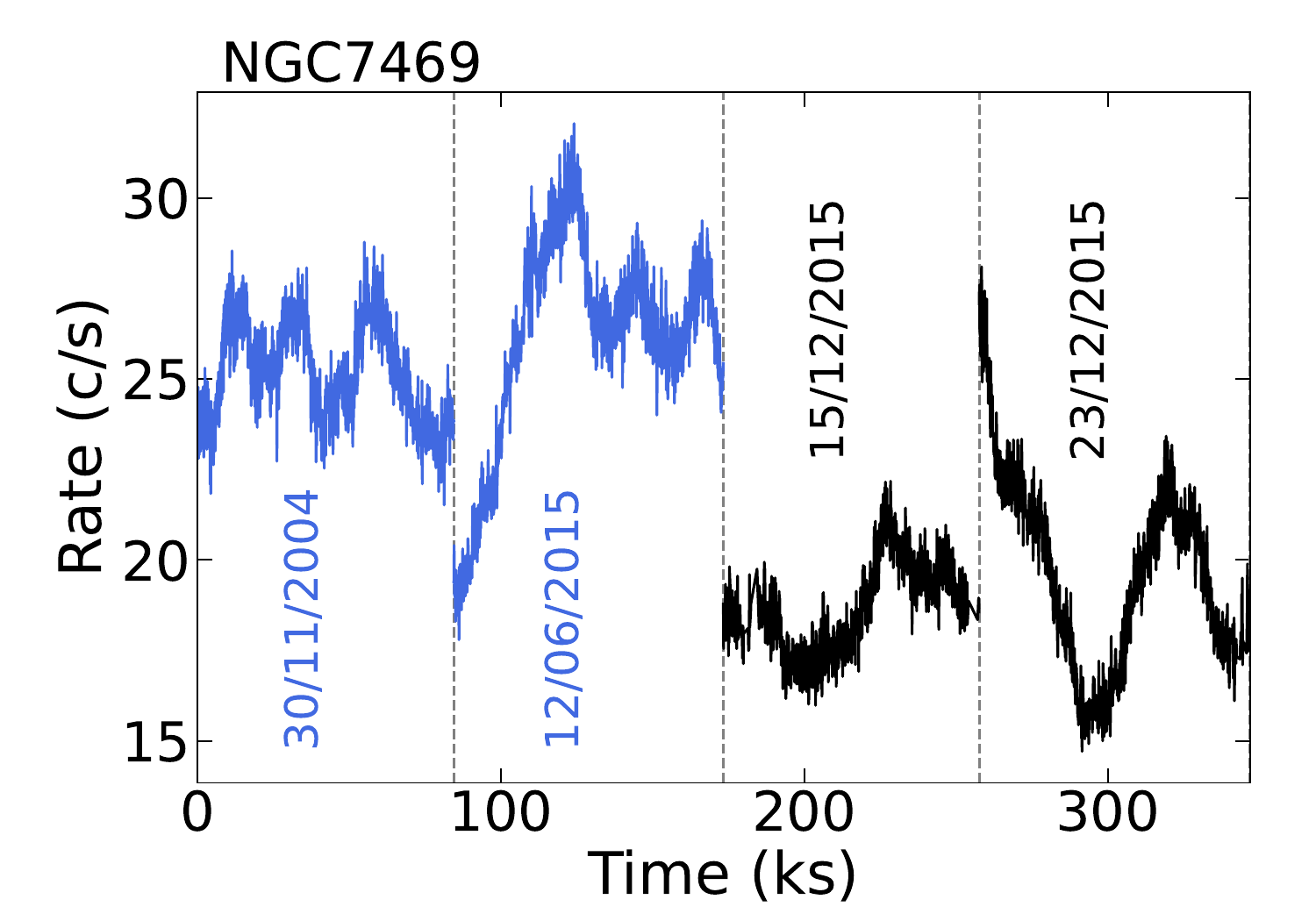}
   	\includegraphics[width=0.33\textwidth]{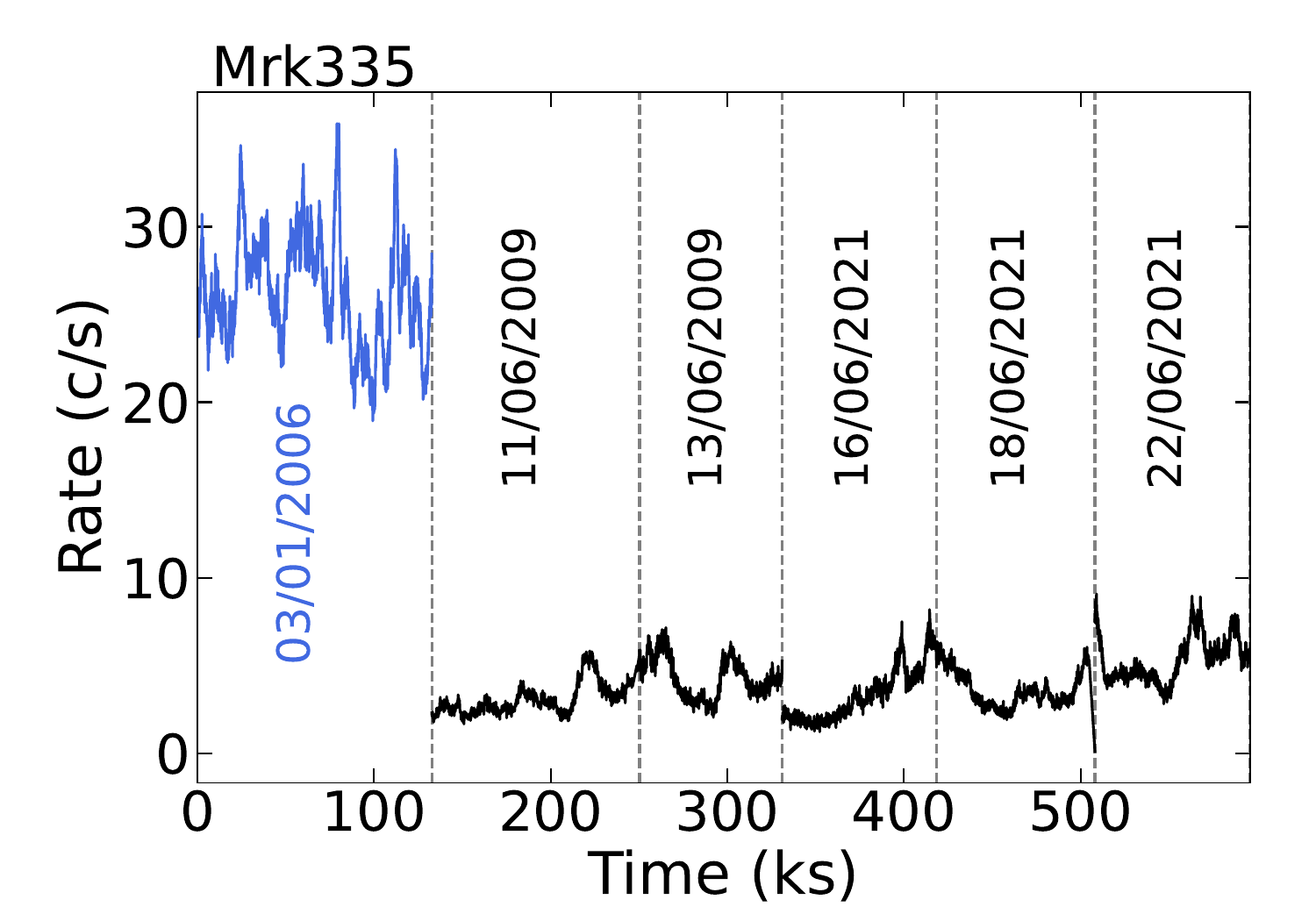}
	\includegraphics[width=0.33\textwidth]{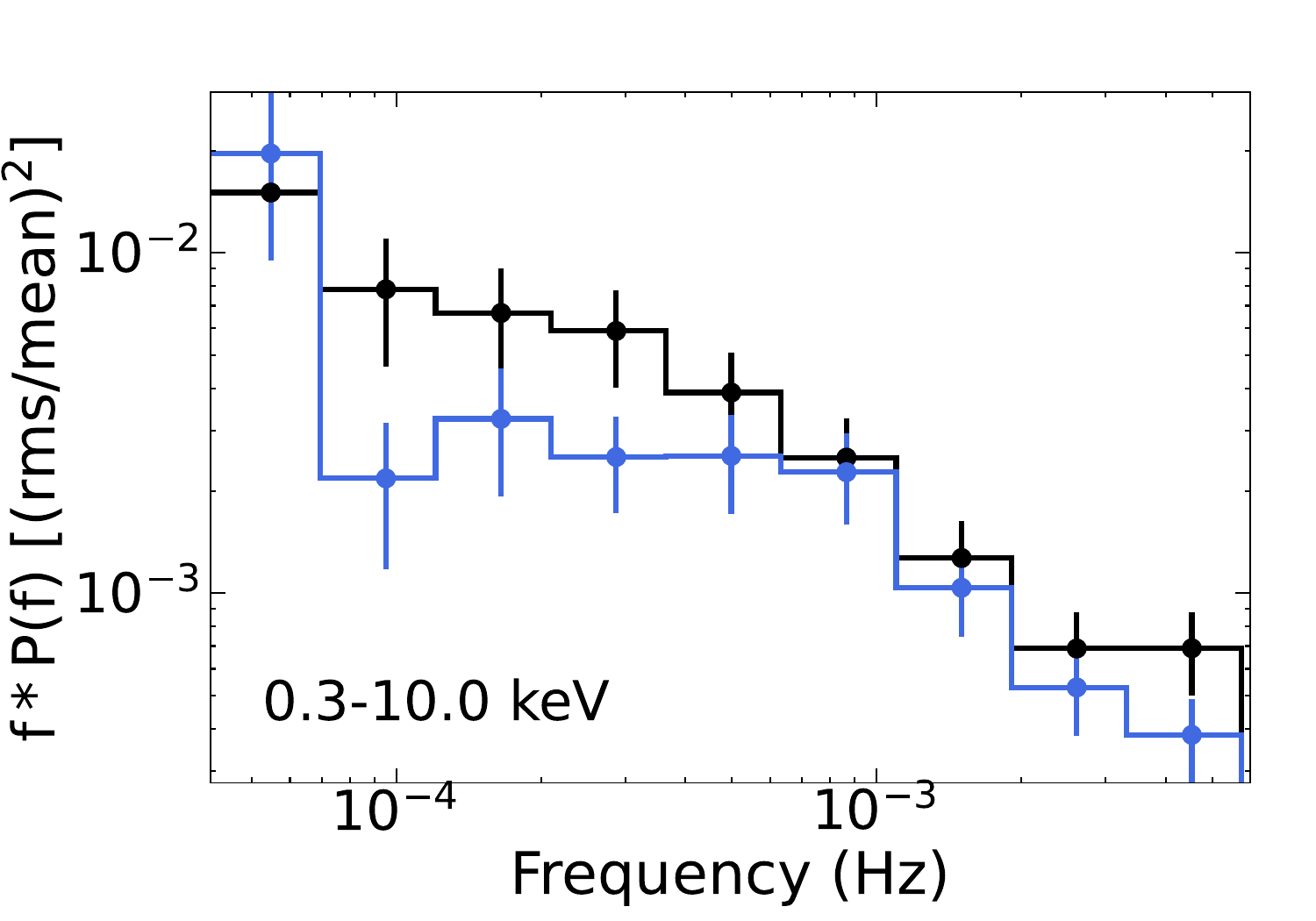}
 	\includegraphics[width=0.33\textwidth]{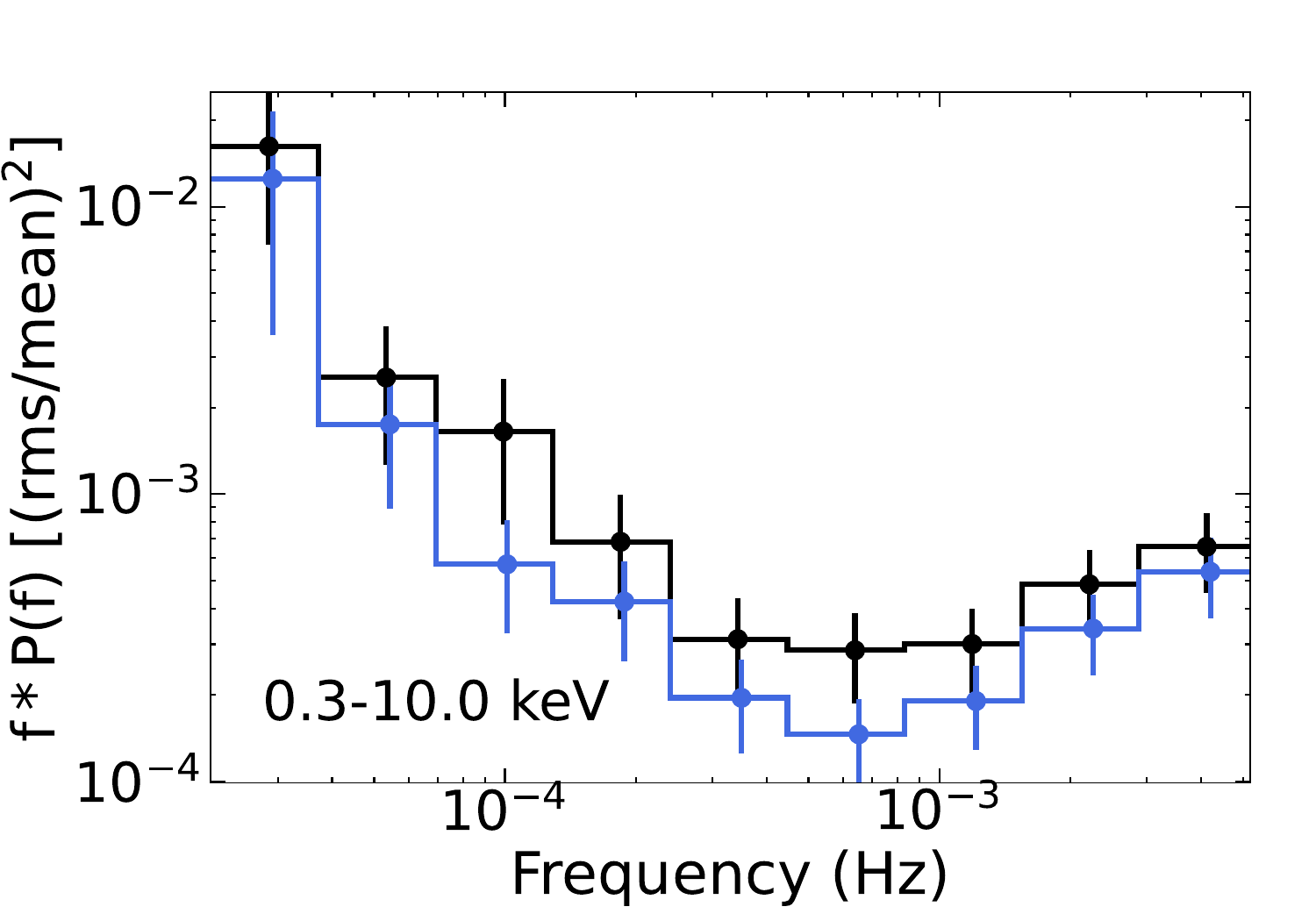}
   	\includegraphics[width=0.33\textwidth]{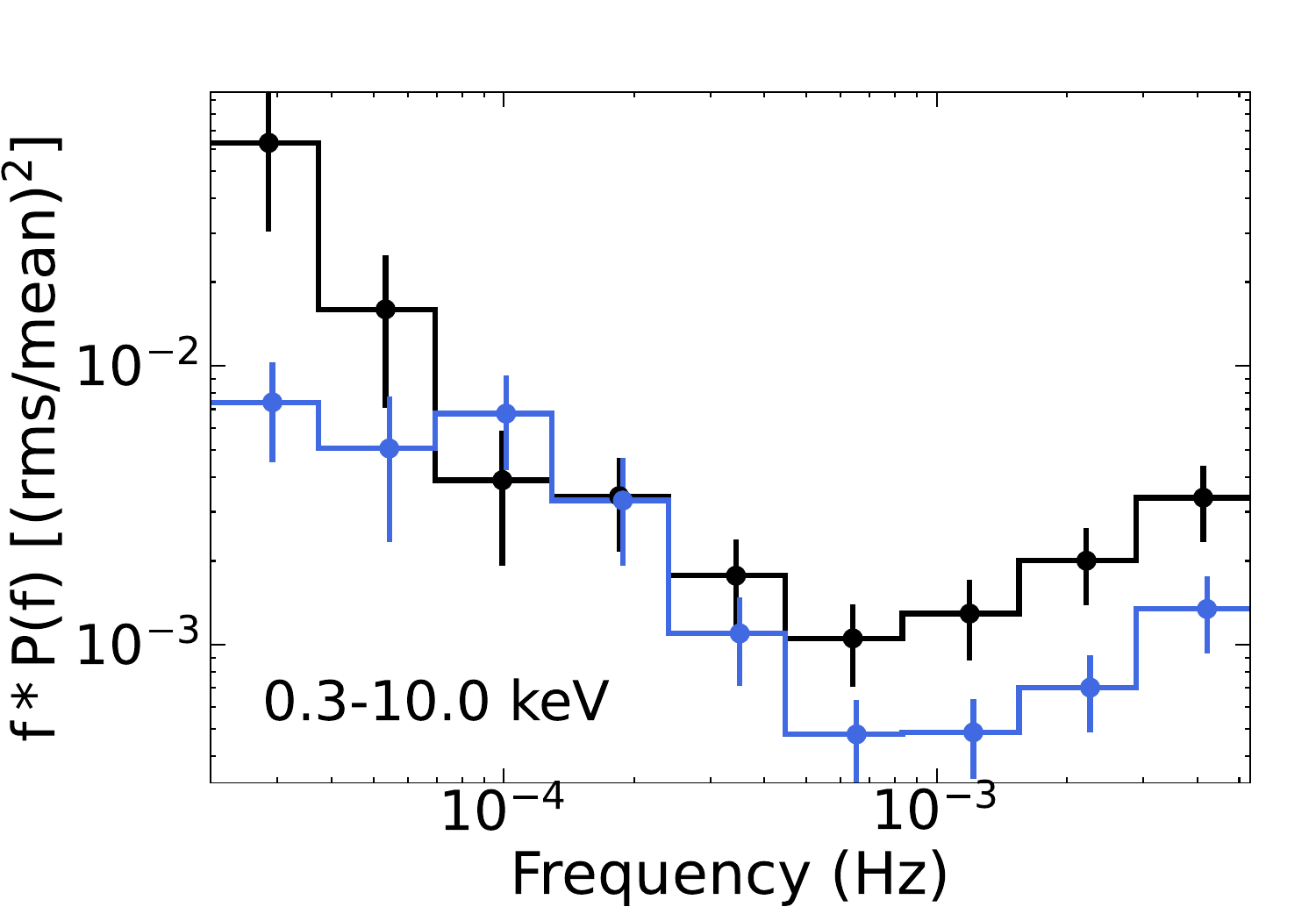}
    \caption{Similar to Figure \ref{fig:lc_pds}. The EPIC-pn light curves (\textit{top}) and power spectral density (PSD, \textit{bottom}) of Ark 564 (\textit{left}), NGC 7469 (\textit{middle}), and Mrk 335 (\textit{right}). The `low-/high-flux' observations are marked by the \textit{black} and \textit{royalblue}, respectively.}
    \label{app:fig:lc_pds}
\end{figure*}

The background-subtracted light curves of AGN without transient UFOs and corresponding periodogram are shown in Figure \ref{app:fig:lc_pds}, similar to Figure \ref{fig:lc_pds}. The frequencies at which the Poisson noise becomes dominant can be obtained from our PSD spectra.

\begin{figure*}[htbp]
    \centering
	\includegraphics[width=0.24\textwidth]{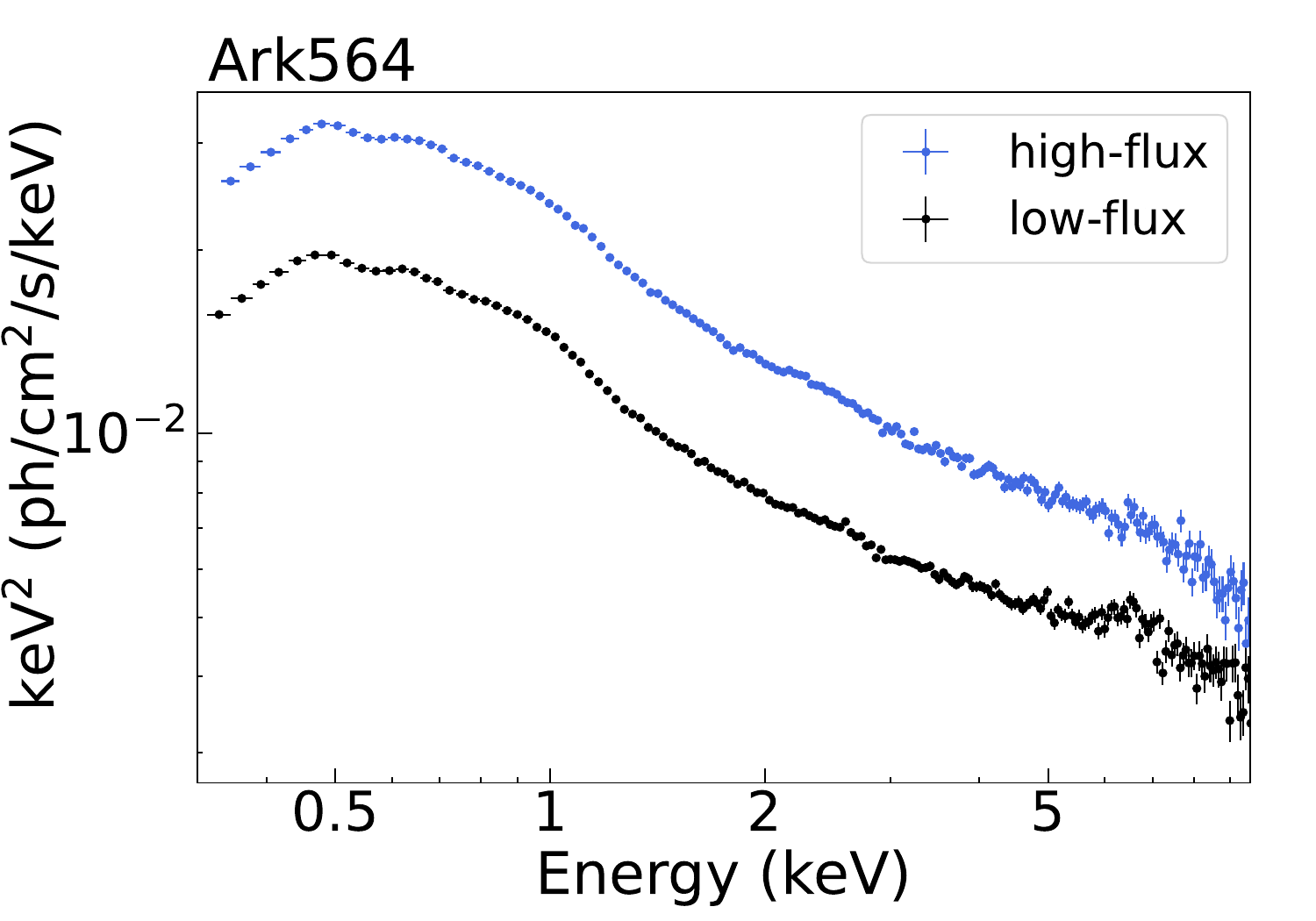}
	\includegraphics[width=0.24\textwidth]{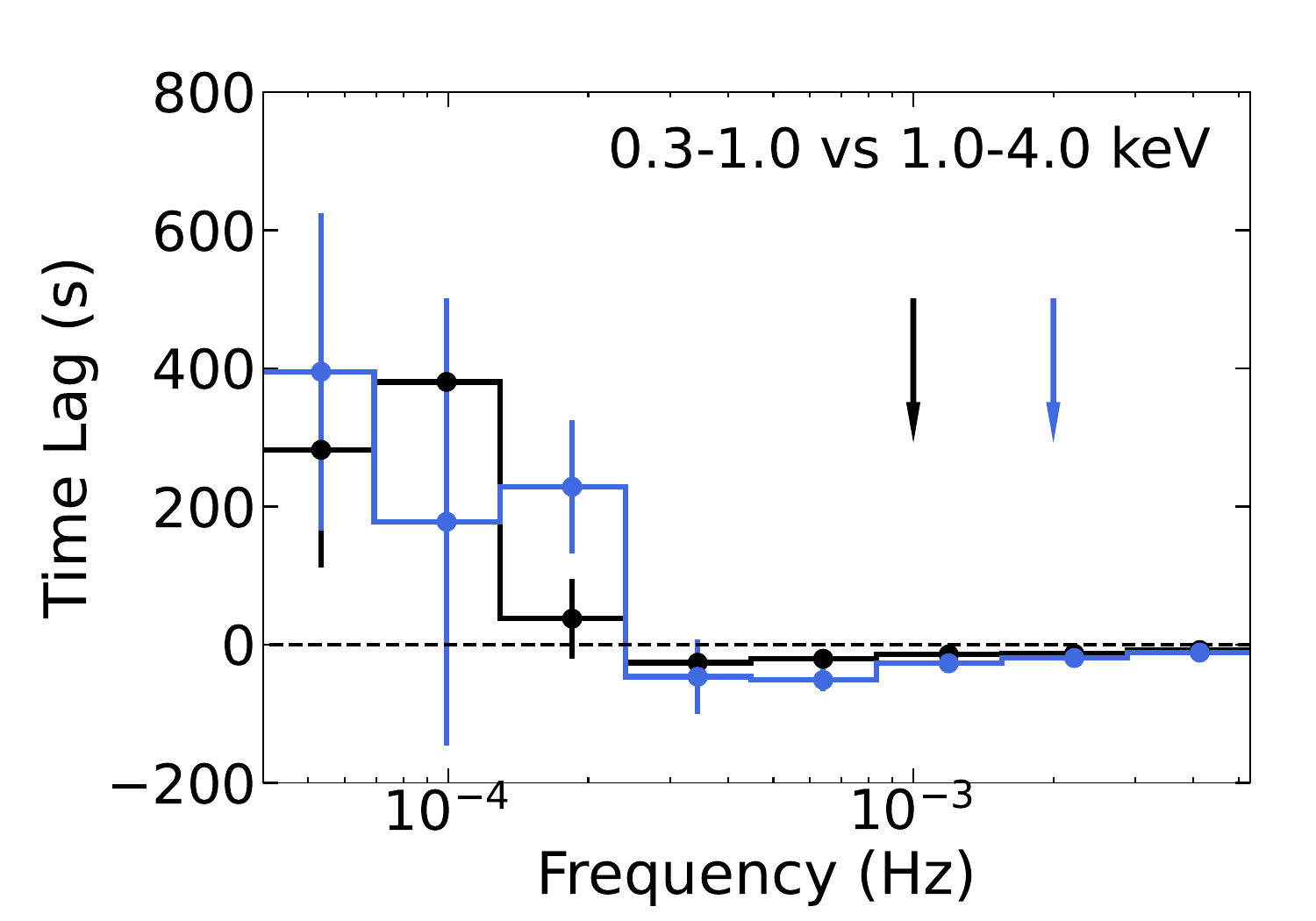}
	\includegraphics[width=0.24\textwidth]{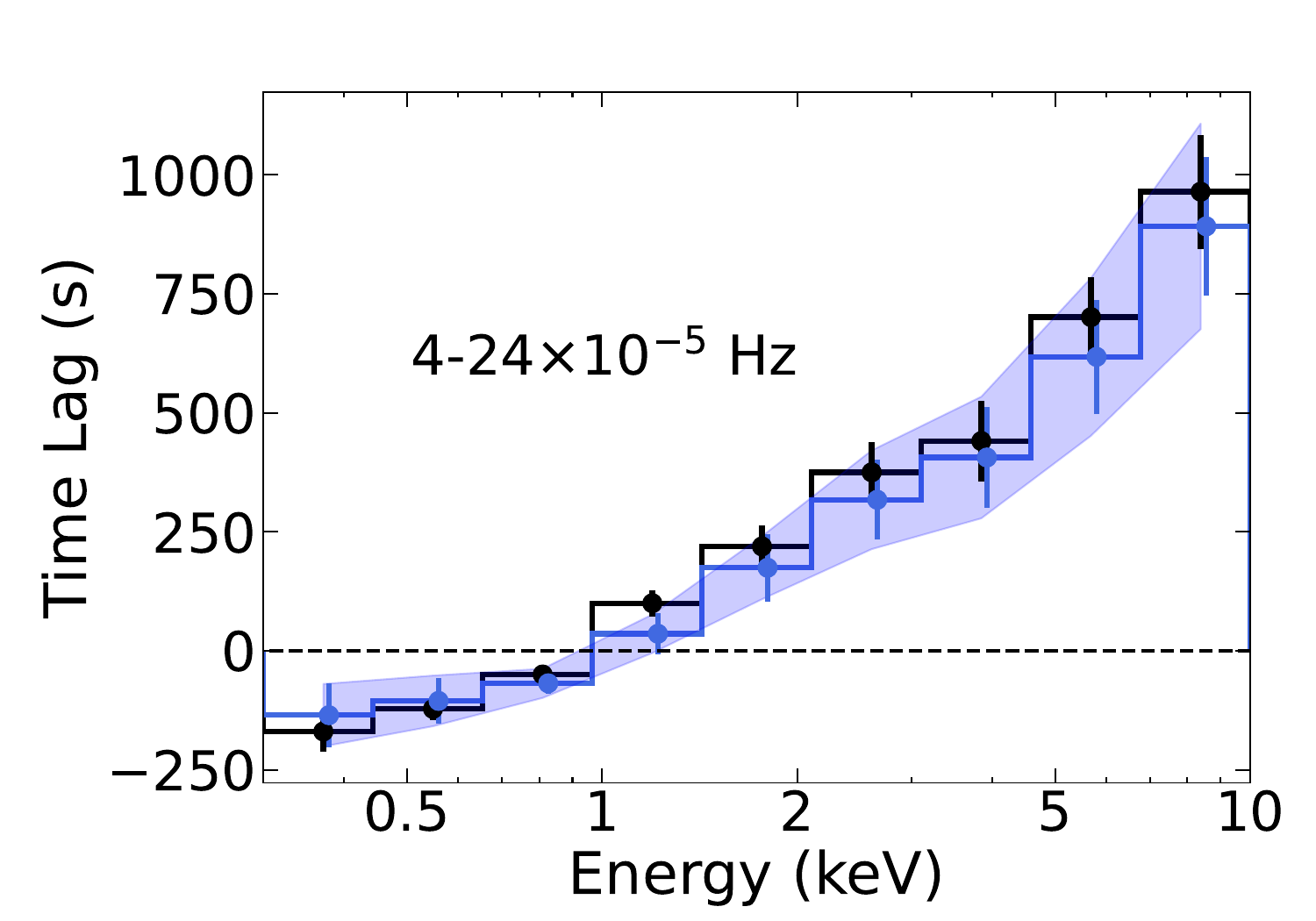}
	\includegraphics[width=0.24\textwidth]{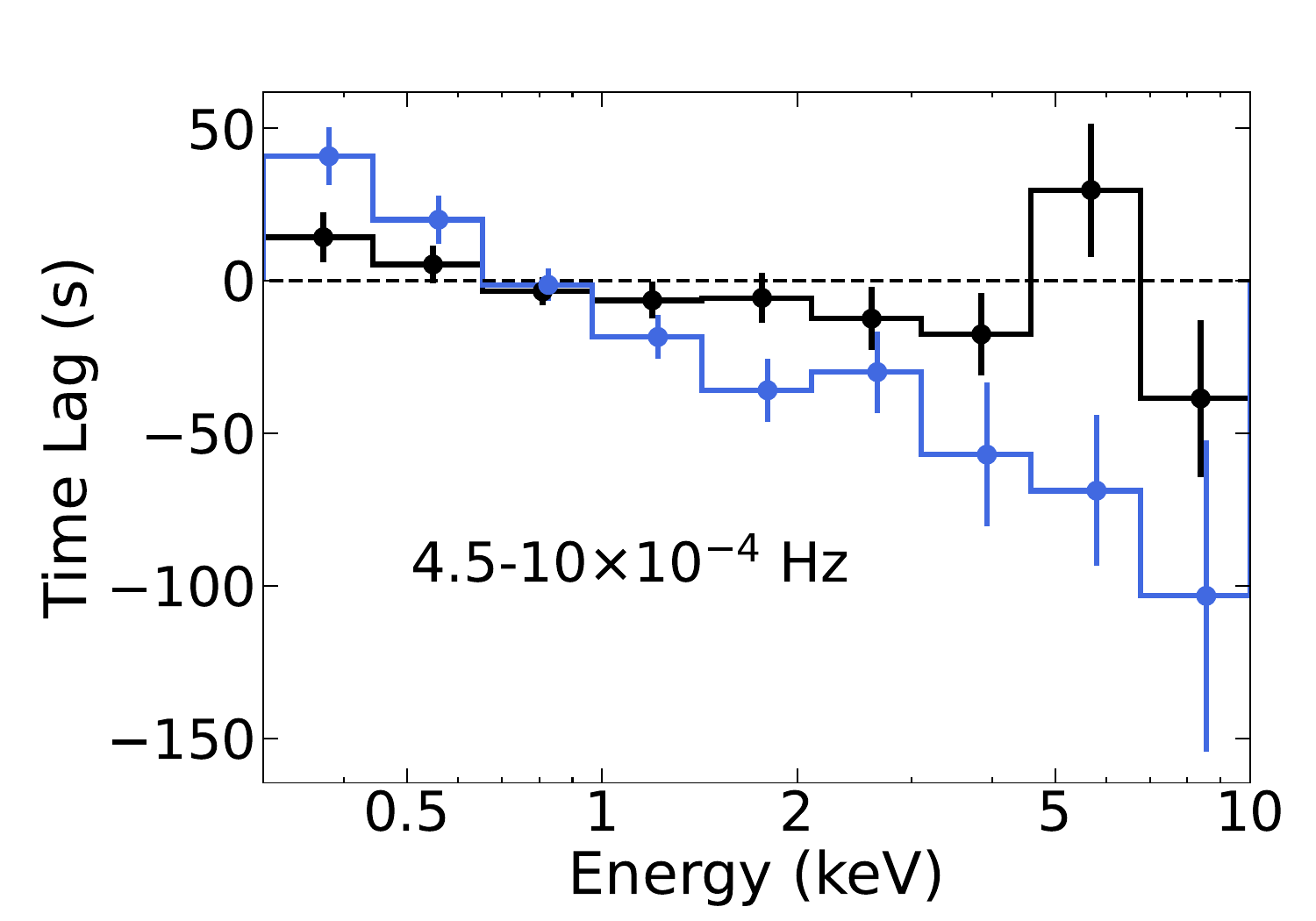}
	\includegraphics[width=0.24\textwidth]{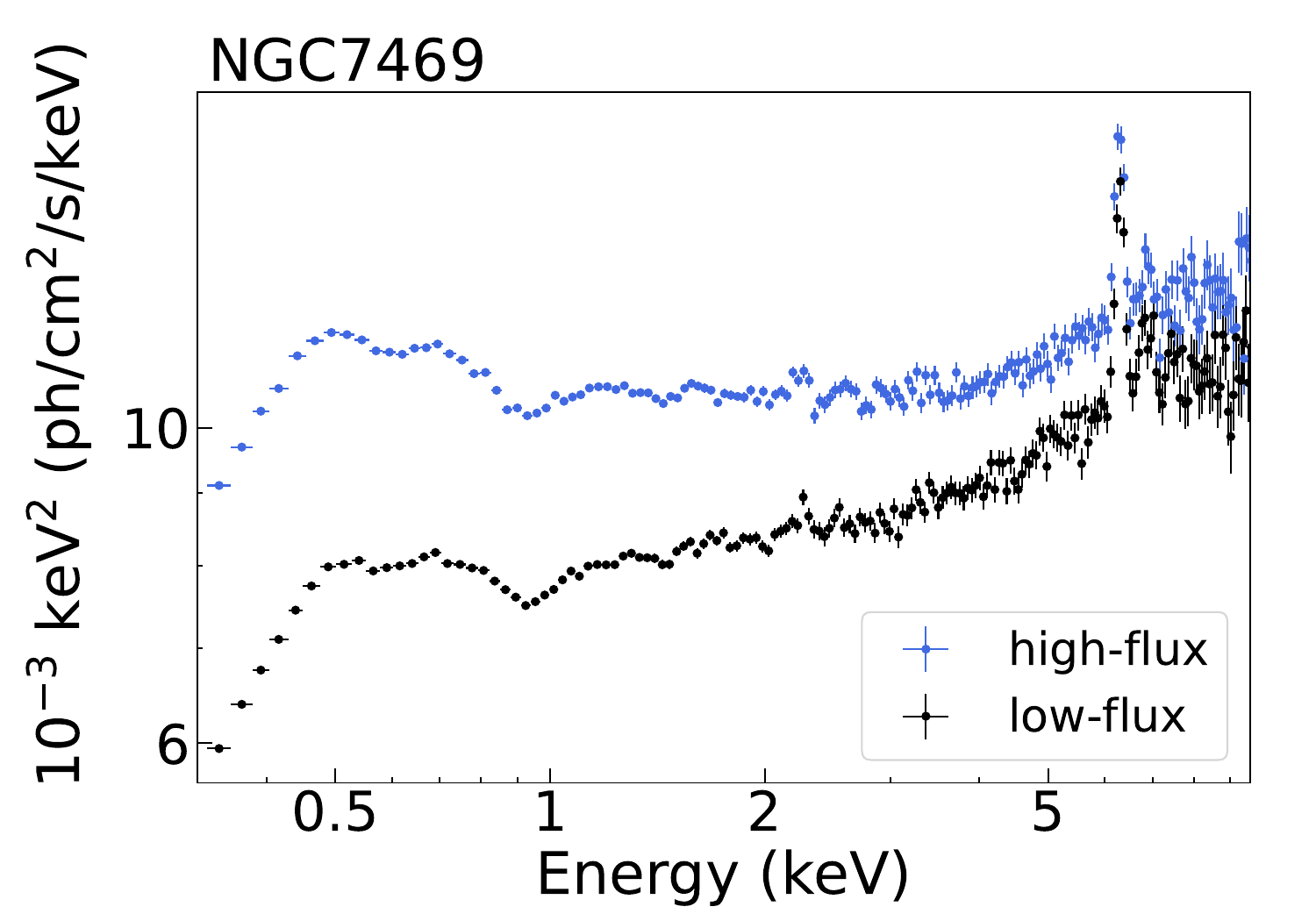}
	\includegraphics[width=0.24\textwidth]{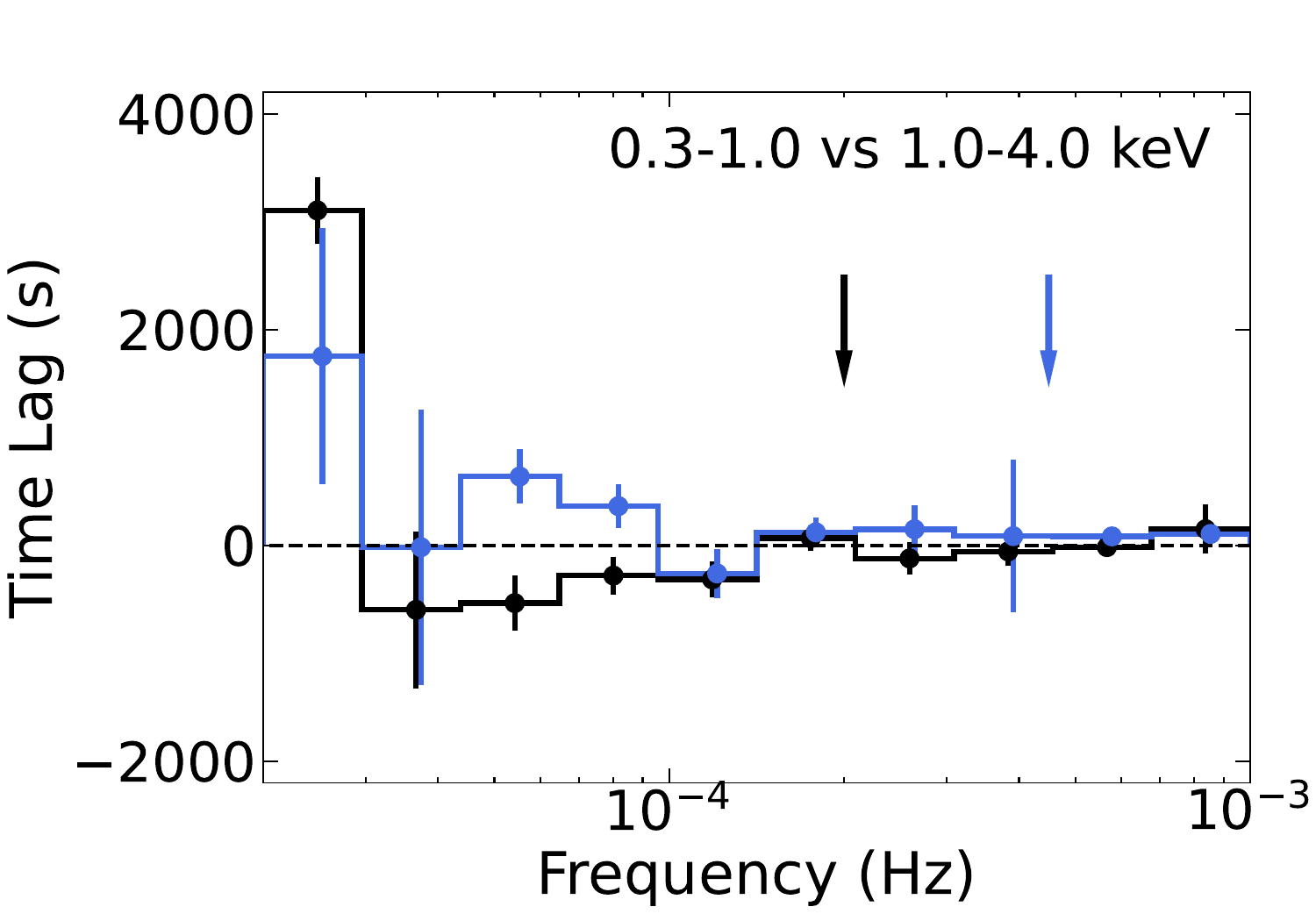}
	\includegraphics[width=0.24\textwidth]{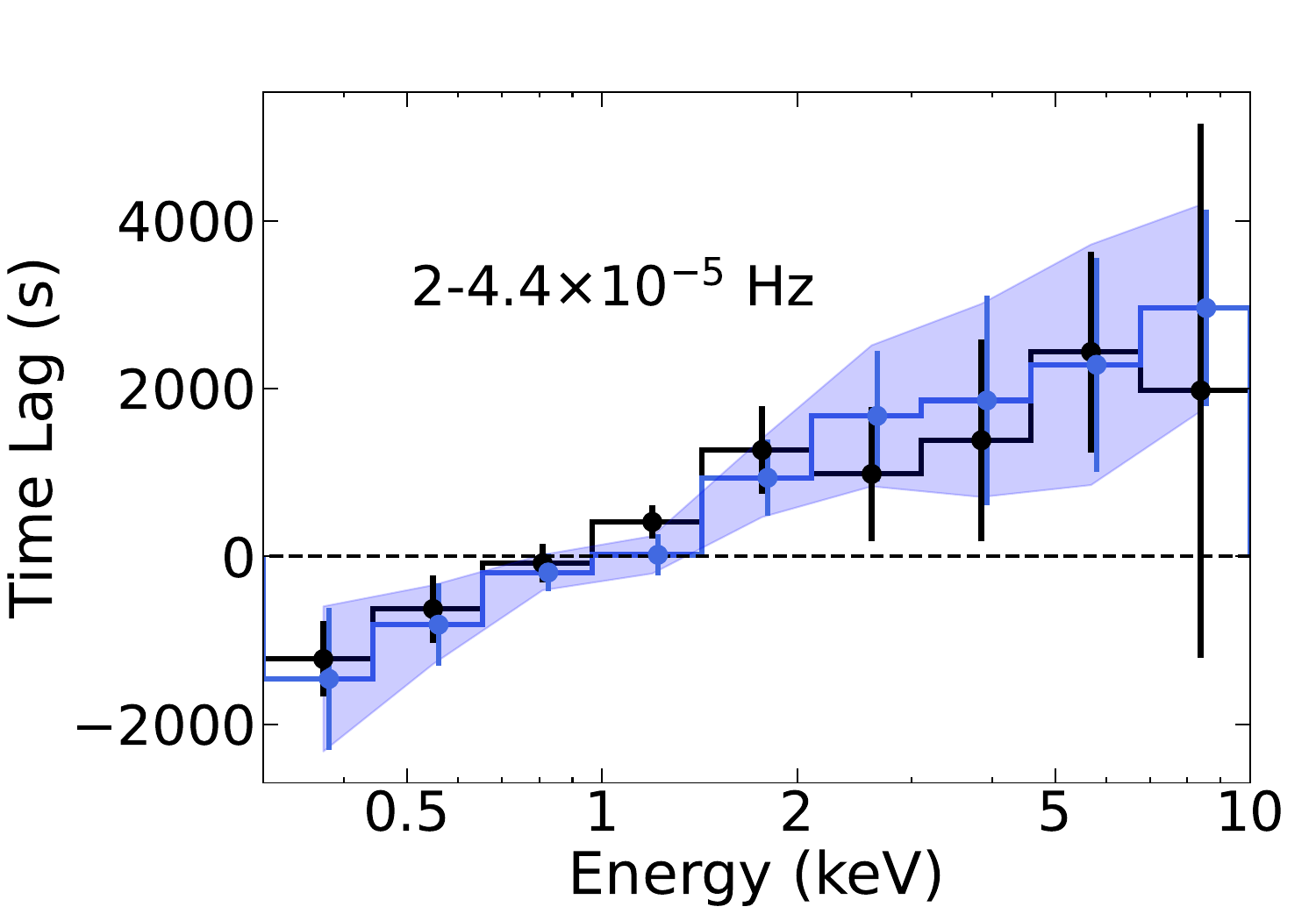}
	\includegraphics[width=0.24\textwidth]{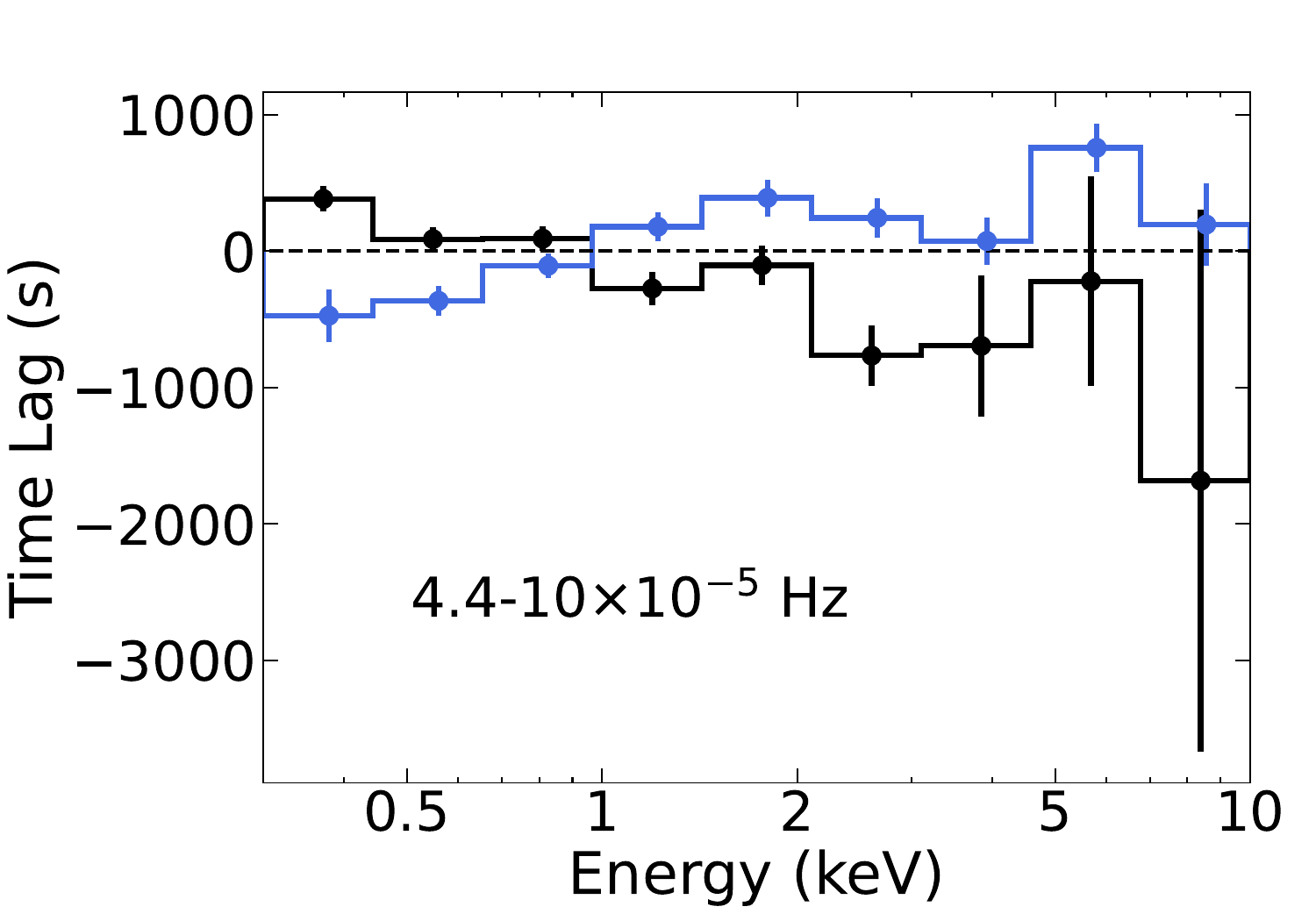}
   	\includegraphics[width=0.24\textwidth]{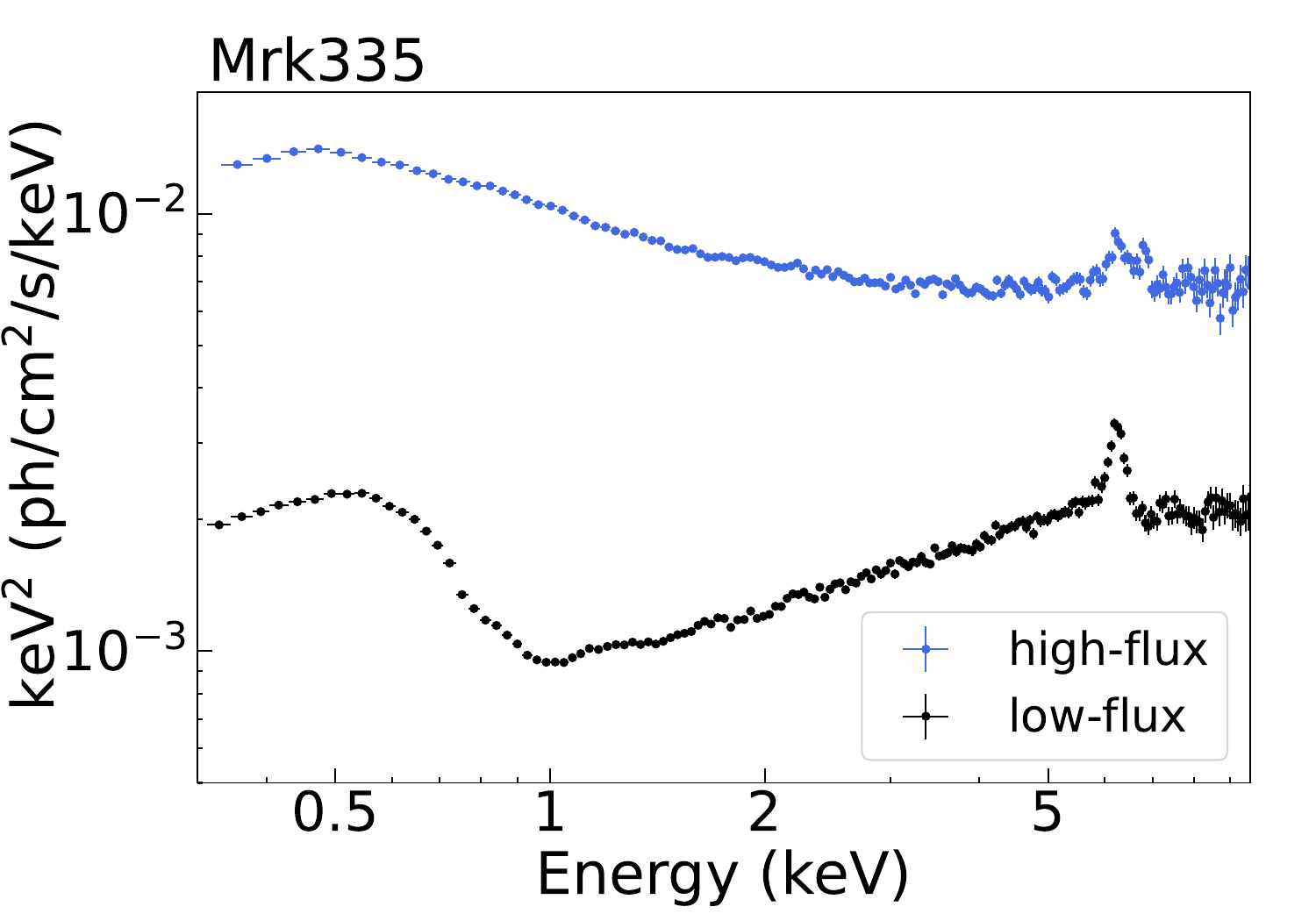}
   	\includegraphics[width=0.24\textwidth]{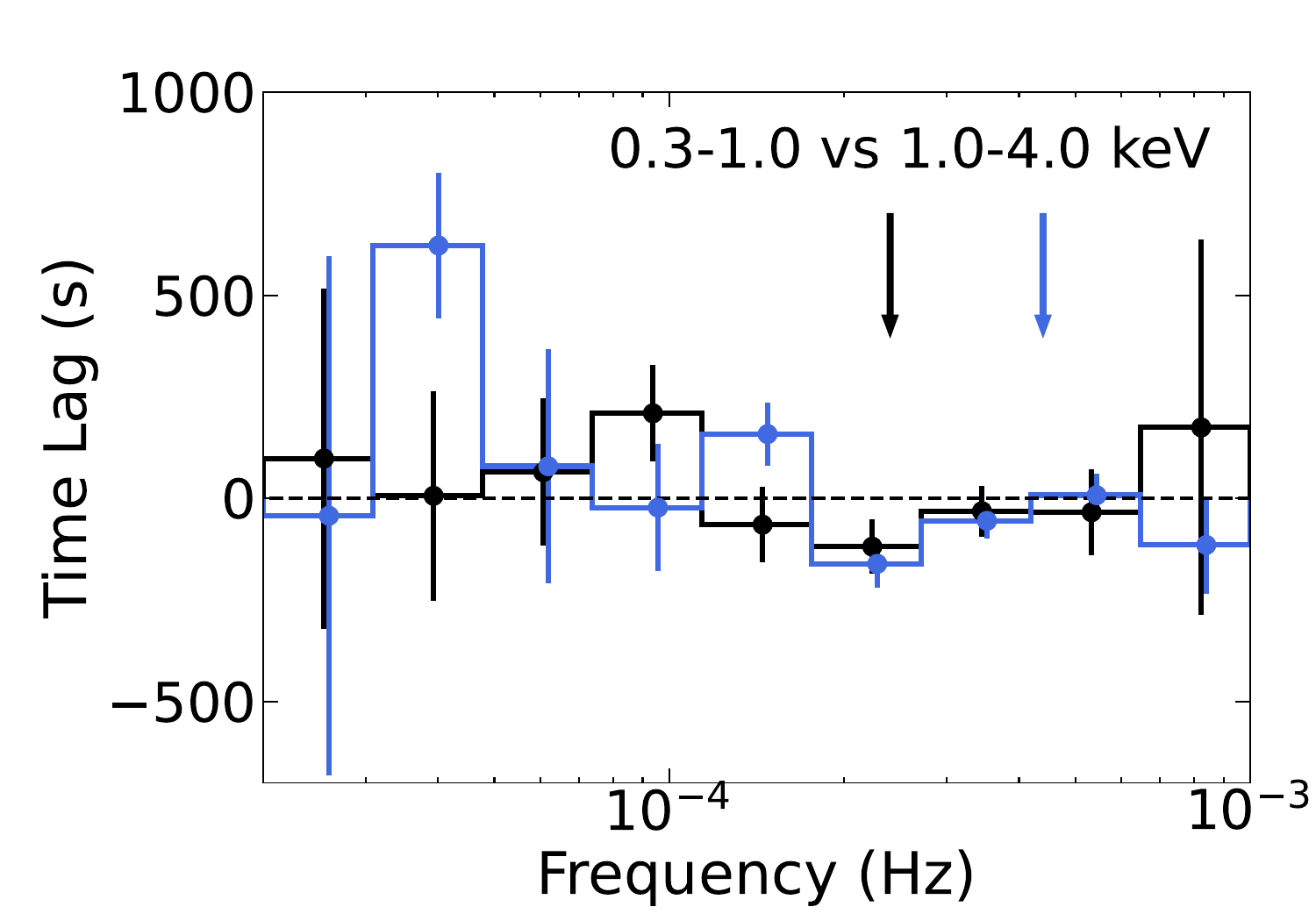}
   	\includegraphics[width=0.24\textwidth]{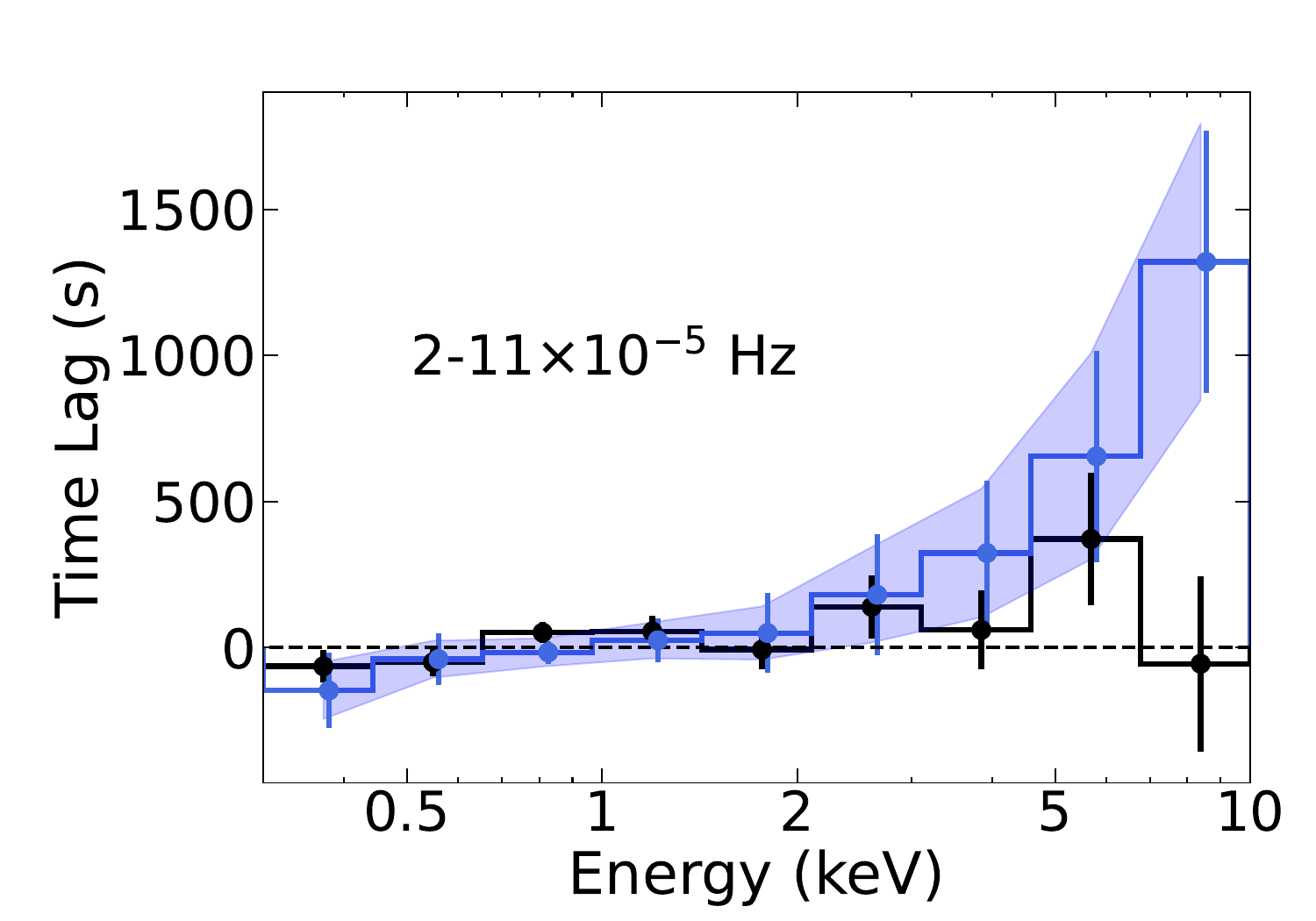}
   	\includegraphics[width=0.24\textwidth]{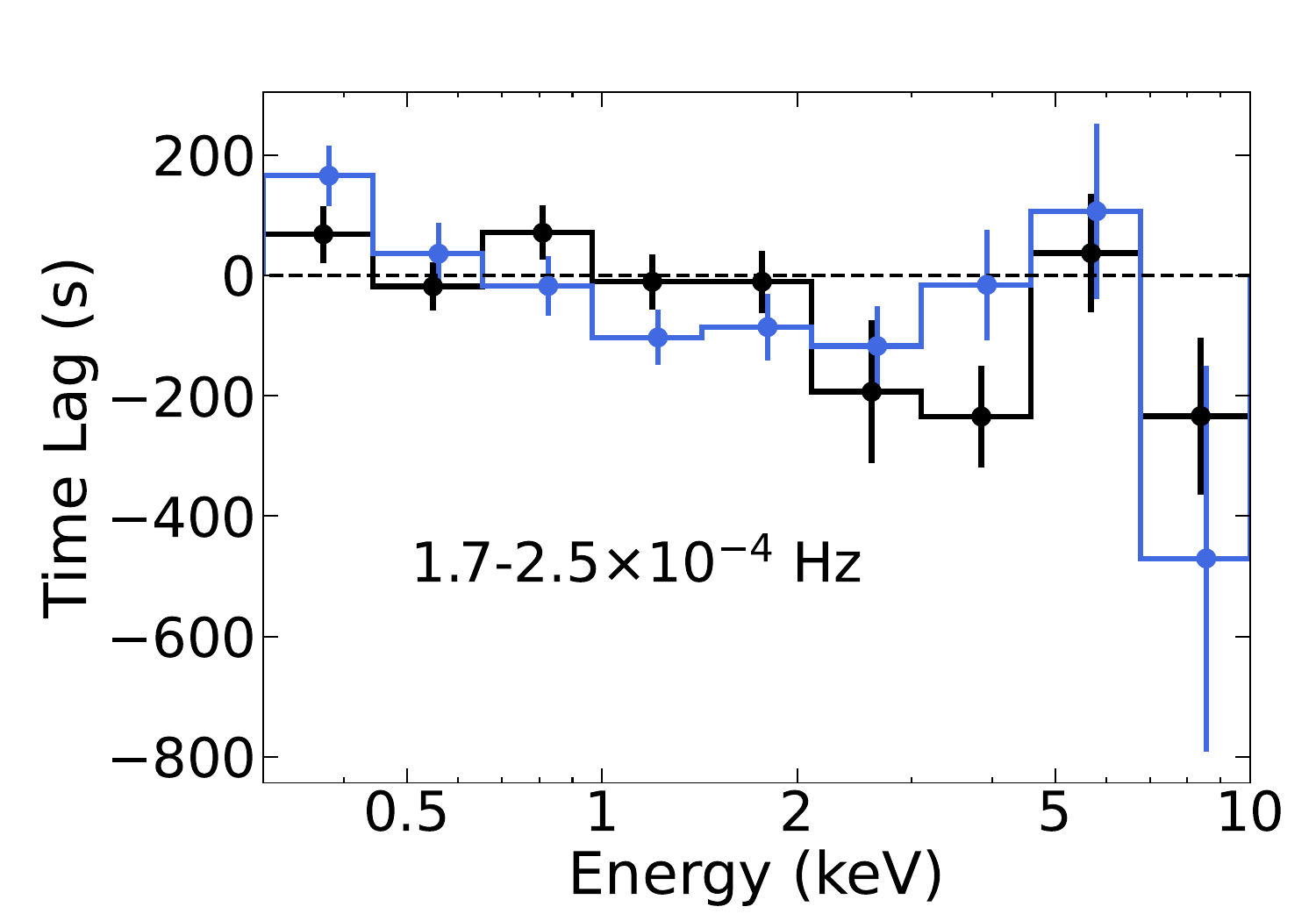}
    \caption{Similar to Figure \ref{fig:lags}. The X-ray time lags for Ark 564 (\textit{top}), NGC 7469 (\textit{middle}), and Mrk 335 (\textit{bottom}).}
    \label{app:fig:lags}
\end{figure*}

For each AGN, the stacked flux energy spectra of observations at different flux states are shown in the first column of Figure \ref{app:fig:lags}. The spectral variations are different among these AGN. The spectral variation of Ark 564 appears in the entire EPIC band without significant changes in the spectral slope, while that of NGC 7469 only occurs in the soft X-ray band, revealing that the soft excess mainly contributes to its spectral and flux variations. In Mrk 335, changes take place in both the spectral slope and spectral normalization, indicating that the variability originates from both broadband continuum variations and ionized absorbers \citep{2015Gallo,2020Komossa}.

The stacked flux-energy, lag-frequency, and lag-energy spectra are shown in Figure \ref{app:fig:lags}, while the low-frequency lag-energy spectra are discussed in detail in Section \ref{subsec:lag-energy}. The lag-frequency spectra of stacked observations at different flux states are generally consistent within uncertainties, with the exception that lags of NGC 7469 flip in the $4.4\mbox{--}10\times10^{-5}$\,Hz band, illustrated in its high-frequency lag-energy spectrum. In the high-frequency lag-energy spectra, the Fe K reverberation lag occurs in most cases regardless of the brightness, except for the high-flux state of Ark 564. We emphasize that the lags in Figure \ref{app:fig:lags} are the stacked results, and the disappeared Fe K lags do not mean Ark 564 is special since some Fe K reverberation results are from specific observations, and not from all available observations \citep{2016Kara}. For example, when we calculate the lag-energy spectra within the same frequency band for individual high-flux observations, the Fe K lag only occurs in the observation performed on 01/07/2011, and it thus disappears in the averaged lags. The lags in the soft X-ray band are consistent with the lag-frequency spectra, where soft X-ray lags appear in nearly all spectra except for the high-flux spectrum of NGC 7469, the origin of which is beyond the scope of this work. 

\end{appendix}

\end{document}